\pdfoutput=1
\documentclass[preprint,12pt,authoryear]{elsarticle}
\usepackage{graphicx}
\usepackage{natbib}
\usepackage{hyperref}
\usepackage{url} % not crucial - just used below for the URL
\usepackage{tabularx}
\usepackage{listings}
\usepackage{xcolor}

\definecolor{codegreen}{rgb}{0,0.6,0}
\definecolor{codegray}{rgb}{0.5,0.5,0.5}
\definecolor{codepurple}{rgb}{0.58,0,0.82}
\definecolor{backcolour}{rgb}{0.95,0.95,0.92}

\lstdefinestyle{mystyle}{
    backgroundcolor=\color{backcolour},
    commentstyle=\color{codegreen},
    keywordstyle=\color{magenta},
    numberstyle=\tiny\color{codegray},
    stringstyle=\color{codepurple},
    basicstyle=\ttfamily\scriptsize,
    breakatwhitespace=false,
    breaklines=true,
    captionpos=b,
    keepspaces=true,
    numbers=left,
    numbersep=5pt,
    showspaces=false,
    showstringspaces=false,
    showtabs=false,
    tabsize=2
  }
\usepackage{subfig} % used to get side by side figues with 1 figure #
\usepackage{bm} % used to bold mathcal{H}
\usepackage{bbm} % used to bold mathcal{1}
\usepackage{color} % needed for \hl in soul
\usepackage{soul} % highlighting with \hl{}
\soulregister\cite7 % \hl won't break on cite
\soulregister\ref7  % \hl won't break on ref

\def\bs{\boldsymbol}

\usepackage{multirow}
\usepackage{hhline}
\def \N{\mbox{N}}

\def \E{\mbox{E}}
\def \bident{\mbox{\boldmath{I}}}

\def \s2{\sigma^2}
\def \hs2{\hat{\sigma}^2}
\def \siga2{\sigma_{\alpha}^2}
\def \sige2{\sigma_{\epsilon}^2}
\def \sig2{\sigma^2}

\newcommand{\bA}{\mbox{\boldmath $A$}}

\newcommand{\bmB}{\mbox{\boldmath $B$}}

\newcommand{\bmb}{\mbox{\boldmath $b$}}
\newcommand{\bb}{\mbox{\boldmath $b$}}

\newcommand{\bmd}{\mbox{\boldmath $d$}}
\newcommand{\bd}{\mbox{\boldmath $d$}}

\newcommand{\bme}{\mbox{\boldmath $e$}}

\newcommand{\bJ}{\mbox{\boldmath $J$}}

\newcommand{\bmQ}{\mbox{\boldmath $Q$}}

\newcommand{\bms}{\mbox{\boldmath $s$}}

\newcommand{\bmu}{\mbox{\boldmath $u$}}

\newcommand{\bmv}{\mbox{\boldmath $v$}}

\newcommand{\bmw}{\mbox{\boldmath $w$}}

\newcommand{\bmy}{\mbox{\boldmath $y$}}

\newcommand{\bmz}{\mbox{\boldmath $z$}}

\newcommand{\beq}{\begin{equation}}
\newcommand{\eeq}{\end{equation}}
\newcommand{\beqn}{\begin{eqnarray}}
\newcommand{\eeqn}{\end{eqnarray}}

\newcommand{\bmzero}{\mbox{\bf 0}}

\newcommand{\bbeta}{\mbox{\boldmath $\beta$}}

\newcommand{\btheta}{\mbox{\boldmath $\theta$}}

\newcommand{\bmphi}{\mbox{\boldmath $\phi$}}
\newcommand{\bphi}{\mbox{\boldmath $\phi$}}

\newcommand{\bmeta}{\mbox{\boldmath $\eta$}}

\newcommand{\bepsilon}{\mbox{\boldmath $\epsilon$}}

\newcommand{\bSigma}{\mbox{\boldmath $\Sigma$}}

\newcommand*\dif{\mathop{}\!\mathrm{d}}

\hypersetup{
    colorlinks,
    linkcolor={red!50!black},
    citecolor={blue!50!black},
    urlcolor={blue!80!black}
}

\usepackage[english]{babel}
\addto\captionsenglish{}

\newcommand{\singlespace}{\renewcommand{\baselinestretch}{1.0}}

\usepackage{amssymb}
\usepackage{amsmath}
\usepackage{mathtools}

\journal{}

\begin{document}

\begin{frontmatter}


%% use the tnoteref command within \title for footnotes; %% use the tnotetext command for theassociated footnote; %% use the fnref command within \author or \affiliation for footnotes; %% use the fntext command for theassociated footnote; %% use the corref command within \author for corresponding author footnotes; %% use the cortext command for theassociated footnote; %% use the ead command for the email address, %% and the form \ead[url] for the home page: %% \title{Title\tnoteref{label1}} %% \tnotetext[label1]{} %% \author{Name\corref{cor1}\fnref{label2}} %% \ead{email address} %% \ead[url]{home page} %% \fntext[label2]{} %% \cortext[cor1]{} %% \affiliation{organization={}, %% addressline={}, %% city={}, %% postcode={}, %% state={}, %% country={}} %% \fntext[label3]{}

\title{A Statistical Introduction to Template Model Builder: A Flexible Tool for Spatial Modeling}

%% use optional labels to link authors explicitly to addresses: %% \author[label1,label2]{} %% \affiliation[label1]{organization={}, %% addressline={}, %% city={}, %% postcode={}, %% state={}, %% country={}}
%%
%% \affiliation[label2]{organization={}, %% addressline={}, %% city={}, %% postcode={}, %% state={}, %% country={}}

\author[stat]{Aaron Osgood-Zimmerman\corref{cor}}
\cortext[cor]{Email: azimmer@uw.edu}

\author[stat,bios]{Jon Wakefield}

\address[stat]{Department of Statistics}
\address[bios]{Department of Biostatistics}

\address{University of Washington, Seattle}

%\affiliation[label1]{organization={University of Washington}, city={Seattle}, state={Washington}}

\begin{abstract} %% Text of abstract
  The integrated nested Laplace approximation (INLA) is a well-known and popular technique for spatial modeling with a user-friendly interface in the R-INLA package. Unfortunately, only a certain class of latent Gaussian models are amenable to fitting with INLA.  In this paper we describe Template Model Builder (TMB), an existing technique which is well-suited to fitting complex spatio-temporal models. TMB is relatively unknown to the spatial statistics community, but is a highly flexible random effects modeling tool which allows users to define complex random effects models through simple {\tt C++} templates. After contrasting the methodology behind TMB with INLA, we provide a large-scale simulation study assessing and comparing R-INLA and TMB for continuous spatial models, fitted via the Stochastic Partial Differential Equations (SPDE) approximation. The results show that the predictive fields from both methods are comparable in most situations even though TMB estimates for fixed or random effects may have slightly larger bias than R-INLA. We also present a smaller discrete spatial simulation study, in which both approaches perform well. We conclude with an analysis of breast cancer incidence and mortality data using a joint model which cannot be fit with INLA.
\end{abstract}

%\begin{graphicalabstract} %%\includegraphics{grabs}

\begin{highlights}
\item In-depth statistical review of Template Model Builder (TMB) methodology
\item Large-scale continuous spatial simulation study assessing the Stochastic Partial Differential Equations representation of Gaussian Processes in TMB and R-INLA
\item Discrete spatial simulation with conditionally constrained spatial distributions in TMB
\item Novel nonlinear spatial model to jointly estimate cancer incidence and mortality in the European Union
\item Provides example code for fitting popular continuous and discrete spatial models in TMB
\end{highlights}

\begin{keyword} Bayesian inference; Empirical Bayes; Laplace approximation; Stochastic Partial Differential Equations; Hierarchical models; Gaussian Markov random fields


%% MSC codes here, in the form: \MSC code \sep code %% or \MSC[2008] code \sep code (2000 is the default)

\end{keyword}

\end{frontmatter}


\section{Introduction}\label{sec:intro}

Inference for spatial random effects models is notoriously difficult, due to the dimensionality of the parameter space and the dependencies in the likelihood surface. Inference can proceed via either Markov chain Monte Carlo \citep{margossian:etal:20} or analytic approximation methods, including variational methods \citep{blei:etal:17} and those based on the Laplace approximation (LA). In this paper, we focus on the latter, since they are fast and are widely-used in practice. In particular, we consider the integrated nested Laplace approximation (INLA) method implemented in the R-INLA package (\cite{rue2009}, \cite{martins2013}) and the LA within the Template Model Builder (TMB) package (\cite{kristensen2016}).

Since its introduction, INLA has increasingly grown in popularity and is now the method of choice for many spatial and spatio-temporal analyses. The method is very well-documented (\cite{rue2009}, \cite{blangiardo2013}, \cite{lindgren2015}, \cite{rue2017}, \cite{martino2020}) and its popularity is evidenced by a number of book-length treatments specifically for spatial data (\cite{blangiardo2015}, \cite{krainski2018}, \cite{moraga2019}, \cite{gomez2020}) along with numerous applications. However, the {\tt R} implementation of INLA does not offer complete flexibility with respect to the likelihood specification.

In this paper, we discuss the utility of the existing random effects modeling tool TMB, developed by \cite{kristensen2016}. TMB is exceptionally flexible, allowing the user to define custom models within a {\tt C++} template, is computationally efficient, and is applicable to a wide class of sampling models. While base TMB, the focus of this paper, provides deterministic approximations to marginal densities, the release of {\tt tmbstan} by \cite{monnahan:kristensen:18} provides an approach to stochastically sample from target distributions defined with TMB templates using no U-turn sampling (NUTS) MCMC allowing for fully Bayesian inference. We focus on TMB since it is less well known to statisticians, though is popular in the ecological literature (\cite{thorson2016}, \cite{bolstad2017}, \cite{free2019}). We provide a detailed description of the methodology and source code to implement a variety of the most popular spatial models. The spatial simulation studies we use to assess TMB and R-INLA adds to a growing body of literature showing the strengths and limitations of these methods (\cite{taylor2014}, \cite{ferkingstad2015}, \cite{auger2017}). They also extend the limited number of studies validating the stochastic partial differential equations (SPDE) approach to fitting continuous spatial models (\cite{teng2017}, \cite{righetto2020}).

The structure of this paper is as follows. In Section \ref{sec:inferential} we describe the models that we are considering, along with an overview of inferential approaches. Sections \ref{sec:inla} and \ref{sec:TMB} and \ref{sec:differences} describe INLA and TMB and summarize their differences, respectively.  Section \ref{sec:sim} compares the two approaches, via extensive simulations, and Section \ref{sec:euro.bc} considers the modeling of European breast cancer data using an incidence/mortality model that can be fitted in TMB, but not in INLA.  We conclude with a discussion in Section \ref{sec:discussion}.

\section{Inferential Overview}\label{sec:inferential}

We will consider the model:
\begin{eqnarray}
  \bmy | \bbeta, \bmb, \bphi_1 &\sim & p_1(\bmy | \bbeta , \bmb, \bphi_1 )\label{eq:sampling}\\ \bmb | \bphi_2 &\sim & p_2(\bmb | \bphi_2)\label{eq:randomeffects}
\end{eqnarray} where $\bmy$ represent data, $p_1$ and $p_2$ are the likelihood and random effects distributions, respectively, $\bbeta$ represent fixed effects, $\bmb$ random effects and $\bphi=[\bphi_1,\bphi_2]$ variance-covariance parameters, with $\bphi_1$ appearing in the likelihood and $\bphi_2$ in the prior. The random effects may be split into a set of spatial random effects $\bmu$ and non-spatial random effects $\bmv$ so that $\bmb=[\bmu,\bmv]$. We consider Gaussian spatial random effects so that $\bmb|\bphi_2$ falls within the class of latent Gaussian models (LGMs).

A Bayesian approach to inference completes the model (\ref{eq:sampling}) and (\ref{eq:randomeffects}) by adding the hyperprior, $p_3(\bbeta,\bphi)=p_3(\bbeta)p_3(\bphi)$, with the prior for $\bbeta$ assumed to be Gaussian. INLA is generally used to carry out fully Bayesian inference. The automatic differentiation at the heart of TMB allows various inferential approaches. We let,
$$f(\bbeta,\bmb,\bphi) = \log p_1(y | \bbeta , \bmb, \bphi_1 ) + \log p_2(\bmb | \bphi_2).$$ The simplest approach is maximum likelihood estimation (MLE) for the fixed effects and variance-covariance parameters (REML is also available for $\bphi$). The MLEs maximize the marginal likelihood,
\begin{equation}
  \label{eq:tmb.marg.lik} \mathcal{L}(\bbeta,\bphi) = \int \mbox{exp}\left[f(\bbeta,\bmb,\bphi) \right] ~\dif\bmb.
\end{equation} Inference may proceed via the asymptotic normal distribution of the MLE with uncertainty expressed via the observed information evaluated at the MLE, as in TMB's methodological ancestor, ADMB (\cite{fournier2012}). Within TMB one may also maximize the marginal posterior,
\begin{equation}
  p( \bbeta , \bphi | \bmy ) \propto \mathcal{L}(\bbeta,\bphi) \times p_3(\bbeta)\times p_3(\bphi), \label{eq:marg.post}
\end{equation}
to produce marginal maximum a posteriori (MMAP) estimates for the fixed effects and variance parameters, and  where inference will be based on the asymptotic distribution of the posterior.  In either setting, inference for the random effects then occurs through empirical Bayes by maximization of $f(\widehat\bbeta,\bmb,\widehat\bphi)$ having conditioned on the parameter estimates, $\widehat\bbeta$ and $\widehat\bphi$

\section{Integrated Nested Laplace Approximation}\label{sec:inla}

INLA has been well described in a number of statistical venues including recently in \cite{martino2020} and \cite{rue2017}. We provide a brief overview of its details here for comparison against the details of TMB presented in Section \ref{sec:TMB} and to aid in comparing the INLA and TMB simulation results presented in Section \ref{sec:sim}.

We first note that INLA does not use the ``standard" Laplace approximation (which is used in TMB, as described in Section \ref{sec:TMB}), but rather implements the Laplace ratio approximation (LRA) described in Tierney and Kadane (1986, Section 4.1)\nocite{tierney1986} which benefits from some cancellation of approximation error.

INLA was introduced by \cite{rue2009} to provide a quick option for Bayesian computation for the class of additive LGMs (ALGMs). In an ALGM, the general hierarchical model formulation described in (\ref{eq:sampling}) and (\ref{eq:randomeffects}) is restricted to models where the conditional expectation of the observations can be related to a linear combination of the fixed and random effects via a known link function $g$:
\begin{equation}
  \E[y_i| \bbeta, \bmb] = g(\eta_i) = g\left(\beta_0 + \sum_{j=1}^J \beta_j z_{ij} + \sum_{k=1}^K b_i^{(k)}\right) \label{eq:lin.pred}
\end{equation}
for observed covariates, $z_{ij}$, associated with the fixed effects, $\beta_j, j=1,\ldots, J$, random effects, $\{b_i^{(k)},\ k=1, \ldots, K\}$, and with $\eta_i$ the linear predictor for each observation $i$. All the parameters of the linear predictor are assumed to be Gaussian, completing the LGM definition.

We introduce a slight change of model formulation, to accommodate the consolidation of the like terms. Specifically, we collect together the Gaussian fixed and random effects, from (\ref{eq:randomeffects}), to write
$$p_2(\bbeta, \bmb |\bphi_2) = p_2( \bmb | \bphi_2 ) \times p_3(\bbeta).$$ Defining these terms to be $\bmB = \left[\bbeta, \bmb^{(1)},\ldots,\bmb^{(K)}\right]$, in the LGM setting we then have
\begin{equation}
  \bmB|\bphi_2 \sim \N\left(\bmzero, \bmQ^{-1}(\bphi_2)\right) \label{eq:lat.gauss}
\end{equation}
where $\bmQ^{-1}(\bphi_2)$ is the precision matrix for the Gaussian field. This ensures that the linear predictor $\bmeta$ is Gaussian as well. To complete the Bayesian model specification, the hyperprior, $p_3(\bphi)$ is specified.

The primary targets of inference for the INLA algorithm are univariate posterior densities for the latent field parameters, $p(B_i|\bmy)$, and the joint posterior of the hyperparameters, $p(\bphi|\bmy)$. INLA approximates these in three steps:
\begin{enumerate}
\item Explore and discretize $\bphi$-space via an approximation,
  \begin{equation} \widetilde{p}(\bphi|\bmy)\propto \frac{p_1(\bmy|\bbeta, \bmb, \bphi)p_2(\bbeta, \bmb|\bphi)p_3(\bphi)}{\widetilde{p}_G(\bbeta, \bmb|\bphi, \bmy)}\bigg|_{\scriptsize\bbeta=\bbeta^*(\bphi), \bmb=\bmb^*(\bphi)},\label{eq:inla.hyp.approx}
  \end{equation} where $\widetilde{p}_G(\bbeta, \bmb|\bphi, \bmy)$ is the Gaussian approximation to the conditional distribution obtained by numerically finding and matching the mode, $\{\bbeta^*(\bphi), \bmb^*(\bphi)\}$, and curvature at the mode, for given $\bphi$.
  \end{enumerate}

  \noindent The approximation in (\ref{eq:inla.hyp.approx}) is equivalent to the Laplace Ratio Approximation (LRA) for the posterior marginal proposed by \cite{tierney1986}. As the hyperparameter space is explored, (\ref{eq:inla.hyp.approx}) is evaluated at $L$ high-density points to generate an approximate discretization: $\widetilde{p}(\bphi^{(l)}|\bmy)$ at points $\{\bphi^{(1)}, \ldots, \bphi^{(L)}\}$. R-INLA has three available approaches for selecting $\bphi^{(l)}$: the empirical Bayes (EB) option uses only the modal value as a single integration point, the grid method develops a regular grid on the primary orthogonal axis of the hyperparameter space, and the central composite design (CCD) approach which efficiently selects the modal value and a group of `star points' surrounding the center. The default within R-INLA selects the grid option for small numbers of hyperparameters and otherwise selects the CCD method

  \begin{enumerate}[2.]
  \item Approximate $p(B_i|\bphi^{(l)}, \bmy)$ for $l=1, \ldots, L$ using one of three approximations: Gaussian, Laplace, or Simplified Laplace (SL).
  \end{enumerate}

 \noindent   In the Gaussian approximation, $\widetilde{p}_G(B_i|\bphi^{(l)}, \bmy)$ is calculated directly as the marginal of $\widetilde{p}_G(\bbeta, \bmb|\bphi, \bmy)$, from (\ref{eq:inla.hyp.approx}). While very fast, this approximation is often not particularly good \cite[Section 4.7.2]{blangiardo2015}. In the Laplace approximation, a computationally optimized version of the LRA of \cite{tierney1986} is used,
\begin{equation} \widetilde{p}_L(B_i|\bphi^{(l)}, \bmy) \propto \frac{p_1(\bmy|\bbeta, \bmb, \bphi^{(l)})p_2(\bbeta, \bmb|\bphi^{(l)})p_3(\bphi)}{\widetilde{p}_G(\bmB_{-i}|B_i,\bphi^{(l)}, \bmy)}\bigg|_{\scriptsize\bmB_{-i}=\bmB_{-i}^*(B_i, \bphi^{(l)})},\label{eq:inla.param.lap.approx}
\end{equation}
 \noindent where $\widetilde{p}_G(\bmB_{-i}|B_i,\bphi, \bmy)$ is the Gaussian Laplace approximation to $p(\bmB_{-i}|B_i,\bphi, \bmy)$ with mode $\bmB_{-i}^*(B_i, \bphi)$. This approximation often works very well since the conditional distribution of the latent field parameters are generally close to Gaussian, but it is computationally expensive since it must be recomputed for all desired combinations of $\bmB$ and $\bphi$. The SL approximation, $\widetilde{p}_{SL}(B_i|\bphi^{(l)}, \bmy)$, uses a Taylor-series approximation of $ \widetilde{p}_L(B_i|\bphi^{(l)}, \bmy) $. This approximation is quick and accurate for many applications and is the default option within R-INLA. More details on this approximation can be found in \cite[Section 3.2]{rue2009}.

\begin{enumerate}[3.]

\item Approximate the marginal using numerical integration,
\begin{equation} \widetilde{p}(B_i|\bmy) = \sum_{l=1}^L\widetilde{p}(B_i|\bphi^{(l)}, \bmy)\times\widetilde{p}(\bphi^{(l)}|\bmy)\times\Delta_l,\label{eq:inla.num.int}
\end{equation} over the integration points, $\bphi^{(l)}$, appropriately scaled by their associated weights, $\Delta_l$.
\end{enumerate}

Although INLA returns the univariate marginals, in general we may be interested in functions of the parameters and R-INLA provides a method to sample from the approximate joint posterior, implemented in their sampling function, {\tt inla.posterior.sample()}, using the mixture:
\[
  \widetilde{p}(\bbeta, \bmb, \bphi |\bmy)\approx\sum_{l=1}^L\widetilde{p}_G(\bbeta, \bmb, |\bmy, \bphi^{(l)})\times\widetilde{p}(\bphi^{(l)}|\bmy).
\]
For each draw, $d$, first a sample, $\bphi^{(d)}$ is drawn from the discretized hyperparameter posterior, $\widetilde{p}(\bphi^{(l)}|\bmy)$, and then a sample, $\{\bbeta^{(d)},\bmb^{(d)}\}$, is drawn from a Gaussian approximation to the joint conditional latent distribution, $\widetilde{p}_G(\bbeta, \bmb, |\bmy, \bphi^{(d)})$ which is found to match the mode and curvature at the mode conditional on the specific  hyperparameter draw.  While there is no guarantee that this joint approximation will lead to the same approximate univariate marginals from the full INLA algorithm, by default R-INLA corrects both the mean and the skew of the Gaussian marginals sampled from the joint posterior my mapping to a SkewNormal distribution using the better-approximated marginal posteriors (\cite{wakefield2016}). We note that even when the EB method for the hyperparameter integration is chosen in INLA and there is only one `integration point' for $\bphi$, this approximate joint distribution can still account for skew.

R-INLA only allows for a limited set of built-in likelihoods, and while INLA theoretically can be used on any ALGM, it does have some pragmatic suggestions to ensure reasonable computing times. In particular, a reasonably small dimension of $\bphi$ is required (\cite{rue2017} suggest 2-5 and not more than 20) to minimize the computational burden of the numerical integration in (\ref{eq:inla.num.int}). Crucially, sparsity in the precision of the latent field parameters can be leveraged at numerous points in the algorithm by R-INLA to greatly reduce the computational cost and speed up the approximation. The class of ALGMs is the target of the INLA method because they often permit sparsity and because the posteriors can often be well approximated with the LRA. Spatial statistics applications commonly use Gaussian Markov Random Field (GMRF) model specifications which maintain high levels of sparsity in the precision. A multivariate Gaussian random variable is a GMRF on an undirected graph $G=\{V, E\}$ with vertices, $V$, and edges $E$ if non-zeros in the precision matrix correspond to the edges of $G$. GMRF models are used in the following simulations and in the cancer application.

\section{Template Model Builder}\label{sec:TMB}

TMB, rooted in frequentist inference, requires the user to differentiate between random and non-random effects.  We focus our description of TMB in the context of a frequentist treatment while noting the differences for Bayesian inference.  In contrast to INLA, TMB uses a single LA to integrate out the random effects from the full joint distribution to obtain an approximation to the marginal likelihood in (\ref{eq:tmb.marg.lik}). Defining the conditional mode,
\begin{equation}
  \widehat{\bmb}(\bbeta,\bphi) := \underset{\bmb}{\mbox{argmin}}~ -f(\bbeta, \bphi, \bmb),  \label{eq:tmb.re.est}
\end{equation}
TMB approximates (\ref{eq:tmb.marg.lik}) using the LA to marginalize over the random effects:
\begin{eqnarray}
  \mathcal{L}(\bbeta,\bphi) \approx \tilde{\mathcal{L}}(\bbeta,\bphi) = (2\pi)^{n/2}|\bm{\mathcal{H}}(\bbeta,\bphi)|^{1/2}\exp[-f(\bbeta,\widehat{\bmb}(\bbeta,\bphi),\bphi)], \label{eq:tmb.laplace}
\end{eqnarray} where $ \bm{\mathcal{H}}(\bbeta,\bphi)$ is the Hessian of $f(\bbeta,\widehat{\bmb}(\bbeta,\bphi) ,\bphi)$, with $(j,k)$-th element
\begin{eqnarray}
  \frac{\partial^2}{\partial b_j \partial b_k} f( \bbeta, \widehat{\bmb}(\bbeta,\bphi),\bphi). \label{eq:tmb.hess}
\end{eqnarray}
% is the taken with respect to $\bmb$ at the point $\widehat{\bmb}(\bbeta,\bphi)$, and and $|\bm{\mathcal{H}}(\bbeta,\bphi)|$ is the determinant of the Hessian.

Estimation in TMB is performed through a two-stage nested optimization procedure which searches for the vector $(\widehat{\bbeta}, \widehat{\bphi} )$ maximizing $\tilde{\mathcal{L}}(\bbeta, \bphi)$ as defined in (\ref{eq:tmb.laplace}). To evaluate $ \tilde{\mathcal{L}}(\bbeta,\bphi)$, we need to evaluate both $\widehat{\bmb}$ in (\ref{eq:tmb.re.est}) and $\bm{\mathcal{H}}(\bbeta,\bphi)$ in (\ref{eq:tmb.hess}), but neither are usually available in closed-form. While $\widehat{\bmb}(\bbeta,\bphi)$ may be evaluated through nonlinear optimization of $f$, the crux of the work is to evaluate $\bm{\mathcal{H}}(\bbeta,\bphi)$. TMB computes gradients and the Hessian using automatic differentiation (\ref{app:ad} presents an overview) performed via {\tt CppAD} \citep{bell2007} and at the same time can auto-detect sparsity in the model in order to leverage sparse matrix routines. TMB reuses the LA and the automatically generated Hessian to produce estimates of the joint covariance between the fixed effects estimates, $\widehat{\bbeta}$ and $\widehat{\bphi}$, and the random effects predictors, $\widehat{\bmb}$, as described in Section \ref{sec:tmb.var}. Inference then proceeds assuming asymptotic normality for the joint distribution of the estimated fixed effects and predicted random effects.

\subsection{The TMB Estimation Algorithm}\label{sec:tmb.alg}
TMB implements a two-step nested optimization routine to iteratively search for the fixed effects estimates with random effects predictions also being produced. Given initial starting values, the routine performs the following steps at each evaluation in the search:

\begin{enumerate}
\item Given current values of the fixed effects parameters, $(\bbeta^\star, \bphi^\star)$, perform nonlinear optimization to find updated modal values of the random effects, $\widehat{\bmb}(\bbeta^\star, \bphi^\star)$ as in (\ref{eq:tmb.re.est}) and set this to be the current value of the random effects, $\bmb^\star$.
\item Given current values of the random effects parameters, $\bmb^\star$, find the modal values of the Laplace approximation to the marginal likelihood, $\tilde{\mathcal{L}}(\bbeta, \bphi)$, shown in (\ref{eq:tmb.laplace}) and set them to the current value of the fixed effects, $(\bbeta^\star, \bphi^\star)$. In addition, evaluate the gradient of the marginal likelihood to assess stopping conditions.
\item If the maximum gradient component (MGC) of the marginal likelihood is below a stopping threshold the routine stops, otherwise go to step 1.
 \end{enumerate}

 If a stopping criteria was reached, then the final values of the fixed effect and random effects are returned as the fixed effects estimates, $(\widehat\bbeta, \widehat\bphi) = (\bbeta^\star, \bphi^\star)$, and the random effects predictions are updated one final time to yield $\widehat\bmb(\widehat\bbeta, \widehat\bphi)$.

 \subsection{The TMB Estimators}\label{sec:tmb.est}

In models without hyperpriors, TMB produces the marginal MLEs of the $\bbeta, \bphi$, from the approximate marginal likelihood defined in (\ref{eq:tmb.laplace}). The random effects predictions in this setting, taken as the mode of their conditional distribution, $p(\bmb|\widehat{\bbeta}, \widehat{\bphi}_1, \bmy)$, using `plug-in' estimates of the fixed effects MLEs, are empirical Bayes predictors (\cite{devalpine2009}).

In models with hyperpriors, TMB produces the marginal maximum a posteriori estimates of $\bbeta, \bphi$, again using the Laplace approximation from (\ref{eq:tmb.laplace}) applied to the marginal posterior in (\ref{eq:marg.post}). With priors, basing random effects inference on the estimated posterior, $p(\bmb|\widehat{\bbeta},\widehat{\bphi}_1,\bmy)$, where the hyperparameters have been estimated using the data, TMB provides parameteric empirical Bayes (PEB) estimates \cite[Chapter 3.3]{carlin2000}. In essence, this replaces the posterior, $$p(\bmb|\bmy) = \int p(\bmb|\bmy, \bbeta, \bphi)p(\bbeta,\bphi|\bmy)~\dif\bbeta\dif\bphi,$$ with $p(\bmb|\widehat{\bbeta},\widehat{\bphi}_1,\bmy)$ where $\widehat{\bbeta}$ and $\widehat{\bphi}_1$ are MMAP estimates. The EB option for the numerical integration in (\ref{eq:inla.num.int}) in the INLA algorithm makes the same tradeoff: skipping the computational complexity of the integration by not accounting for the uncertainty in the (hyper)parameters.

%  Fully Bayesian approaches include a hyperprior distribution, $p_3(\bbeta, \bphi)$, and compute the posterior distribution by integrating over the hyperparameters:

%  \begin{align}
%    p(\bbeta, \bmb|\bmy) &= \frac{\int p_1(y|\bbeta, \bmb, \bphi_1)p_2(\bmb|\bphi_2)p_3(\bbeta, \bphi)~\dif\bphi}{\int\int p_1(y|\bbeta, \bmb, \bphi_1)p_2(\bmb|\bphi_2)p_3(\bbeta, \bphi)~\dif\bbeta\dif\bmb\dif\bphi} \nonumber \\
%    &= \int p(\bbeta, \bmb|\bmy, \bphi)p(\bphi|\bmy)~\dif\bphi. \label{eq:bayes.mix}
%  \end{align}

% \noindent The Bayesian posterior is a mixture of conditional posteriors, $p(\bbeta, \bmb|\bmy, \bphi)$, for fixed values of the hyperparameters mixed over their posterior marginal distribution, $p(\bphi|\bmy)$.

While the random effects predictors are fast to compute via optimization having conditioned on the fixed effects estimates, with the number of REs increasing proportionally to the number of data observations, EB predictions of random effects generally have no guarantees of consistency \citep{thorson2016}. In order to improve the predictions for the random effects and differentiable functions of the mixed effects, $\bmd(\bbeta, \bphi, \bmb)$, TMB implements a generic bias-correction adjustment termed the ``epsilon'' method in \cite{thorson2016}. The correction is applicable to quantities which depend on random effects, $\bmd(\widehat\bbeta, \widehat\bphi, \bmb)$, and has no effect on the fixed effects estimates. This correction is based an approximation to moment-generating functions, using $M(\bepsilon)=E[\text{exp}\{\bepsilon \bs\cdot \bd(\widehat\bbeta, \widehat\bphi,\bmb)\}]$, with $\bs\cdot$ denoting the dot product, proposed by \cite{tierney1989} (Section 3.1).  Introducing an auxiliary parameter vector, $\bepsilon$, of dimension equal to the dimension of $\bmd()$, consider a new function with the auxiliary parameters:
\begin{eqnarray}
  \label{eq:tmb.nuisance}
  \bme(\widehat\bbeta, \widehat\bphi, \bmb, \bepsilon |\bmy) =  \text{log} \Big( \int \text{exp}\big(f(\widehat\bbeta, \widehat\bphi, \bmb) + \bepsilon\bs\cdot\dif(\widehat\bbeta, \widehat\bphi, \bmb) \big)~\dif\bmb \Big).
\end{eqnarray}
Employing the namesake method of moment-generating functions, the form of the bias-corrected estimator is found by differentiating with respect to $\bepsilon$ and then setting $\bepsilon=\bs 0$:
\begin{eqnarray}
  \frac{\partial}{\partial\bepsilon}\left(\bme(\widehat\bbeta, \widehat\bphi, \bmb, \bepsilon |\bmy) \right)\biggm\lvert_{\bepsilon = \bs 0} &= &\frac{\int\text{exp}\left(f(\widehat\bbeta, \widehat\bphi,\bmb)  + \bepsilon\bs\cdot\bmd(\widehat\bbeta, \widehat\bphi, \bmb)\right)\bmd(\widehat\bbeta, \widehat\bphi, \bmb)~\dif\bmb}{\int\text{exp}\left(f(\widehat\bbeta, \widehat\bphi,\bmb) + \bepsilon\bs\cdot\bmd(\widehat\bbeta, \widehat\bphi, \bmb)\right)~\dif\bmb} \biggm\lvert_{\bepsilon = \bs 0}\nonumber \\ %\label{eq:tmb.bias.correct1} \\
&= &\frac{\int\text{exp}\left(f(\widehat\bbeta, \widehat\bphi,\bmb) \right) \bmd(\widehat\bbeta, \widehat\bphi, \bmb)~\dif\bmb}{\int\text{exp}\left(f(\widehat\bbeta, \widehat\bphi,\bmb) \right)~\dif\bmb}  \label{eq:tmb.bias.correct2} \\
  &= &\text{E}[\bmd(\widehat\bbeta, \widehat\bphi, \bmb)\lvert \bmy]. \nonumber
\end{eqnarray}

\noindent This identity allows TMB to improve the predictions for the random effects ($\bmd(\cdot)$ is the identity) and nonlinear functions of the mixed effects using the gradient of the approximate marginal likelihood with respect to the auxiliary parameters, $\bepsilon$, leveraging its available tools: {\tt CppAD} to automatically evaluate derivatives and the LA to approximate the integrals. \cite{thorson2016} note that this is only a bias correction algorithm because the LA used to evaluate both numerator and denominator in (\ref{eq:tmb.bias.correct2}) will be inexact unless the likelihood, conditional on the fixed effects, is multivariate Gaussian.

\subsection{Variance of TMB Estimators}\label{sec:tmb.var}

% The TMB algorithm returns the joint modal vector as the parameter estimates and predictions:
% % $\hat{\bs\mu}_{\hat{\bs \theta}, \hat{ \bs b}}=
% \left(\widehat{\bs \theta}, \hat{ \bs b}(\widehat{\bs\theta})\right)$. The default random effects predictions provided by TMB are the final iteration of the modal random effects values, conditional on the estimated fixed effects, $\hat{\bs b}(\hat{\bs\theta})$, as returned from the TMB algorithm described in Section~\ref{sec:tmb.alg}. While these predictions are fast to compute, neither the hierarchical-likelihood predictions from a frequentist model nor the empirical Bayes predictors have guarantees of consistency.

%% Furthermore, even if they were unbiased, for a nonlinear function,
%% $d()$, of some or all of the random effects, the oft-used plug-in
%% estimator of $d(\bs b)$, $d(\hat{\bs b})$, will be biased.

TMB approximates the covariance of the fixed effects using the inverse of the observed Hessian of the log-likelihood:

\begin{equation}
   \bSigma_{\scriptsize \widehat\bbeta, \widehat\bphi}= \text{Cov}(\widehat\bbeta, \widehat\bphi) = \left(-\nabla^2 \text{log}\tilde{\mathcal{L}}(\widehat\bbeta,\widehat\bphi)\right)^{-1}.
\end{equation}

In models that include random effects, the joint covariance of the fixed and random effects is approximated using an application of the law of total variance and a linearization:
\begin{eqnarray}
  \label{eq:tmb.joint.var}
   \bSigma_{\scriptsize \widehat\bbeta, \widehat\bphi, \widehat\bmb}= \text{Cov}\begin{pmatrix} \widehat\bbeta, \widehat\bphi\\ \hat{\bmb}\end{pmatrix}
  &=&\text{E}\left[\text{Cov}\left(\widehat\bbeta, \widehat\bphi, \hat{\bmb} \middle\vert \, \widehat\bbeta, \widehat\bphi \right)\right] +
    \text{Cov}\left[\text{E}\left(\widehat\bbeta, \widehat\bphi, \hat{\bmb} \middle\vert \, \widehat\bbeta, \widehat\bphi \right)\right] \nonumber\\
  &\approx&
    \begin{pmatrix}
      0 & 0\\
      0 & \bm{\mathcal{H}}_{bb}^{-1}(\bbeta, \bb, \bphi)
    \end{pmatrix} +
    \bs J \bSigma_{\scriptsize \widehat\bbeta, \widehat\bphi}\bs J^{\text{T}}
\end{eqnarray}

\noindent where $\bm{\mathcal{H}}_{bb}(\bbeta, \bb, \bphi)$ is the random effects sub-matrix of the full joint Hessian of $f(\bbeta, \bb, \bphi)$, and $\bJ$ is the Jacobian of the vector $\left(\btheta, \hat{\bmb}(\btheta) \right)^{\text{T}}$ with respect to $\bs\theta$. The $\delta$-method is used to find the joint covariance of differentiable functions of the mixed effects:
% \begin{eqnarray}
%   \label{eq:tmb.joint.delta}
%   \text{Cov}\left(\bs d\begin{pmatrix} \widehat\bbeta, \widehat\bphi\\ \hat{\bs b}\end{pmatrix}\right)
%   &= \nabla \bs d\, \text{Cov}\begin{pmatrix} \hat{\bs b}\\ \widehat\bbeta, \widehat\bphi \end{pmatrix} \nabla \bs d^{\text{T}}.
% \end{eqnarray}
  \begin{eqnarray}
  \label{eq:tmb.joint.delta}
  \text{Cov}\left(\bmd (\widehat\bbeta, \widehat\bphi, \hat{\bmb})\right)
    &= \nabla \bmd\,\bSigma_{\scriptsize \widehat\bbeta, \widehat\bphi, \hat{\bmb}} \nabla \bmd^{\text{T}},
  \end{eqnarray}
\noindent where $\nabla \bmd$ is the Jacobian of $\bmd$. For models with only fixed effects, this simplifies to
\begin{eqnarray}
  \label{eq:tmb.fe.cov}
  \text{Cov}\left(\bs d(\widehat\bbeta, \widehat\bphi)\right)=\nabla \bmd\    \bSigma_{ \scriptsize \widehat\bbeta, \widehat\bphi} \nabla \bmd^{\text{T}}.
\end{eqnarray}

Much like the correction for the random effects predictions, TMB can improve the covariance estimator of the random effects prediction and functions of the mixed effects shown in (\ref{eq:tmb.joint.delta}) using the second derivative of $\bs e(\widehat\bbeta, \widehat\bphi, \bmu, \bepsilon |\bmy)$ from (\ref{eq:tmb.nuisance}). The form of the improved variance estimator again uses the law of total variance:

% \begin{eqnarray}
%   \label{eq:tmb.var.correct}
%   \text{Cov}\left(\bs d\begin{pmatrix} \widehat\bbeta, \widehat\bphi\\ \hat{\bs b}\end{pmatrix}\right) =
%   &\biggl[\frac{\partial^2}{\partial^2\bs\epsilon}\left(\bs e(\widehat\bbeta, \widehat\bphi, \bs b, \bs\epsilon |\bs y) \right) + \nonumber \\
%   &\frac{\partial}{\partial\bs\theta} \frac{\partial}{\partial\bs\epsilon}\left(\bs e(\widehat\bbeta, \widehat\bphi, \bs b, \bs\epsilon |\bs y) \right)^{\text{T}}\,
%     \left(  \bm{\mathcal{H}}( \widehat\bbeta, \widehat\bphi ) \right)^{-1}\,
%     \frac{\partial}{\partial\bs\theta} \frac{\partial}{\partial\bs\epsilon}\left(\bs e(\widehat\bbeta, \widehat\bphi, \bs b, \bs\epsilon |\bs y) \right)\biggr]_{\bs\epsilon = \bs 0},
% \end{eqnarray}

\begin{eqnarray}
  \label{eq:tmb.var.correct}
  \text{Cov}\left(\bs d (\widehat\bbeta, \widehat\bphi, \widehat{\bmb})\right) &=&
  \biggl[\frac{\partial^2}{\partial^2\bs\epsilon}\left(\bs e(\widehat\bbeta, \widehat\bphi, \bs b, \bs\epsilon |\bs y) \right) + \nonumber \\
&&\frac{\partial}{\partial\bs\theta} \frac{\partial}{\partial\bs\epsilon}\left(\bs e(\widehat\bbeta, \widehat\bphi, \bs b, \bs\epsilon |\bs y) \right)^{\text{T}}~
       \bSigma_{\scriptsize \widehat\bbeta, \widehat\bphi}~
    \frac{\partial}{\partial\bs\theta} \frac{\partial}{\partial\bs\epsilon}\left(\bs e(\widehat\bbeta, \widehat\bphi, \bs b, \bs\epsilon |\bs y) \right)\biggr]_{\bs\epsilon = \bs 0},
\end{eqnarray}

\noindent where the first term on the right-hand side works out to be the standard variance estimator for random effects conditional on the fixed effects (demonstrated analogously to the derivation in (\ref{eq:tmb.bias.correct2})), and the second term accounts for having conditioned on the fixed effects estimators. This is an improvement over the naive EB variance estimators which ignore the conditioning on the fixed effects \cite[Chapter 3.5]{carlin2000}.

% Since TMB provides a means to calculate the joint covariance between the fixed and random effects, further derived quantities of interest can be calculated by transforming samples from a multivariate Gaussian distribution using the fixed effects estimates and random effects predictions as the mean.

\section{Contrasting TMB and INLA}
\label{sec:differences}

For clarity, Table~\ref{tab:differences} provides a summary of the primary differences - and similarities - between TMB and INLA that were discussed in detail in the previous two sections.

TMB and INLA will yield the most similar results under the following conditions:
\begin{itemize}
\item TMB is coded to optimize the marginal posterior using the same priors and with the same internal parameter representations used by R-INLA,
\item TMB uses both bias and variance corrections to functions of REs,
\item R-INLA uses the Gaussian approximations and the empirical Bayes `integration' strategy,
\item R-INLA does not use the bias and skew corrections when drawing posterior samples.
\end{itemize}

\begin{table}[h!]
  \begin{centering}

    \begin{tabular}{p{.31\textwidth}|p{.34\textwidth}|p{.34\textwidth}|}
      \cline{2-3}
                                            & \hspace{4em} TMB                           & \hspace{3.7em} INLA \\
      \hline

      \multicolumn{1}{|l|}{\multirow{1}[15]{*}{Inferential Method: FE}}  & Frequentist: MLEs from marginal likelihood (or MMAP from marginal posterior) & \multirow{2}[15]{*}{Full Bayesian (or PEB)} \\
      \cline{1-2}
      \multicolumn{1}{|l|}{\multirow{1}{*}{Inferential Method: RE}} & Frequentist: EB (or PEB) &  \\

      \hline \hline

      \multicolumn{1}{|l|}{\multirow{1}[5]{*}{Laplace approximation}} & Standard Gaussian approximation & LRA from \cite{tierney1986}\\
      \hline
      \multicolumn{1}{|l|}{\multirow{1}[5]{*}{Bias Corrections}} & Yes, to functions of REs & Applied to align joint posterior samples with marginal distributions\\
      \hline
      \multicolumn{1}{|l|}{\multirow{1}{*}{Var. Corrections}} & Yes, to functions of REs  & No \\
      \hline
      \multicolumn{1}{|l|}{\multirow{1}{*}{Skew Corrections}} & No & Applied to align joint posterior samples with marginal distributions \\
      \hline
      \multicolumn{1}{|l|}{\multirow{1}{*}{Hyperpar. Integration}} & No                            & Yes (no) \\

      \hline \hline

      \multicolumn{1}{|c|}{\multirow{1}[18]{*}{Inferential sampling}} & Gaussian centered at point estimates and covariance from observed information and linearization & Gaussian approximation to joint posterior with mean and skew corrections applied to marginals\\

      \hline \hline

      \multicolumn{1}{|c|}{\multirow{1}{*}{Hessian Evaluation}} & Automatic Differentiation & Finite Differences \\
      \hline
      \multicolumn{1}{|c|}{\multirow{1}{*}{Sparse Matrices}} & Yes & Yes \\
      \hline
      \multicolumn{1}{|c|}{\multirow{1}{*}{Parallelization}} & Yes & Yes \\
      \hline
    \end{tabular}

  \end{centering}
  \caption{Summarization of primary differences between TMB and INLA. Entries in (parentheses) indicate outcomes from TMB for models that include priors and indicate outcomes from INLA under eb `numerical integration' over the hyperparameters. The table is split into sections corresponding to methods, approximations, post-model sampling, and computation. FE=fixed effects, RE=random effects, MMAP=marginal maximum a posterior estimates, PEB = parametric Empirical Bayes, LRA = Laplace ratio approximation.}  \label{tab:differences}
\end{table}

\section{Spatial Simulation Study}
\label{sec:sim}

We performed two simulation studies on popular continuous and discrete spatial smoothing models to assess the ability of TMB and R-INLA to estimate the total spatial field effects, the parameters and hyperparameters, and how their computational performance scales as data volume and random effects dimension increase.

The second purpose of the continuous simulation study was to perform a thorough assessment the popular stochastic partial differential equations (SPDE) representation fitting GPs which uses a finite element method over a triangulation to solve a particular SPDE whose solution is known to have Mat\'ern covariance \citep{stein:1999}. For details on the SPDE approximation, see \cite{lindgren2010} and \cite{miller2020}. The discrete simulation study was included in part to assess and verify inference in TMB using hard constraints on the random effects parameters.

For each study, a grid of experiments, shown in Tables~\ref{tab:sim_params} and \ref{tab:disc_sim_params}, was defined. Each level of each experiment was replicated 25 times to obtain Monte Carlo errors on the validation metrics. For each replicate within each experiment, completely new spatial fields, sampling locations, and observations were generated. For each of these, TMB and R-INLA algorithms were run, and inference and validation was performed using 500 joint samples drawn from each model, projected to 5$\times$5 km$^2$ raster grids in the case of the continuous simulation, and compared against against the truth. For both models, the internal representation of parameters in TMB were coded to align with those used by R-INLA. TMB was always run using its bias correction method and the improved variance estimates, and R-INLA joint estimates were generated using its available mean bias correction. All of the simulation analyses use the empirical distribution taken across the 25 replicates of each experimental level.

\subsection{Continuous Spatial Simulation}
\label{sec:sim.cont}

This simulation was designed with respect to three governing motivations that dictated the choice of models, covariates, and true parameter ranges: (1) vetting TMB and INLA inference while (2) assessing the SPDE approximation in a variety of settings (3) using simulated risk fields and data akin to those commonly see in public health settings. Although motivated by public health applications, the broad range the parameters, such as the maximum number of simulated data locations, greatly extend this study's applicability beyond any one applied domain.

We consider Gaussian process (GP) models with mean $\mu(\bms)$ at location $\bms$, and the Mat\'ern function where the covariance between two spatial locations distance $||\bms_i - \bms_j||$ from one another is:
\begin{equation}
  C(u(\bms_i), u(\bms_j)) = \frac{\sigma_\text{m}^2 }{2^{\nu-1}\Gamma(\nu)}\left( \kappa||\bms_i - \bms_j|| \right)^{\nu}K_{\nu}\left(\kappa||\bms_i - \bms_j|| \right), \label{eq:matern.app}
\end{equation} where $\sigma_\text{m}^2$ is the variance, $\kappa > 0$ is a scaling parameter related to the range, $r_m=\frac{\sqrt{8\nu}}{\kappa}$, defined to be the distance at which the spatial correlation drops to 0.1, $\nu > 0$ is related to the smoothness of the field, and $K_\nu$ is the modified Bessel function of the second kind. For any finite collection of locations within the domain, $\{\bms_1, \ldots, \bms_{n}\}\in\mathcal{D}$, the random vector $\bmu = [u(\bms_1),\ldots,u(\bms_{n})]$ has multivariate Gaussian distribution with precision matrix $\bmQ$.

The simulated data, observed at locations $\bms_i,\ i=1, \ldots, n_s,$ within Nigeria and selected using a stratified spatial sampling design, arise from the following hierarchical model:
\begin{eqnarray*}
  \bmy |\bbeta, \bmb ,\bmphi_1&\sim& p_1(\bmy |\bbeta, \bmb,\bmphi_1) \label{eq:sim_hier} \\
  \mbox{E}[y_i|\bbeta,u_i , v_i] &=& g^{-1}\left(\alpha+\bmz^{\tiny{T}}_i \bbeta+ u_i + v_i\right)\nonumber\\
  \bmu &\sim &\mbox{N}(\bmzero, \bmQ(r_\text{m}, \sigma^2_\text{m}))\nonumber\\
  \bmv &\sim& \mbox{N}(\bmzero,\bident_{n_s} \sigma^2_\text{clust}). \nonumber%% \\ %%\bs Q &= precision\ from\ Mat\acute ern. \nonumber
\end{eqnarray*} where $\alpha$ is the intercept and is fixed to $-1$ across all simulations, $\bident_{n_s}$ is the $n_s \times n_s$ identity matrix and the last two lines correspond to $p_2(\bmb | \bmphi_2)$ with $\bmb=[\bmu,\bmv]$ and $\bmphi_2=[\sigma^2_\text{m},r_\text{m},\sigma^2_\text{clust}]$.  The precision for the spatial GP, $\bmQ$ is Mat\'ern with range $r_\text{m}$ and standard deviation $\sigma_\text{m}$. In some models, two spatially varying covariates were included: access time to heath care (\cite{weiss2018}) and malaria incidence (\cite{weiss2019}). While the SPDE representation is used to fit the spatial fields (using three different triangulation resolutions shown in Figure \ref{fig:meshes}), the true fields are simulated directly on a high resolution regular grid using the {\tt RandomFields R} package \citep{schlather2015}.

The domain, covariates and stratified sampling scheme were chosen to represent public health datasets, such as the Demographic and Health Surveys (DHS) and UNICEF Multiple Indicator Cluster Surveys, which are increasingly being used to predict continuous spatial(-temporal) maps of health outcomes. Our stratified cluster sampling design mimics the one used by DHS which stratifies by regions (administrative level 1 units) and urban/rural status and usually collects observations at 250-750 clusters, with typically 25-35 households sampled within each cluster. The covariates, access (travel time) to health care and malaria incidence, are both reasonable choices to be correlated with heath risks and were further chosen for their different spatial characteristics, and the magnitude of the GP and total field were selected to yield  moderately rare outcomes. Both Binomial and Gaussian data are commonly collected in a variety of applications (Poisson likelihoods are examined in the discrete simulation), and the binary data provide an extra challenge for both TMB and INLA. The Mat\'ern range was selected to represent medium-small and medium-large spatial ranges over the approximate $10\times 10$ (degrees latitude-longitude) domain. The Mat\'ern variance takes two values representing small- and large-scale spatial effects relative to the covariate effects and the small, medium, and large values of the $iid$ cluster variance and Gaussian observation variance. The number of vertices in the SPDE triangulation were selected to represent medium, fine, and very fine meshes, relative to the domain and resolution of the $5\times 5$km raster representation, and to assess the computational scaling of TMB and R-INLA. While the INLA method with empirical Bayes integration and Gaussian approximations are closest to the methods in TMB, we also evaluate some of the more accurate options in order to asses the default (and better) approximations available in R-INLA.

The (nearly) full combinatorial grid of simulation parameters shown in Table~\ref{tab:sim_params} comprised the set of 16128 experiments (R-INLA with 8000 clusters and the full Laplace approximation was excluded due to computational time constraints and the Gaussian variances were not varied for Binomial observations). We note that the internal representation of parameters in TMB were coded to align with those used by R-INLA. Additional details including prior specification, the stratified spatial sampling scheme, and SPDE mesh generation may be found in \ref{app:sim.cont}.

\begin{table}[htb!]
  \begin{footnotesize}
    \begin{tabular}{ll}
%      \makebox[1 \mboxwidth][c]{ %centering table
%      \resizebox{1\mboxwidth}{!}{ %resize table
      \begin{tabularx}{1.0\textwidth}{|l|X|}
        \hline
        Parameter & Simulation Values \\
        \hline \hline
        Data Observations & Binomial, Gaussian \\ \hline
        Gaussian Observation Variance, $\sigma_{obs}^2$ & $0.1^2, 0.2^2, 0.4^2$\\ \hline
        Covariates & None, $(-.25\times \mbox{access} + .25\times\mbox{Malaria Incid.})$ \\ \hline
        Number of Clusters, $n_{\bs s}$ & 250, 500, 750, 1000, 2000, 4000, 8000 \\ \hline
        Expected Samples per Cluster, E$[n_i]$ & 35\\ \hline
        Spatial Range (lat-lon degrees) & 1, $\sqrt{8}$ \\ \hline
        Spatial Variance & $0.25^2, 0.5^2$\\ \hline
        Cluster Variance, $\sigma_{clust}^2$  & 0, $0.1^2, 0.2^2, 0.4^2$\\ \hline
        Num. Nodes in SPDE Mesh & 3631, 7922, 13869 (low, medium, high resolution)\\ \hline
        R-INLA Integration Strategy & Empirical Bayes ({\tt EB}), Central Composite Design ({\tt CCD})\\ \hline
        R-INLA Approximation Strategy & {\tt G}aussian, {\tt S}implified Laplace, {\tt L}aplace\\ \hline
      \end{tabularx}
%      }} % makebox resize
    \end{tabular}
    \caption{ \footnotesize{ Parameters varied across the continuous simulation experiments. The total number of experiments was 16128 (Gaussian variance was not varied for Binomial experiments), each replicated 25 times.
        \label{tab:sim_params}}}
  \end{footnotesize}
\end{table}

\subsubsection*{Selected Continuous Simulation Results}
\label{sec:sim.cont.results}

The simulation comparisons presented in this section compare results from TMB against those from R-INLA using the default options: the simplified Laplace approximation and the CCD numerical integration scheme with mean and skew corrections to the marginals of the joint posterior sampling distribution (see Section~\ref{sec:inla}). While using the grid numerical integration scheme and the full Laplace approximations could offer an overall better approximation, and the empirical Bayes integration with Gaussian approximations would be closest to the approximations in TMB, the default INLA options provide a nice balance between computation and accuracy and are what many users explicitly or implicitly choose to use. For these reasons we felt this setting to be a useful and fair benchmark for which to contrast TMB. The extended online results also include comparisons with other combinations of INLA approximations and a number of additional selections are included in \ref{app:sim.cont}.

Additional continuous simulation results for all experiments in Table~\ref{tab:sim_params} can be explored via an interactive R-Shiny web application by visiting \href{https://faculty.washington.edu/jonno/software.html}{faculty.washington.edu/jonno/software.html}.

\begin{figure}[hb!] \centering \includegraphics[width=\linewidth]{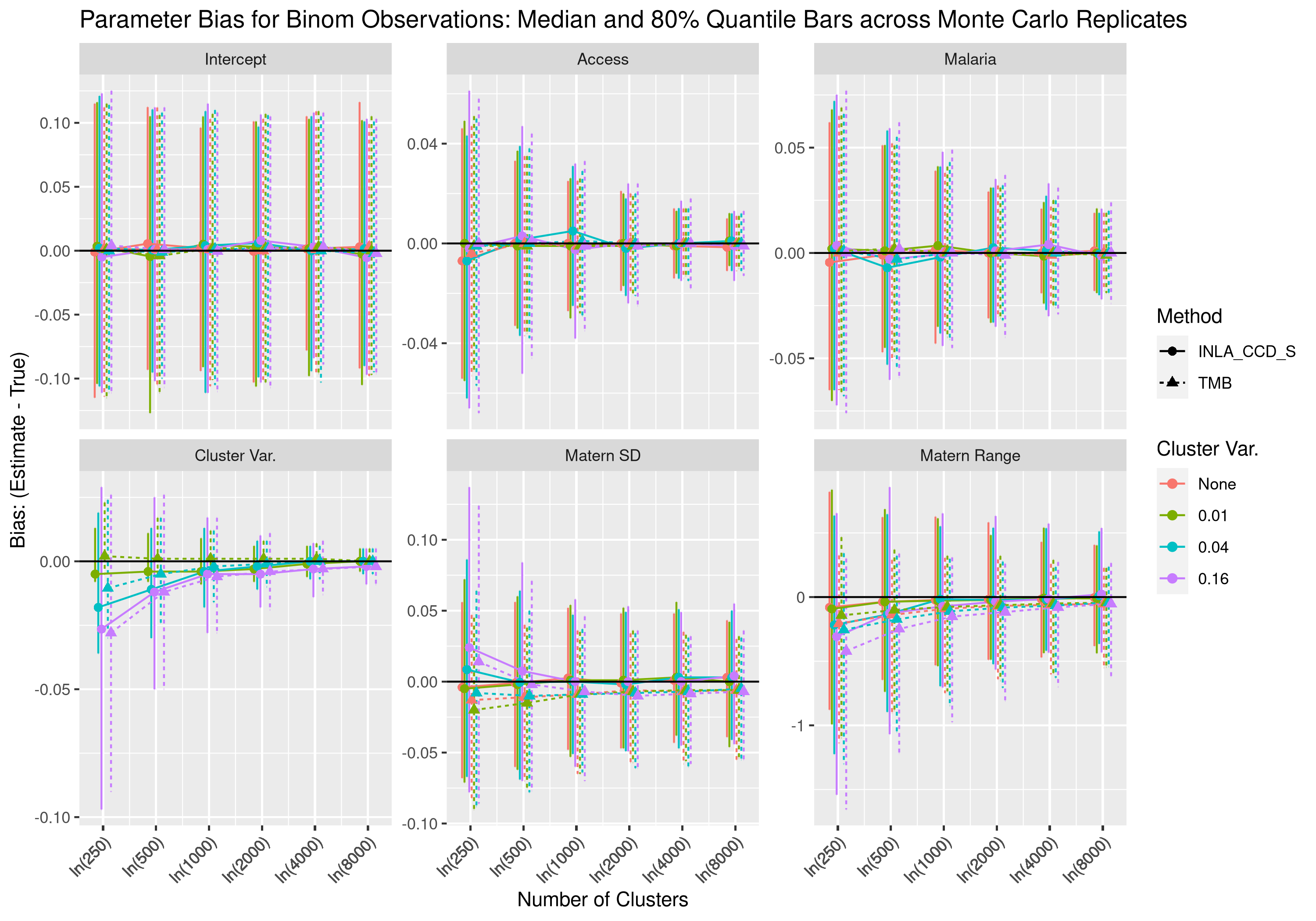} % \singlespace
\caption{ \footnotesize{ Comparison of the estimated parameter bias from TMB (dashed lines) and R-INLA using CCD hyperparameter integration and simplified Laplace approximations (solid lines) plotted against the number of cluster observations for Binomial observation experiments. Colors represent different cluster (i.i.d nugget) variances used in an experiment. Each point is the median bias of 3 experiments (coarse, medium, and fine SPDE triangulation), calculated across 75 replicates, and the bars represent the middle 80\% quantile range of the bias across replicates. \label{fig:binom.par.bias}}}
\end{figure}

\begin{figure}  \vspace{-3em} \centering \includegraphics[width=\linewidth]{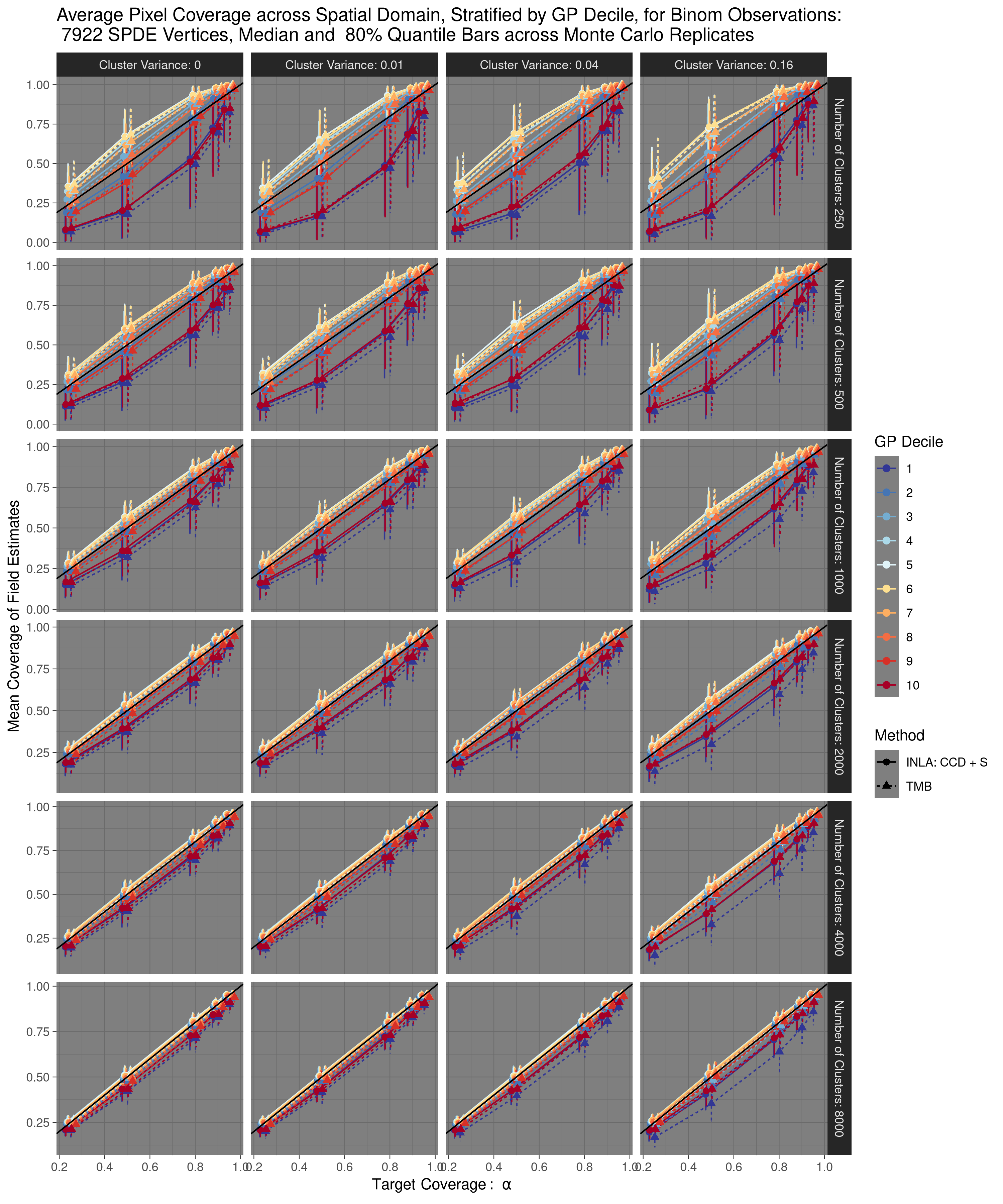} %\singlespace
  \caption{ \footnotesize{ Comparison of the average estimated field coverage of the simulated truth, faceted by cluster (i.i.d. nugget) variance and the number of clusters, from TMB (dashed lines) and R-INLA using CCD hyperparameter integration and simplified Laplace approximations (solid lines) plotted against the target nominal coverage, $\alpha$, for Binomial observation experiments with the medium resolution SPDE triangulation. Colors stratify pixels included in the average coverage calculation by the decile of the true GP for the experiment replicate. Each point is the median average coverage of an experiment, calculated across 25 replicates, and the bars represent the middle 80\% quantile range of the average coverage across replicates.\label{fig:binom.dist.cov}}}
\end{figure}

\begin{figure} \vspace{-3em}
  \centering \includegraphics[width=\linewidth]{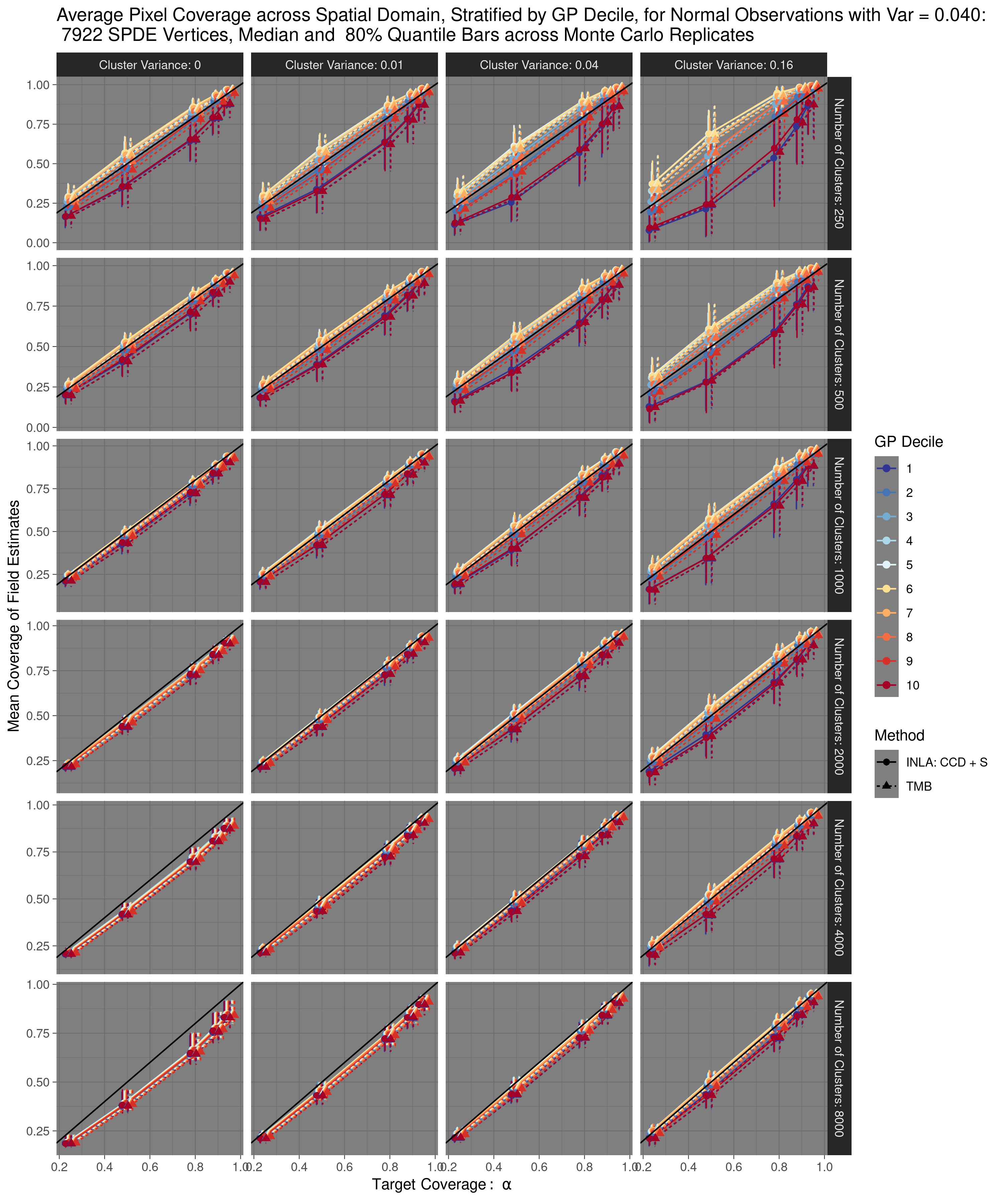} %\singlespace
   \caption{ \footnotesize{ Comparison of the average estimated field coverage of the simulated truth, faceted by cluster (i.i.d. nugget) variance and the number of clusters, from TMB (dashed lines) and R-INLA using CCD hyperparameter integration and simplified Laplace approximations (solid lines) plotted against the target nominal coverage, $\alpha$, for Gaussian observation experiments with $\sigma^2 =0.04$ and the medium resolution SPDE triangulation. Colors stratify pixels included in the average coverage calculation by the decile of the true GP for the experiment replicate. Each point is the median average coverage of an experiment, calculated across 25 replicates, and the bars represent the middle 80\% quantile range of the average coverage across replicates.\label{fig:normal.dist.cov}}}
\end{figure}

The following four figures display:
\begin{itemize}
\item Figure \ref{fig:binom.par.bias}: Binomial scenarios' parameter bias for the intercept, access and malarian incidence coefficients, cluster variance, and the Mat\'ern standard deviation and range.
\item Figure \ref{fig:binom.dist.cov}: Binomial, medium-resolution mesh, scenarios' mean pixel coverage, stratified by the value of the true GP, and faceted by cluster variance and number of spatial observations.
\item Figure \ref{fig:normal.dist.cov}: Normal $\sigma_{obs}^2=0.04$, medium-resolution mesh, scenarios' mean pixel coverage, stratified by the value of the true GP, and faceted by cluster variance and number of spatial observations.
%\item Figure \ref{fig:time} Single core (non-parallelized) fit, predict, and total timing comparisons between TMB and R-INLA.
\end{itemize}

\clearpage

While the overall results are quite similar, TMB generally has larger bias in the fixed effects estimators, particularly hyperparameters which may deviate further from Gaussianity, as shown in the continuous Binomial experiments, Figures \ref{fig:binom.par.bias}, and the continuous Gaussian experiments, Figure \ref{fig:normal.par.bias}. This trend appears consistently across a variety of INLA options, Figures \ref{fig:binom.par.bias.eb.g}-\ref{fig:binom.par.bias.ccd.l}.

In spatial statistics settings, the hyperparameters (and sometimes all parameters) may not be of inferential interest. Figures \ref{fig:binom.dist.cov} and \ref{fig:normal.dist.cov} demonstrate that TMB consistently yields results very similar to those from R-INLA at the spatial field level, and that the results are similar across all ranges of the spatial effect. Figures \ref{fig:binom.pix.cov} and \ref{fig:normal.pix.cov} show these results collapsed across the GP magnitude. In contrast to R-INLA, TMB seems to consistently have slightly lower coverage which could be attributed to the lack of integration over the hyperparameters even though the covariance estimator in (\ref{eq:tmb.var.correct}) attempts to account for this.

In the Binomial experiment, we saw no notable differences in the spatial field coverage across the different resolutions of the SDDE triangulation suggesting that the approximation was appropriately resolved. In the Gaussian data setting, we observed that the coarser meshes undercovered the field estimates in experiments with small $\sigma^2_{clust}$ and large sample sizes - but this was mostly remedied at the finer triangulation resolution. Interestingly, we saw more severe field undercoverage for larger numbers of spatial observations. These patterns were observed in results from both TMB and INLA. See the lower left plots of Figures \ref{fig:normal04.dist.cov.c}-\ref{fig:normal04.dist.cov.f}.

\clearpage
%%%%%%%%%%%%%%%%%%%%%%%%%%%%%%%%%%%%%%%%%%%%%%%%%%%%%%%%%%%%%%%%%%%%%%
%%%%%%%%%%%%%%%%%%%%%%%%%%%%%%%%%%%%%%%%%%%%%%%%%%%%%%%%%%%%%%%%%%%%%%
%%%%%%%%%%%%%%%%%%%%%%%%%%%%%%%%%%%%%%%%%%%%%%%%%%%%%%%%%%%%%%%%%%%%%%
\subsection{Discrete Spatial Simulation}
\label{sec:sim.disc}

For the discrete simulation study, we considered the BYM2 model (a modern formulation of the classic Besag-York-Mollie model developed by \cite{riebler2016}). This model requires a sum-to-zero constraint a hard constraint, using appropriate conditional densities \cite[Section 12.1.7.4]{gelfand2010}, was implemented in TMB to match the linear constraint used in the R-INLA BYM2 model formulation.

The discrete model was implemented over the 37 regions (first-level administrative units) of Nigeria, and neighbors were defined by to be regions with shared boundaries. Within each region, the population, $n_s$, for that area was first sampled from iid Poisson distributions. Conditional on the population, poisson observations were simulated from the following hierarchical model:
\begin{eqnarray}
  y_i|n_s, \eta_i &\sim &\text{Poisson}(n_s\times \eta_i)   \label{eq:disc.sim.hier} \\
  \eta_i &= &\text{exp}\left(\alpha + \bs b_i \right)\nonumber\\
  \bs b &= &\frac{1}{\sqrt{\tau}}\left( \sqrt{1-\varphi}\boldsymbol v + \sqrt{\varphi}\boldsymbol u_{\star} \right) \nonumber \\
  \bs v &\sim &N\left(\bs 0, \bs I\right)\nonumber \\
  \bs u_\star &\sim &N\left(\bs 0, \bs Q_{\star}^{-1} \right), {\text s.t. }   \sum_{i=1}^{37} u_{\star_i} = \bs 0, \nonumber
\end{eqnarray}
\noindent with $\alpha$ the GMRF intercept, fixed at -3 across simulations, BYM2 field $\bs b$ with total variance $\tau^{-1}$, mixing parameter $\varphi$ controlling the contribution of $\bs v$, the unstructured i.i.d. portion of the BYM2 field, and $\bs u_{\star}$, the scaled spatially structured component of the BYM2. The structured portion of the BYM2 is specified with precision $\bs Q_\star$, a scaled version of the precision from the classic BYM ICAR model, and is constrained to sum to zero.

The full combinatorial grid of simulation parameters is shown in Table~\ref{tab:disc_sim_params} consisted of a set of 20 experiments, with each experiment replicated 25 times to obtain Monte Carlo errors on the validation metrics. Additional details including prior specification and the constrained random effects density may be found in \ref{app:sim.disc}.

\begin{table}[h!]

\makebox[1 \textwidth][c]{       %centering table
\resizebox{1\textwidth}{!}{   %resize table

  \begin{tabularx}{1.\textwidth}{|l|X|}
    \hline
    Parameter    &  Simulation Values \\
    \hline
    \hline
    Data Observations & Poisson \\
    \hline
    Mean Observations per Region, $\text{E}[n_{\bs s}]$ & 16, 36, 49, 100, 400\\
    \hline
    BYM2 $\varphi$ & 1, $0.25, 0.5, 0.75, 0.9$ \\
    \hline
    BYM2 Variance, $(\tau^{-1})$ & $0.5$\\
    \hline
    GMRF Intercept & -3\\
    \hline
    R-INLA Integration Strategy  & Central Composite Design ({\tt CCD})\\
    \hline
    R-INLA Approximation Strategy  & {\tt S}implified Laplace\\
    \hline
  \end{tabularx}
}}  % makebox resize
  \singlespace \caption{ \footnotesize{ Parameters varied across the discrete simulation
      experiments. The total number of experiments was 20, and each
      was replicated 25 times.
      \label{tab:disc_sim_params}}}
\end{table}

\subsubsection*{Discrete Simulation Results}
\label{sec:sim.disc.results}

\begin{figure}
  \centering
  \includegraphics[width=\linewidth]{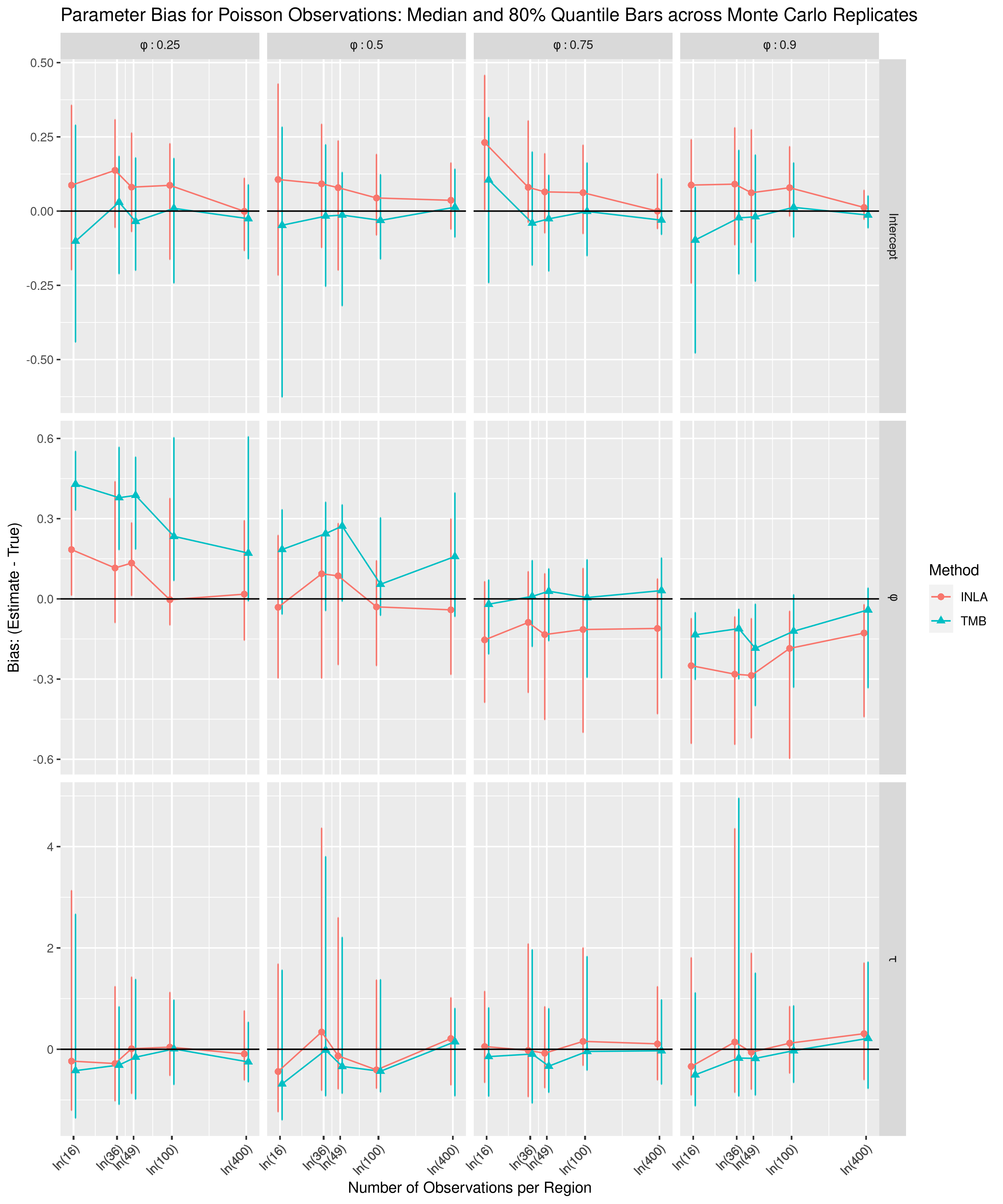}
  \singlespace \caption{ \footnotesize{ Comparison of the estimated
      parameter bias from TMB (red) and R-INLA (blue) plotted
      against the number of observations per region for Poisson data
      experiments with varying values of the true BYM2 $\varphi$. Each point
      is the median bias of 1 experiments, calculated across 25
      replicates, and the bars represent the middle 80\% quantile
      range of the bias across replicates. \label{fig:disc.bias}}}
\end{figure}

\begin{figure}
  \centering
  \includegraphics[width=\linewidth]{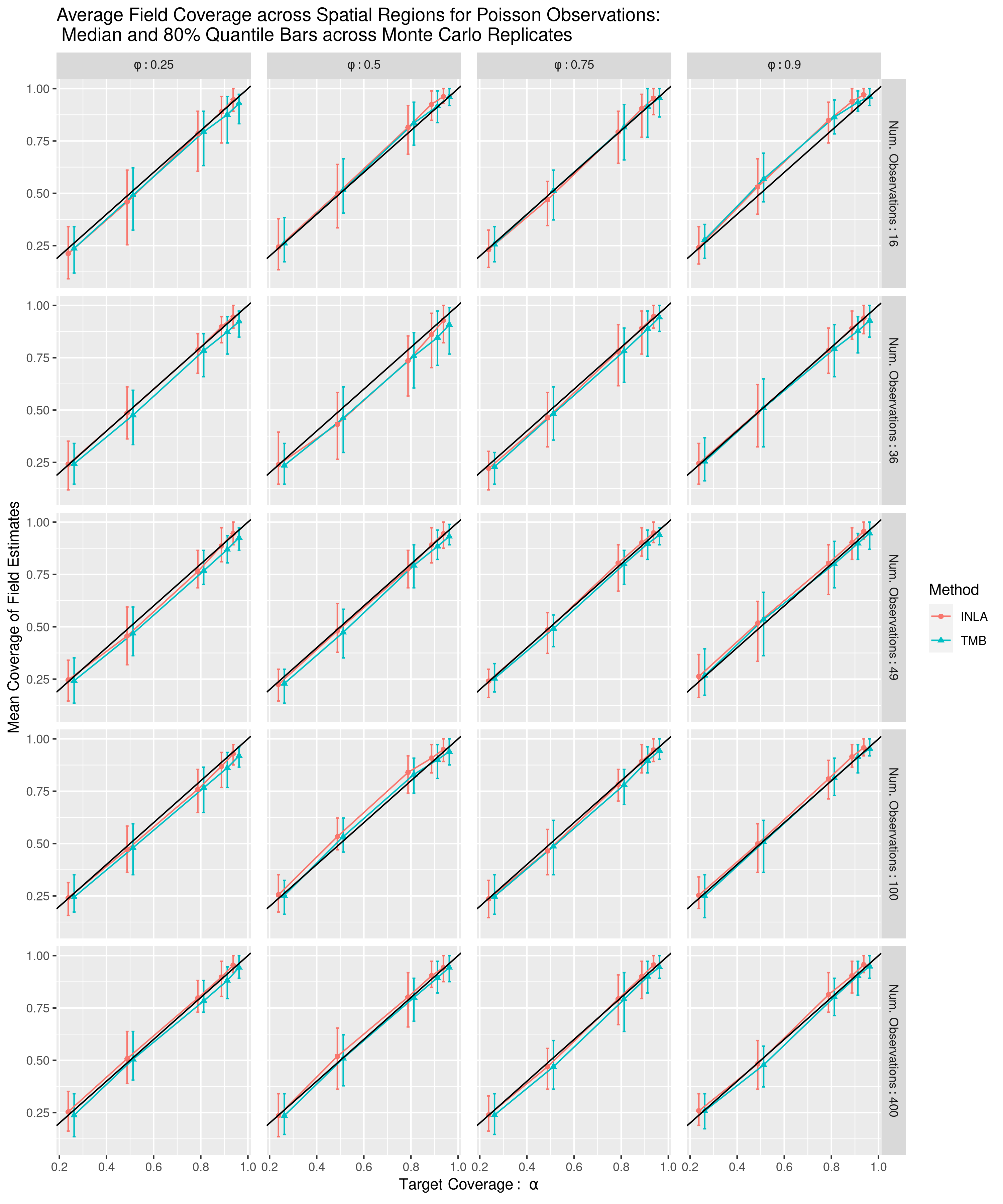}
  \singlespace \caption{ \footnotesize{ Comparison of the average
      estimated region coverage of the simulated truth, faceted by
      values of the true BYM2 $\varphi$ and number of observations per
      region, from TMB (red) and R-INLA (blue), plotted against
      the target nominal coverage, $\alpha$, for Poisson observation
      experiments. Each point is the median average coverage
      of an experiment, calculated across 25 replicates, and the bars
      represent the middle 80\% quantile range of the average coverage
      across replicates. \label{fig:disc.cov}}}
\end{figure}

The results from the discrete simulations, summarized in term of parameter bias in Figure \ref{fig:disc.bias} and in terms of spatial field coverage in Figure \ref{fig:disc.cov}, share many similarities to those observed in the continuous simulations of Section \ref{sec:sim.cont.results}. Both methods produce very similar spatial field estimates though TMB appears more likely to have larger bias for some parameters.

\clearpage % make latex flush all floats (tables/images) before this next section

\section{European Breast Cancer Application}
\label{sec:euro.bc}

Data quality for cancer monitoring vary significantly across the European Union (EU) and range from complete registry coverage and high quality mortality estimates from vital registration to countries with no data. We use TMB to fit a similar model to one previously detailed by \cite{mercer2016} and fitted with user-written MCMC code. We describe a two-level nonlinear model which first assumes a Poisson model for cancer incidence and then, conditional on cancer counts, models deaths as a binomial outcome. This model appropriately handles the variety of cancer data present in the EU in part because the conditional two-level Poison-Binomial form induces a Poisson model for unconditional mortality to account for countries without incidence data from registries. Using data provided by the International Agency for Research on Cancer (IARC), we implement a Bayesian spatial smoothing model to borrow strength between countries to provide estimates of national incidence and mortality, along with measures of uncertainty.

The approach directly models mortality (which is more universally available) and the mortality-incidence (MI) ratio, to estimate both incidence and mortality for all countries. The model synthesizes four types of country data: type I countries with national incidence and mortality data, type II countries with sub-national incidence and mortality data (from registries) in addition to national mortality, type III countries with only national mortality, and type IV countries with no available data. While there is reason to think cancer incidence may be spatially correlated across countries, for example due to environmental and lifestyle risks, different preventions and screening strategies may result in large variability between nearby countries. For this reason, the BYM2 model presented in Section \ref{sec:sim.disc} was used to model a combination of spatial and iid country effects. We would also expect some smoothness in mortality over space due to similarities in GDP, and therefore healthcare, in close by countries.

\begin{figure}[h!]  \centering \includegraphics[width=\linewidth]{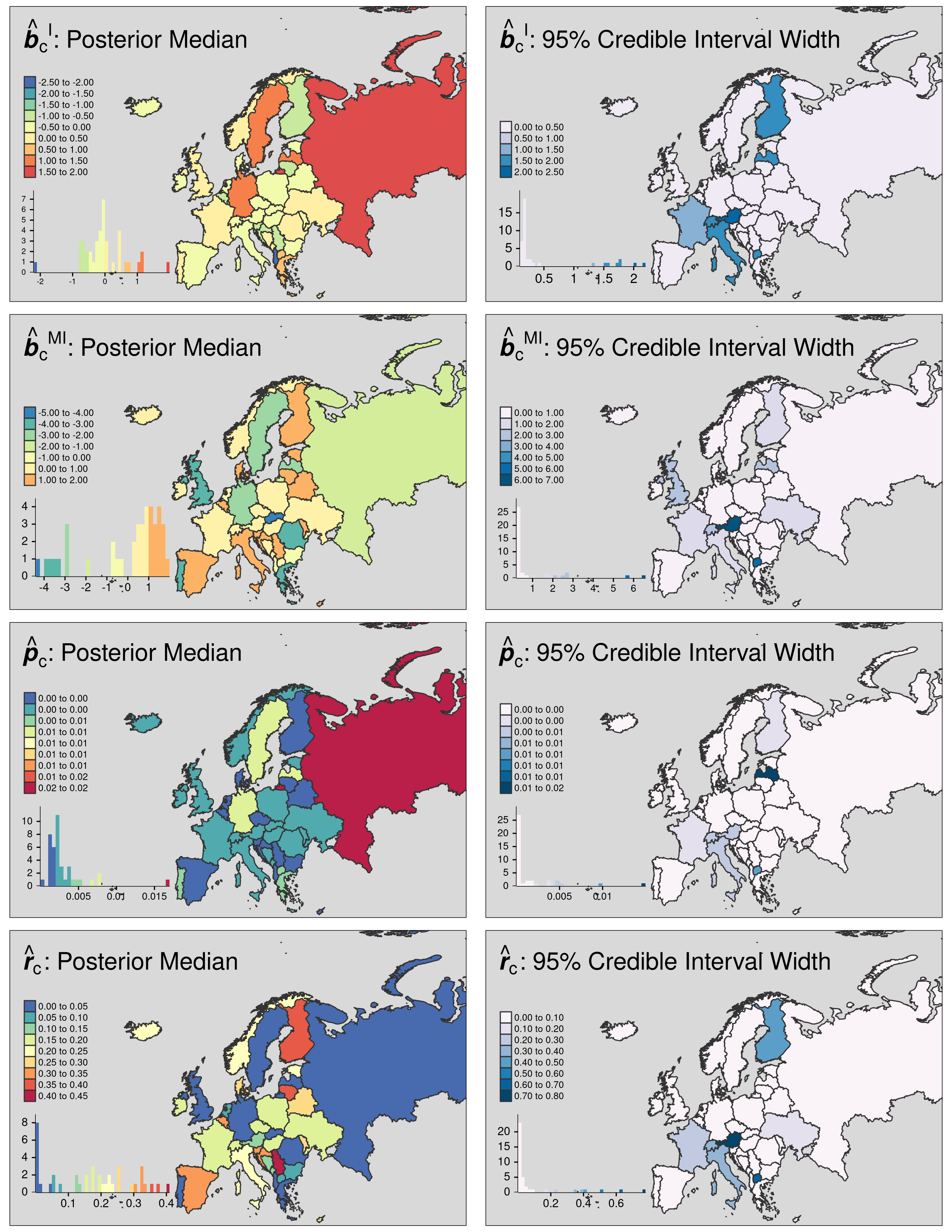} %\singlespace
   \caption{ \footnotesize{ Each row consists of a pair of posterior median and 95\% credible interval widths for the estimated quantity. Row 1: the BYM2 country random effects for incidence, row 2: the BYM2 country random effects for MI ratio, row 3: estimates country incidence rates, and row 4: estimated country mortality|incidence probabilities.}}
\label{fig:euro.bc.maps}
\end{figure}

For countries, $c$, that have both national mortality and incidence data (type I) we assume a Poisson process with rate $p_c$ for cancer incidence, where $Y_c$ is the number of reported individuals with breast cancer from a population of $N_c$. Conditional on having cancer, we model total mortality, $Z_c$, as a binomial outcome with probability of death $r_c$ for each of the $Y_c$ individuals with cancer. This induces a Poisson process for mortality when incidence is unobserved with rate $p_c\times r_c$. For illustration, we work with only with data from 2008 in women aged 50-54. Our base model for type I countries is:
\begin{eqnarray} Y_c|N_c,p_c &\sim& \mbox{Poisson}(N_c\times p_c),\hspace{1em} p_c = \mbox{exp}(\alpha_c^I)\label{eq:1} \\
  Z_c|Y_c,r_c &\sim& \mbox{Binomial}(Y_c, r_c),\hspace{1em} r_c = \frac{\mbox{exp}(\alpha_c^{MI})}{1+\mbox{exp}(\alpha_c^{MI})}\label{eq:2}
  \end{eqnarray} which implies the unconditional mortality model:
  \begin{eqnarray}
    Z_c|N_c,p_c &\sim \mbox{Poisson}(N_c q_c),\hspace{1em} q_c = p_c\times r_c\label{eq:3}.
  \end{eqnarray}
  We assume a log- and logit-linear model for incidence and conditional mortality and we assign the following forms:
  \begin{eqnarray}
    \alpha_c^I &=& \alpha^I + b_c^I \label{eq:4}\\ \alpha_c^{MI} &=& \alpha^{MI} + b_c^{MI} \label{eq:5}
  \end{eqnarray}
  where $p_c$ is the reported incidence, $r_c$ is the reported mortality, $\alpha^I$ and $\alpha^{MI}$ are global intercepts, and $b_c^I$ and $b_c^{MI}$are country random effects that are assumed to have BYM-2 structure comprising of a spatially correlated term as well as an unstructured (iid) country specific term. Additional details on the data and the model may be found in \ref{app:euro.bc}. Maps of $\hat{b}_c^I$, $\hat{b}_c^{MI}$, $\widehat{p}_c$, and $\widehat{r}_c$, with measures of uncertainty, are presented in Figure~\ref{fig:euro.bc.maps}.

The nonlinear unconditional mortality rate, $q_c = p_c\times r_c$, is necessary to include countries with incomplete or missing mortality data and prohibits this model from being fit within INLA. A similar model that used country-level fixed effects without the spatial random effect would be possible in INLA \citep{meehan20} but without complete data in each country this is not feasible.

\clearpage

\section{Discussion}\label{sec:discussion}

We were pleasantly surprised to find near concurrence in spatial field estimator distributions in TMB and R-INLA -- in both continuous and linearly constrained discrete model settings -- across a wide range of simulation parameters. The generally smaller parameter bias in INLA results may be due to the integration over the fixed effects, similar to restricted maximum likelihood (REML) inference which is known to reduce bias, in contrast to the ML inference performed in TMB. One possible remedy for this could be to effectively enable REML in TMB by adding all linear fixed effects parameters to the list of `random' parameters input by the modeler.

The field coverage figures clearly demonstrate that random effects coverage is a function of the magnitude of the effect (\cite{yu2018}) and should serve as a warning for those trying to interpret coverage of specific portions of estimated spatial fields. The discrete Poisson experiment showed excellent recovery of the BYM2 field and the continuous models, with the SPDE approximation to the GP performed nearly as well. We see that the SPDE approximation in the binomial scenarios appears more robust than in the Gaussian and the undercoverage under extremely large data volumes warrants further investigation. Even so, the SPDE approximation performs very well across a wide range of experiments and by increasing the density of the mesh until the results no longer change the user should be able to confidently determine when the approximation is resolved. We believe these to be the largest-scale simulation studies on the SPDE approach to fitting GPs in TMB or in R-INLA and we hope that the demonstrated success encourages continued use of this convenient approach.

The simulations also provided an opportunity to assess the relative computational burdens of TMB and R-INLA. \ref{app:timing} describes both the serial (one CPU thread) and parallelized timing experiments. Figures \ref{fig:time} and \ref{fig:time.parallel} demonstrate that TMB scales extremely well, even in comparison to R-INLA which was designed, in part, to be a computationally efficient and quick alternative to MCMC sampling. We note that timing tests attempting to replicate `real world' experience are quite difficult and these timing results only provide some crude measures of computational cost. Nonetheless, TMB appears to perform quite favorably and, at minimum, it could be used to quickly test and iterate models before final inference is performed using the modeler's method of choice.

One of the main limitations of this study is the lack of tuning of any particular simulation and inference. Due to the sheer number of experiments, a set of default model parameters (for example, the starting values) were used across simulations. The simulation code was built to catch and relaunch convergence issues in both TMB and R-INLA that may, with human interaction, have been remedied. While sensible tuning decisions were attempted, this also implies a lack of sensitivity analysis on the prior selection for any one simulated dataset or experiment. Furthermore, these simulation studies use correctly specified models and do not attempt to study either tool under model misspecification.

The novel incidence-mortality model from Section \ref{sec:euro.bc} demonstrates the utility of the flexibility TMB provides. Simulation results for this model shown in Figure \ref{fig:simres} indicate that TMB is capable of recovering parameter estimates from nonlinear mixed effects models that cannot be fit within R-INLA. One of the main tradeoffs in return for this flexibility is relative difficulty of constructing the {\tt C++} templates. There has already been efforts, such as the {\tt glmmTMB} {\tt R}-package (\cite{glmmTMB}) which allows users to run {\tt TMB} from within {\tt R} using standard {\tt R} notation for mixed effects models, to streamline the use of TMB, but none will likely be able to offer the freedom of directly coding within the {\tt C++} templates. We note that there are numerous tutorial documents and examples available on the TMB github page and the TMB authors have worked to make it easier for {\tt R} coders to use. Example code for fitting spatial GPs via the SPDE approach is available in \ref{app:code} to provide a sense of modeling in TMB and R-INLA, and the full code used in this study is available online to serve as starting points for those interested in spatial modeling in TMB. Furthermore, others have previously used TMB in spatial settings, such as classic spatial models (for example, \cite{dwyer2016}), and more recent work to account for missing spatial information (\cite{wilson:wakefield:20}, \cite{wilson:wakefield:21} and \cite{marquez:wakefield:21}), and their code is available too.

The primary restriction of TMB's applicability lies in the assumptions underlying the LA. Although many models may be fit within the TMB framework, it is clear that the LA will not perform equally well on all of them - though intelligently chosen reparameterization, like those used for the hyperparameters in both simulations, can help significantly. While the quality of the LA may be difficult to assess, the recent availability of the {\tt tmbstan} package permits MCMC sampling from TMB models - with or without the integration of the random effects performed by LA. By comparing results run with TMB and the LA against those run with MCMC without the LA, it is possible to assess the quality of the LA in particular applications. The speed of TMB makes it a useful option for iterating through models during exploration phases of research and it may provide opportunities to fit models that could not otherwise be fit in reasonable amounts of time or at all.

After conducting these extensive simulations, we have found TMB to perform more than adequately in comparison to existing well-trusted tools. We hope that this unified and detailed explanation of TMB and the additional new model demonstrations will improve understanding, instill confidence, and generate further interest and use in a compelling and generally unknown statistical computational tool.

\section*{Acknowledgments}
\label{sec:ack}
We thank the International Agency for Research on Cancer (IARC) for access to the cancer data used in this study. This work was partially supported by the Institute for Health Metrics and Evaluation (IHME), and we thank them for use of the computing cluster needed to run the continuous spatial simulations and for their support.

\bibliographystyle{format_style/chicago} \bibliography{library}

%\begin{thebibliography}{00}
%
%%% \bibitem[Author(year)]{label} %%% Text of bibliographic item
%\end{thebibliography}

\appendix

\clearpage

\section{Automatic Differentiation}
\label{app:ad}

This appendix is provided as a brief introduction and overview for those unfamiliar with automatic differentiation (AD), also known as algorithmic differentiation. AD comprises a set of computational techniques used to evaluate the derivate of functions within a computer framework. The methods generally work by noting that computers must break down even the most complicated functions into elementary or unary arithmetic operations in order to evaluate them, and that by (repeatedly) applying the chain rule to these operations, derivatives of different orders may also be numerically evaluated. Furthermore, it is well-established that the computational cost to evaluate the derivatives will be no more than a small multiplicative factor above the cost to evaluate the original function. To help provide some intuition with AD, we will use simple linear regression to provide an example of the fundamental AD process.

For $i \in \{1, \ldots, n\}$, assume a simple linear regression model:
\begin{eqnarray*} y_i &= a + b\cdot x_i + \epsilon_i \\ \epsilon_i &\sim N(0, \sigma^2)
\end{eqnarray*} where the common least squares the values for $a$ and $b$ are determined by minimizing the objective function:

\begin{eqnarray}
  \label{eq:rss} RSS = \sum_{i=1}^n\bigg(y_i - (a+b\cdot x_i) \bigg)^2.
\end{eqnarray}

As a minimization problem, the solution may be determined by symbolically differentiating the function in (\ref{eq:rss}) with respect to both $a$ and $b$, setting both equations to zero and simultaneously solving for the two unknowns:

\begin{eqnarray}
  \label{eq:rss_partials} \frac{\partial RSS}{\partial a} &= \sum_{i=1}^n-2(y_i - (a + b\cdot x_i)) = 0 \nonumber \\ \frac{\partial RSS}{\partial b} &= \sum_{i=1}^n-2x_i\cdot(y_i - (a + b\cdot x_i)) = 0.
\end{eqnarray}

For a computer to compute these derivatives through AD, it would generate a list of all the unary operations needed to evaluate (\ref{eq:rss}), as well as the derivatives for each unary operation, which can then be combined using the chain rule to arrive at the forms of the partial derivatives shown in (\ref{eq:rss_partials}). The elementary steps taken to do this are shown in Table~\ref{tab:ad} and (\ref{eq:chain_rule}).

\begin{table}
\begin{center}
  \begin{tabular}{cccc} Unary Evaluation & Unary Operation & Symbolic Value & Partials \\
    \hline \hline
    \multirow{2}{*}{1} & \multirow{2}{*}{$u_1 = b\cdot x_i$} & \multirow{2}{*}{$b\cdot x_i$} & $\frac{\partial u_1 }{\partial b} =x_i$ \\
    & & & $\frac{\partial u_1 }{\partial x_i} = b$\\ \hline
    \multirow{2}{*}{2} & \multirow{2}{*}{$u_2 = a + u_1$} & \multirow{2}{*}{$a+b\cdot x_i$} & $\frac{\partial u_2}{\partial a} = 1$\\
    & & & $\frac{\partial u_2}{\partial u_1} = 1$ \\ \hline
    \multirow{2}{*}{3} & \multirow{2}{*}{$u_3 = y_i - u_2$} & \multirow{2}{*}{$y_i - (a+b\cdot x_i)$} & $\frac{\partial u_3}{\partial y_i} = 1$ \\
    & & & $\frac{\partial u_3 }{\partial u_2} = -1$ \\ \hline
    \multirow{2}{*}{4} & \multirow{2}{*}{$u_4 = u_3^2$} & \multirow{2}{*}{$\bigg(y_i - (a+b\cdot x_i)\bigg)^2$} & \multirow{2}{*}{$\frac{\partial u_4 }{\partial u_3} = 2u_3$}\\
    & & & \\ \hline
    \multirow{2}{*}{5} & \multirow{2}{*}{$RSS_i = u_4$} & \multirow{2}{*}{$\bigg(y_i - (a+b\cdot x_i)\bigg)^2$} & \multirow{2}{*}{$\frac{\partial RSS_i }{\partial u_4} = 1$}\\
    & & & \\ \hline
\end{tabular}
\end{center} %\singlespace
\caption{\footnotesize{Unary operations taken to evaluate the objective function defined in (\ref{eq:rss}) as well as the numeric evaluation of the first partial derivatives shown in (\ref{eq:rss_partials}).}}\label{tab:ad}
\end{table}

Once the function has been broken into its elementary operations and the partials have been derived for each of the basic elementary operations, the partials of the complete function can be quickly evaluated through the chain rule:

\begin{eqnarray}
  \label{eq:chain_rule}
  \frac{\partial RSS}{\partial a} &= \sum_{i=1}^n \frac{\partial RSS_i}{\partial a} = \sum_{i=1}^n\frac{\partial RSS_i}{\partial u_4}\frac{\partial u_4}{\partial u_3}\frac{\partial u_3}{\partial u_2}\frac{\partial u_2}{\partial a} \nonumber \\
                                  & = \sum_{i=1}^n 1\cdot 2 u_3 \cdot -1 \cdot 1 = \sum_{i=1}^n -2 \bigg(y_i - (a+b\cdot x_i)\bigg) \nonumber\\
  \frac{\partial RSS}{\partial b} &= \sum_{i=1}^n \frac{\partial RSS_i}{\partial b} = \sum_{i=1}^n\frac{\partial RSS_i}{\partial u_4}\frac{\partial u_4}{\partial u_3}\frac{\partial u_3}{\partial u_2}\frac{\partial u_2}{\partial u_1}\frac{\partial u_1}{\partial b} \nonumber \\
                                  & = \sum_{i=1}^n 1\cdot 2 u_3 \cdot -1 \cdot 1\cdot x_i = \sum_{i=1}^n -2x_i \bigg(y_i - (a+b\cdot x_i)\bigg).
\end{eqnarray}

Once this formulation of the derivative has been built, the derivative may be quickly evaluated for different values of the parameters, $a$, and $b$, conditional on the data, $\bs y$ and $\bs x$. Of course, higher order derivatives may then be calculated by iteratively applying AD.

When TMB compiles a function to quickly evaluate a function, often a likelihood, that has been coded up in a {\tt C++} template, it also returns a function, generated through AD, that can quickly evaluate the gradient of the function, as well as the Hessian. The gradient can then be used to more efficiently find the minimum of the function, or it could be used to sample directly from the posterior in, e.g., a Hamiltonion MCMC. The Hessian, as in (\ref{eq:tmb.hess}), can thus be quickly calculated with AD without the need for human generated symbolic differentiation and explicit coding - which can be tedious and error prone - and can then be used in TMB's Laplace approximation shown in (\ref{eq:tmb.laplace}).

See \cite{fournier2012} and \cite{kristensen2016} for further details about the AD methods used in TMB and \cite{griewank2008} for more general AD theory.

\clearpage

\section{Continuous Spatial Simulation Study}
\label{app:sim.cont}

An overview of the continuous simulation details is provided in Section \ref{sec:sim.cont}. Here we provide an in-depth version with all remaining details and an additional selection of results.

\subsection{Continuous Spatial Simulation Details}
\label{app:sim.cont.details}

We consider Gaussian process (GP) models with mean $\mu(\bms)$ at location $\bms$, and the Mat\'ern function \citep{stein:1999} where the covariance between two spatial locations distance $||\bms_i - \bms_j||$ from one another is:
\begin{equation}
  C(u(\bms_i), u(\bms_j)) = \frac{\sigma_\text{m}^2 }{2^{\nu-1}\Gamma(\nu)}\left( \kappa||\bms_i - \bms_j|| \right)^{\nu}K_{\nu}\left(\kappa||\bms_i - \bms_j|| \right), \label{eq:matern.app}
\end{equation} where $\sigma_\text{m}^2$ is the variance, $\kappa > 0$ is a scaling parameter related to the range, $r_m=\frac{\sqrt{8\nu}}{\kappa}$, defined to be the distance at which the spatial correlation drops to 0.1, $\nu > 0$ is related to the smoothness of the field, and $K_\nu$ is the modified Bessel function of the second kind. For any finite collection of locations within the domain, $\{\bms_1, \ldots, \bms_{n}\}\in\mathcal{D}$, the random vector $\bmu = [u(\bms_1),\ldots,u(\bms_{n})]$ has multivariate Gaussian distribution with precision matrix $\bmQ$.

\begin{figure}[h!]
  \includegraphics[width=\linewidth]{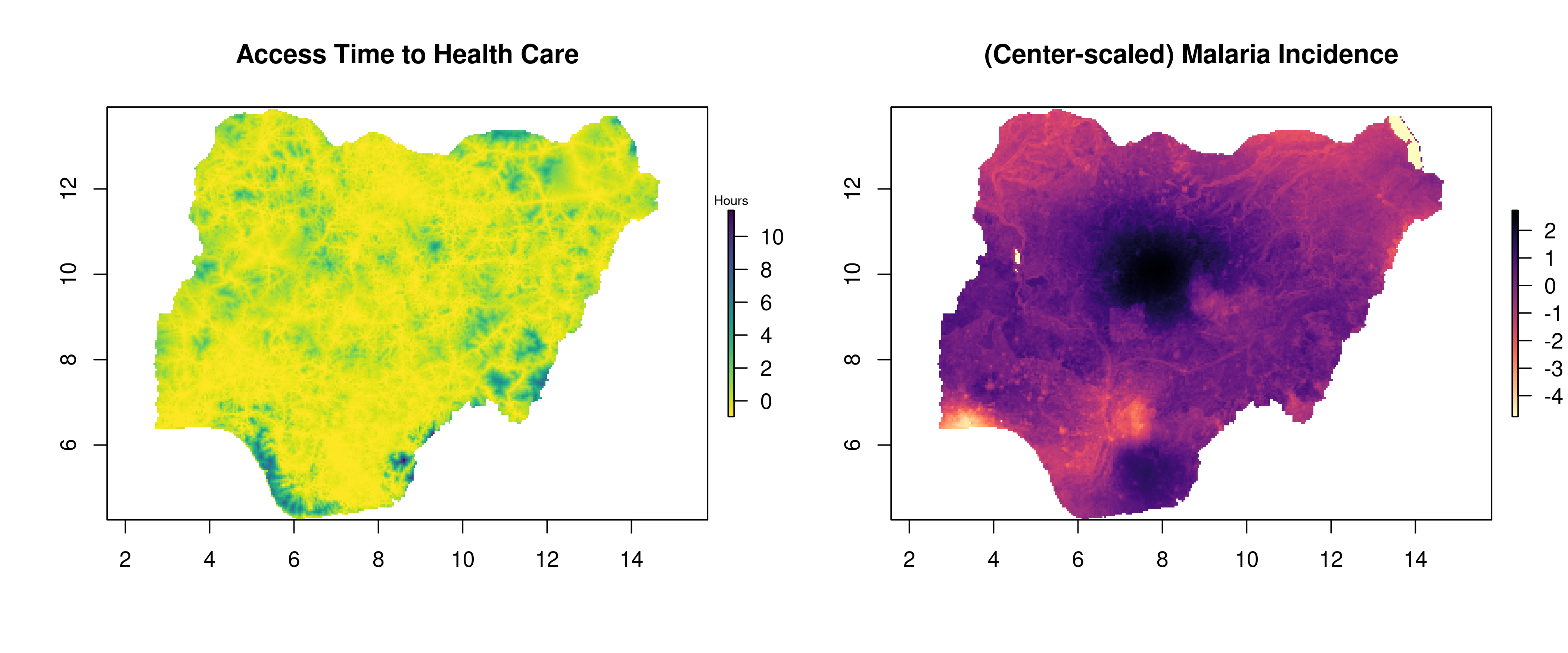}
  \singlespace \caption{\footnotesize{Covariates used in the continuous simulation studies: access (time in hours) to healthcare (\cite{weiss2018}) and malaria incidence (\cite{weiss2018}) in Nigeria.}}
  \label{fig:covs}
\end{figure}

The simulated data, sampled at locations $\bms_i,\ i=1, \ldots, n_s$, arise from the following hierarchical model:
\begin{eqnarray*}
  \bmy |\bbeta, \bmb ,\bmphi_1&\sim& p_1(\bmy |\bbeta, \bmb,\bmphi_1) \label{eq:sim_hier.app} \\
  \mbox{E}[y_i|\bbeta,u_i , v_i] &=& g^{-1}\left(\bmz^{\tiny{T}}_i \bbeta+ u_i + v_i\right)\nonumber\\
  \bmu &\sim &\mbox{N}(\bmzero, \bmQ(r_\text{m}, \sigma^2_\text{m}))\nonumber\\
  \bmv &\sim& \mbox{N}(\bmzero,\bident_{n_s} \sigma^2_\text{clust}). \nonumber%% \\ %%\bs Q &= precision\ from\ Mat\acute ern. \nonumber
\end{eqnarray*} where $\bident_{n_s}$ is the $n_s \times n_s$ identity matrix and the last two lines correspond to $p_2(\bmb | \bmphi_2)$ with $\bmb=[\bmu,\bmv]$ and $\bmphi_2=[\sigma^2_\text{m},r_\text{m},\sigma^2_\text{clust}]$.  The precision for the spatial GP, $\bmQ$ is Mat\'ern with range $r_\text{m}$ and standard deviation $\sigma_\text{m}$. In some models, two spatially varying covariates, shown in Figure~\ref{fig:covs}, were included: access time to heath care (\cite{weiss2018}) and malaria incidence (\cite{weiss2019}) While the SPDE representation is used to fit the spatial fields (using three different triangulation resolutions shown in Figure \ref{fig:meshes}, the true fields are simulated directly on a high resolution regular grid using the {\tt RandomFields R} package \citep{schlather2015}.

To complete the model specification, the following priors are specified:
\begin{eqnarray*} \bbeta &\sim &\mbox{N}(0, 5^2) \\ r_\text{m}, \sigma_\text{m} &\sim &\mbox{PCspde}(u_r= 10, \alpha_r=.95, u_\sigma =1, \alpha_\sigma = .05) \\ \tau_\text{clust} = \sigma^{-2}_\text{clust} &\sim& \mbox{PCprec}(u=.5, \alpha=.05)
\end{eqnarray*} The penalized complexity (PC) priors \citep{simpson2017,fuglstad2019} shrink towards a base model and are set such that $\mbox{Prob}(r_\text{m} < 10^\circ) = .95$, $\mbox{Prob}(\sigma_\text{m} > 1) = .05$ and $\mbox{Prob}(\sigma_\text{clust} > .5) = .05$.

Internally, TMB was coded to use the same internal representations of parameters used in R-INLA: $\text{log}(\kappa)$, and $\text{log}(\tau_*)$.

Simulations are performed with $n_i$ observations taken at each spatial location $\bms_i,\ i=1, \dots, n_s$. Experiments are run on Gaussian data with $\mbox{Var}(y_i) = \sigma_\text{obs}^2/n_i$, using an identity link function, and with the same PCprec prior on the observation variance, $\sigma^2_\text{obs}\in\bphi_1$, as is used for $\sigma^2_\text{clust}$. Experiments run on binomial data have no $\sigma^2_\text{obs}$ and use a logit link function.

\begin{figure}[h!]
  \centering
  \subfloat[Stratified spatial observations]{{\includegraphics[width=.5\textwidth]{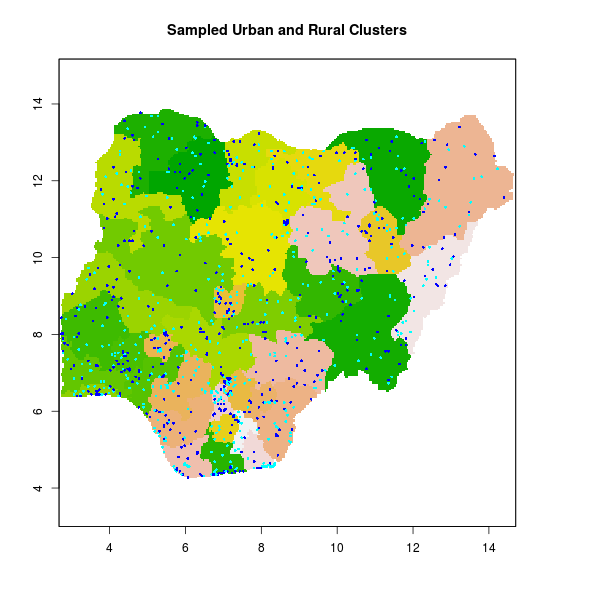} }}
  %% \qquad
  \subfloat[`True' latent surface]{{\includegraphics[width=.5\textwidth]{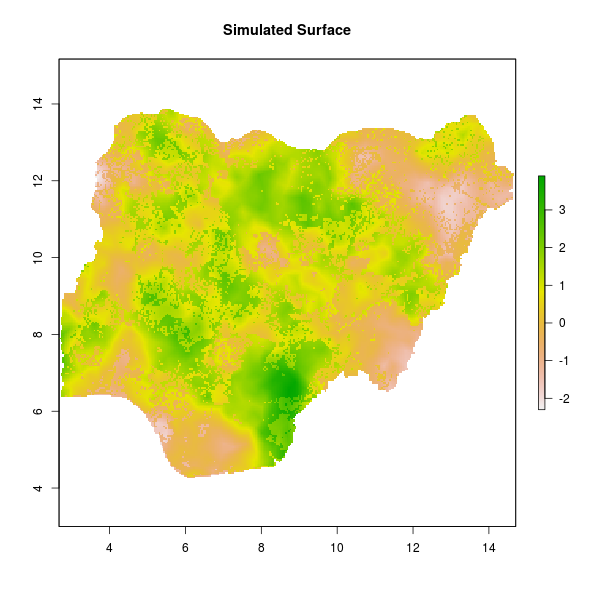} }}

  \singlespace \caption{ \footnotesize{ Examples of (a) simulated
      cluster locations and (b) a simulated latent surface comprised
      of covariates and GP.}}
  \label{fig:sim.locs.field}
\end{figure}

\subsection{SPDE Details}
\label{app:spde.details}

Spatial models can be notoriously difficult to fit at scale and in this study we use the SPDE finite element method representation to approximate the GPs. \cite{lindgren2010} prove that specific discretely indexed GMRF models defined on triangulations can approximate continuous spatial GPs with Mat\'ern covariance (\ref{eq:matern.app}). This relationship relies on the fact that solutions to a specific SPDE class have the Mat\'ern covariance and that their GMRF models are approximate solutions to the SPDE and allows the use of fast and efficient sparse matrix operations permitted on GMRFs to applied to GP models. The specific SPDE of interest takes the form:
\begin{equation}
(\kappa^2 - \Delta)^{\alpha/2}(\tau x(\bms)) = \mathcal{W}(\bms), \label{eq:spde}
\end{equation}
where $\Delta$ is the Laplacian, $\alpha$ controls the smoothness, $\kappa>0$ the scale, $\tau$ the variance, and $\mathcal{W}(\bms)$ is a Gaussian spatial white noise process. \cite{whittle1954} showed that the exact stationary solution to this SPDE is the stationary Gaussian field $x(\bms)$. There is a well-defined relationship between the parameters in (\ref{eq:spde}) and \ref{eq:matern.app}) \cite[Section 6.5]{blangiardo2015}. The GMRF approximation uses a finite element method basis function representation defined on a tringulation over the domain:
\begin{equation}
x(\bms) \approx \sum_{i=1}^V \psi(\bms)w_i \nonumber
\end{equation}
with compact deterministic basis functions $\psi(\bms)$ with weights $w_i$ summed over $V$ vertices in the triangulation. For specific basis functions, the vector of weights, $\bmw$, are Gaussian with mean zero and sparse precision matrix that depends on the parameters in (\ref{eq:spde}). See \cite{lindgren2010} for details.

One of the key aspects of this approximation, as demonstrated by \cite{righetto2020}, is the choice of triangulation. Unlike their work, we have chosen to generate the triangulations to be constant density over the domain without using any of the data locations. The reason for this choice was two-fold. First, it allowed us to fix the mesh across the experiments and replications, removing one source of known variability. This allows us to more readily study the effect of the mesh density which has been shown to be a driver of oversmoothing (\cite{teng2017}). Secondly, the discretization error varies by size of the triangles and triangulations with pronounced variability in resolution may lead to different effects of the spatial field parameters across the domain.

\begin{figure}[!h]
  \includegraphics[width=\linewidth]{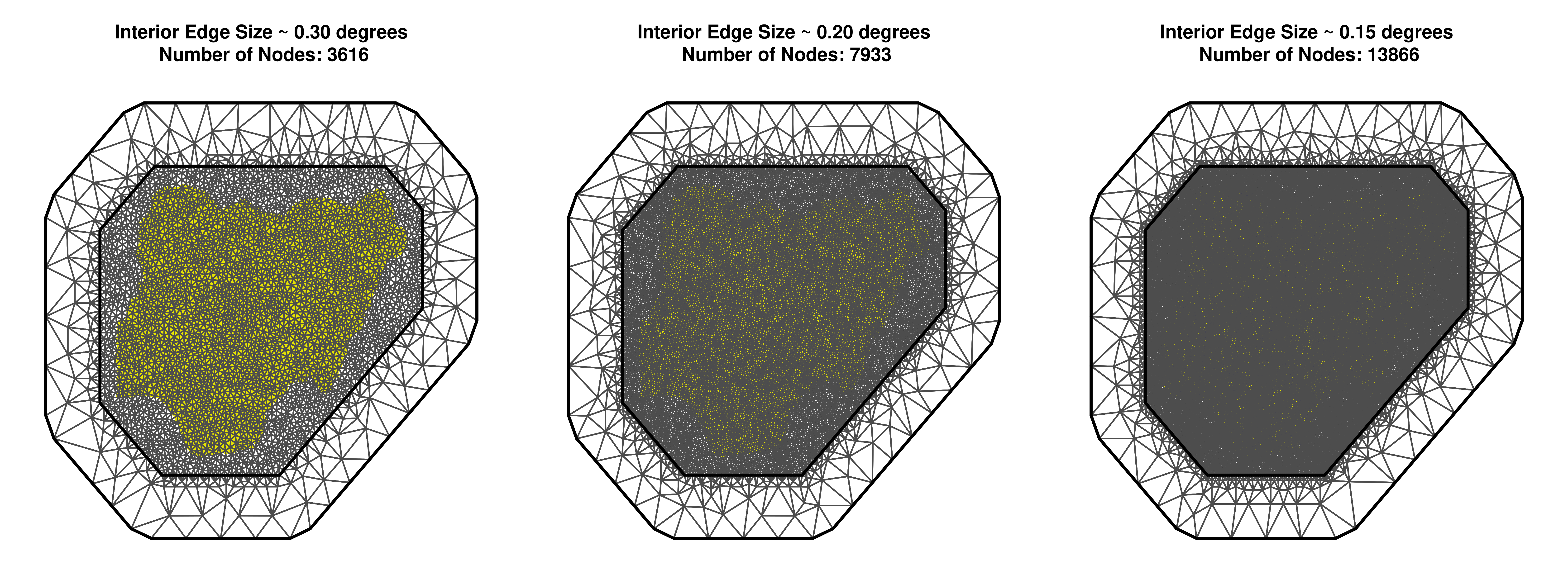}
  \singlespace \caption{\footnotesize{Coarse, medium, and fine
      resolution Delauney triangulations used in the SPDE
      approximation to the GP. The outline of the spatial domain,
      Nigeria, is shown beneath the mesh for reference.}}
  \label{fig:meshes}
\end{figure}

In the continuous spatial simulation we used three different mesh resolutions characterized by the largest allowed triangle edge length: 0.15, 0.2, and 0.3 degrees which corresponded to 3616, 7933, and 13866 vertices, respectively. The triangulations were generated using the {\tt inla.messh.2d} function from the R-INLA package. The centroid of every 5x5km pixel within Nigeria, the modeling domain, were used supplied {\tt loc.domain} argument to define the extent, and the maximum allowed edge length outside the extent was set to 5 degrees. All other arguments were left as the defaults. Plots of the three triangulation meshes are shown in Figure~\ref{fig:meshes}.

\vspace{1em}

\clearpage

\subsection{Additional Continuous Spatial Simulation Results}
\label{app:sim.cont.results}

\begin{figure} \vspace{-3em} \centering \includegraphics[width=\linewidth]{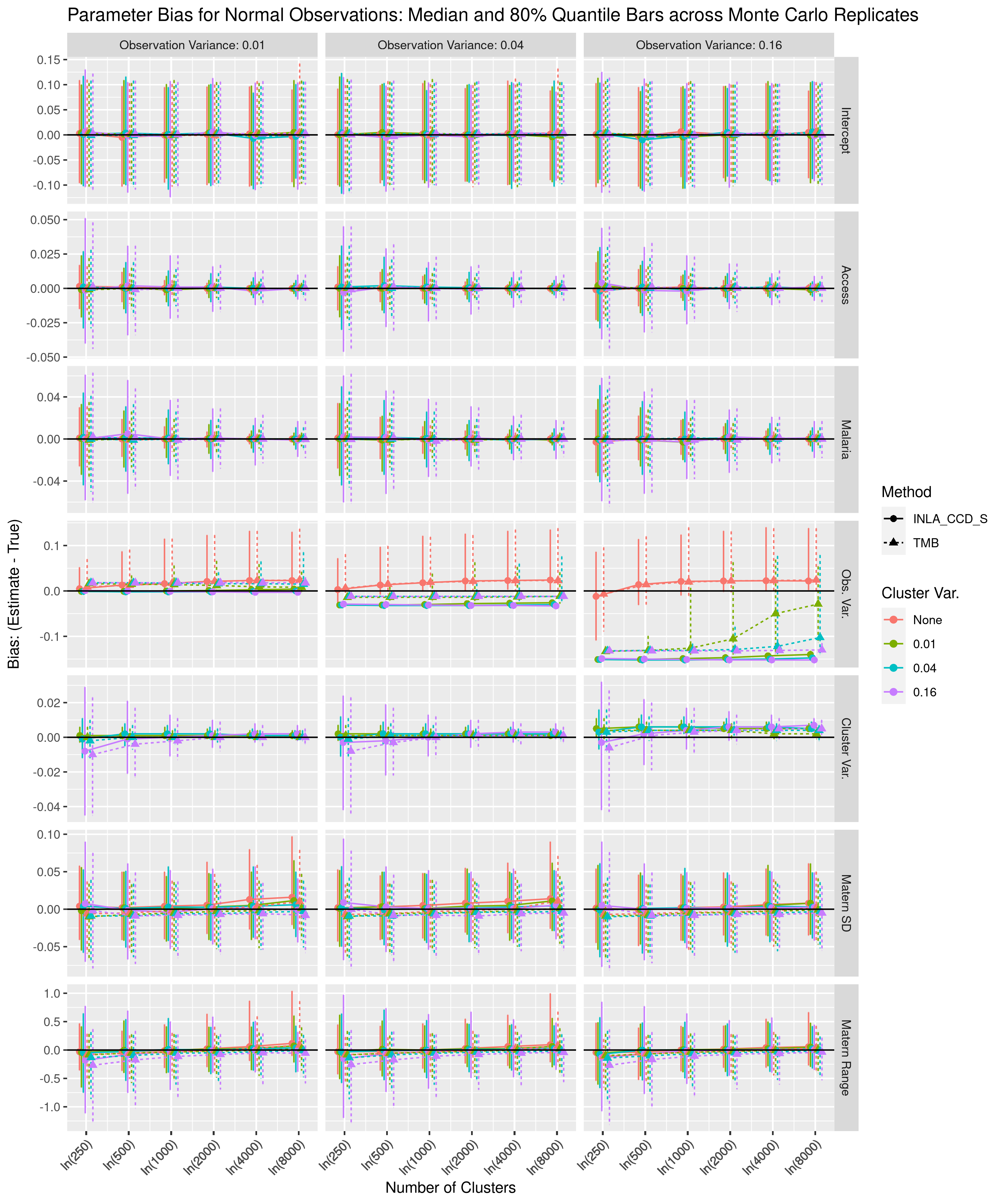} % \singlespace
\caption{ \footnotesize{ Comparison of the estimated parameter bias from TMB (dashed lines) and R-INLA using CCD hyperparameter integration and simplified Laplace approximations (solid lines) plotted against the number of cluster observations for the Gaussian data experiments with varying observation variances. Colors represent different cluster (i.i.d nugget) variances used in an experiment. Each point is the median bias of 3 experiments (coarse, medium, and fine SPDE triangulation), calculated across 75 replicates, and the bars represent the middle 80\% quantile range of the bias across replicates.  \label{fig:normal.par.bias}}}
\end{figure}

\begin{figure} \centering \includegraphics[width=\linewidth]{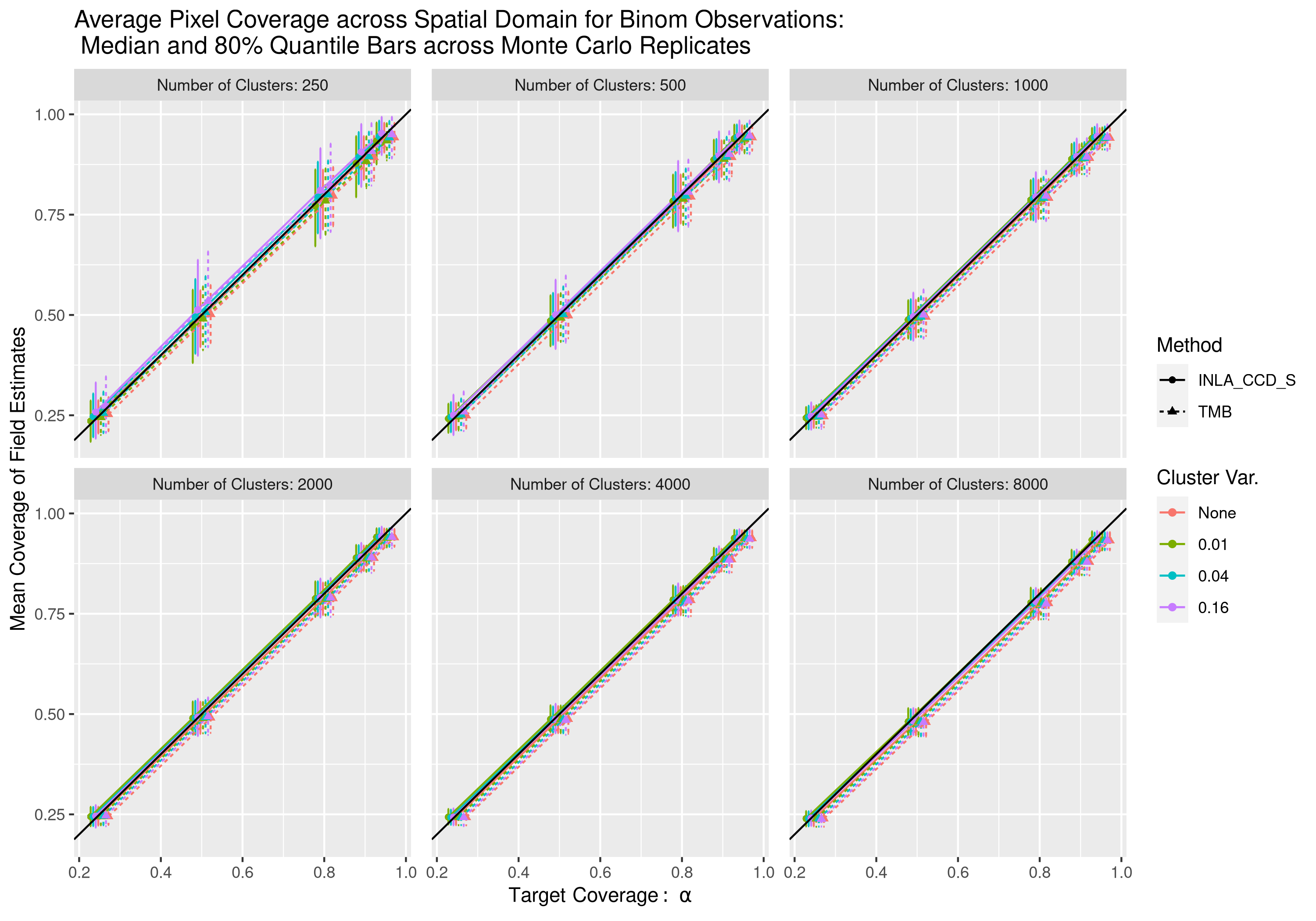} % \singlespace
 \caption{ \footnotesize{ Comparison of the average estimated field coverage of the simulated truth from TMB (dashed lines) and R-INLA using CCD hyperparameter integration and simplified Laplace approximations (solid lines) plotted against the target nominal coverage, $\alpha$, for Binomial observation experiments.  Colors represent different cluster (i.i.d nugget) variances used in an experiment. Each point is the median average coverage of 3 experiments (coarse, medium, and fine SPDE triangulation), calculated across 75 replicates, and the bars represent the middle 80\% quantile range of the average coverage across replicates. \label{fig:binom.pix.cov}}}
\end{figure}

\begin{figure}    \vspace{-3em}  \centering \includegraphics[width=\linewidth]{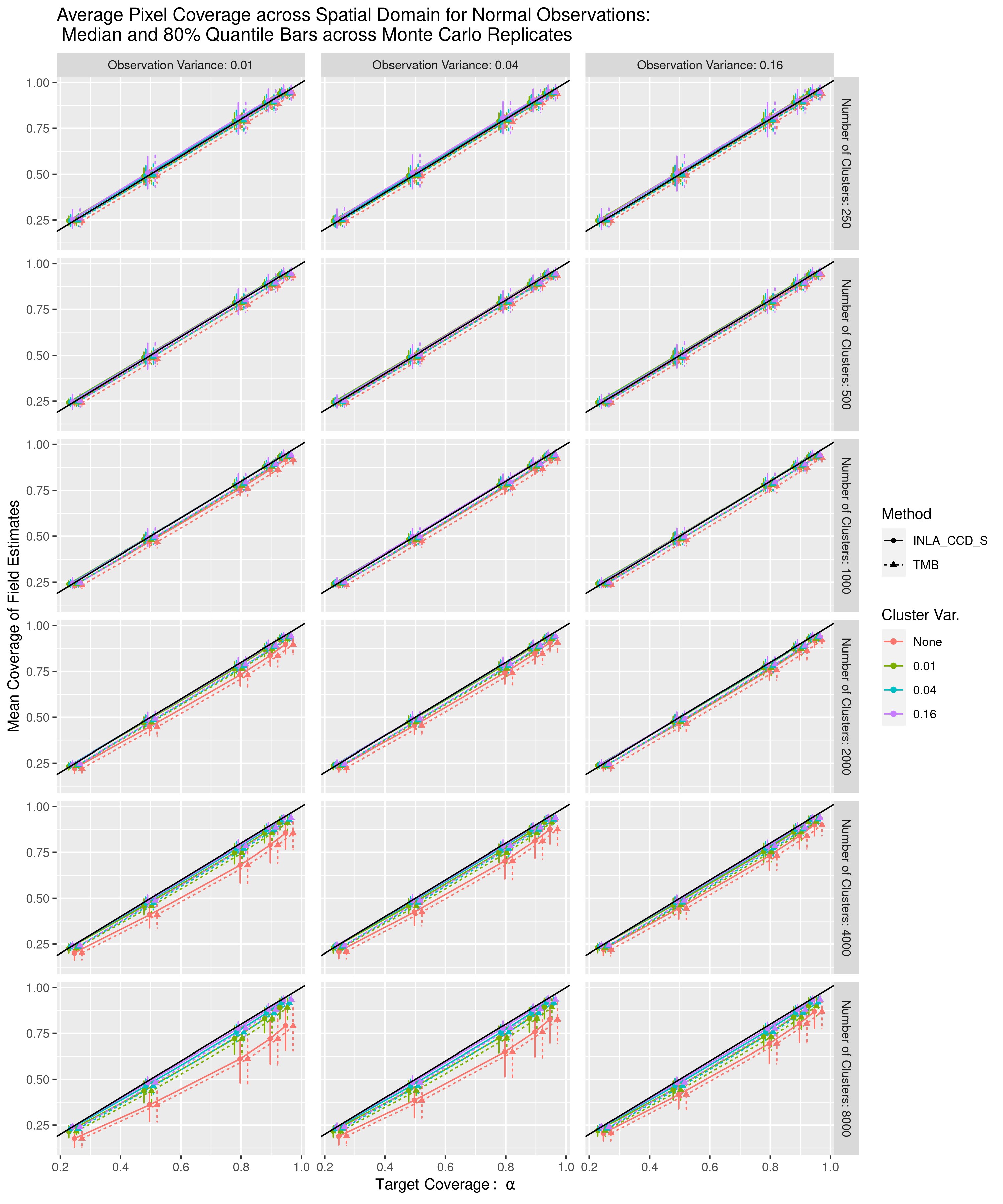} % \singlespace
 \caption{ \footnotesize{ Comparison of the average estimated field coverage of the simulated truth from TMB (dashed lines) and R-INLA using CCD hyperparameter integration and simplified Laplace approximations (solid lines) plotted against the target nominal coverage, $\alpha$, for Gaussian observation experiments with varying observation variances. Colors represent different cluster (i.i.d nugget) variances used in an experiment. Each point is the median average coverage of 3 experiments (coarse, medium, and fine SPDE triangulation), calculated across 75 replicates, and the bars represent the middle 80\% quantile range of the average coverage across replicates. \label{fig:normal.pix.cov}}}
\end{figure}

We provide a few additional plots, extending those show in Section ~\ref{sec:sim.cont.results}, contrasting TMB against the default INLA option results. Figure~\ref{fig:normal.par.bias} shows the parameter bias from these experiments analogous to Figure~\ref{fig:binom.par.bias} in the main text. Figures \ref{fig:binom.pix.cov} and \ref{fig:normal.pix.cov} show collapsed versions of Figures \ref{fig:binom.dist.cov} and \ref{fig:normal.dist.cov} where averaging has been performed over the entire spatial field (as opposed to stratifying by the magnitude of the true GP).

\vspace{2em}

\ref{app:sim.cont.inla.comp} includes parameter bias figures from the Binomial data setting for all combinations of INLA numerical integration and marginal approximations implemented in this study.

\vspace{2em}

\ref{app:sim.cont.mesh.diff} contains three figures showing severe undercoverage of the spatial field estimates in certain Gaussian data experiments that (mostly) disappears as the resolution of the mesh is increased. This pattern is reflected across all combinations of INLA options, but the figures presented here plot the highest quality options we evaluated, CCD integration and full Laplace approximations, demonstrating that the issue is pervasive across both TMB and INLA results.

\vspace{2em}

\ref{app:timing} describes two different timing experiments, one using serial computation and one with parallelization,  used to compare the computational burden of TMB and R-INLA and includes figures summarizing the timing results.

\vspace{2em}

Interactive online plots allow the interested reader to study versions of these figures for all levels of the experiment defined in Table~\ref{tab:sim_params}. A link to the interactive web app may be found at: \\
\href{https://faculty.washington.edu/jonno/software.html}{https://faculty.washington.edu/jonno/software.html}.

\clearpage

\clearpage

\subsubsection{Various INLA options}
\label{app:sim.cont.inla.comp}

All figures shown in this section compare bias results from Binomial data experiments. Bias from TMB (dashed lines) and INLA results (solid lines), under a variety of approximation options, are plotted against the number of cluster observations for Binomial observation experiments. Colors represent different cluster (i.i.d nugget) variances used in an experiment. Each point is the median bias of 3 experiments (coarse, medium, and fine SPDE triangulation), calculated across 75 replicates, and the bars represent the middle 80\% quantile range of the bias across replicates.

\begin{figure}[h] \centering \includegraphics[width=.85\linewidth]{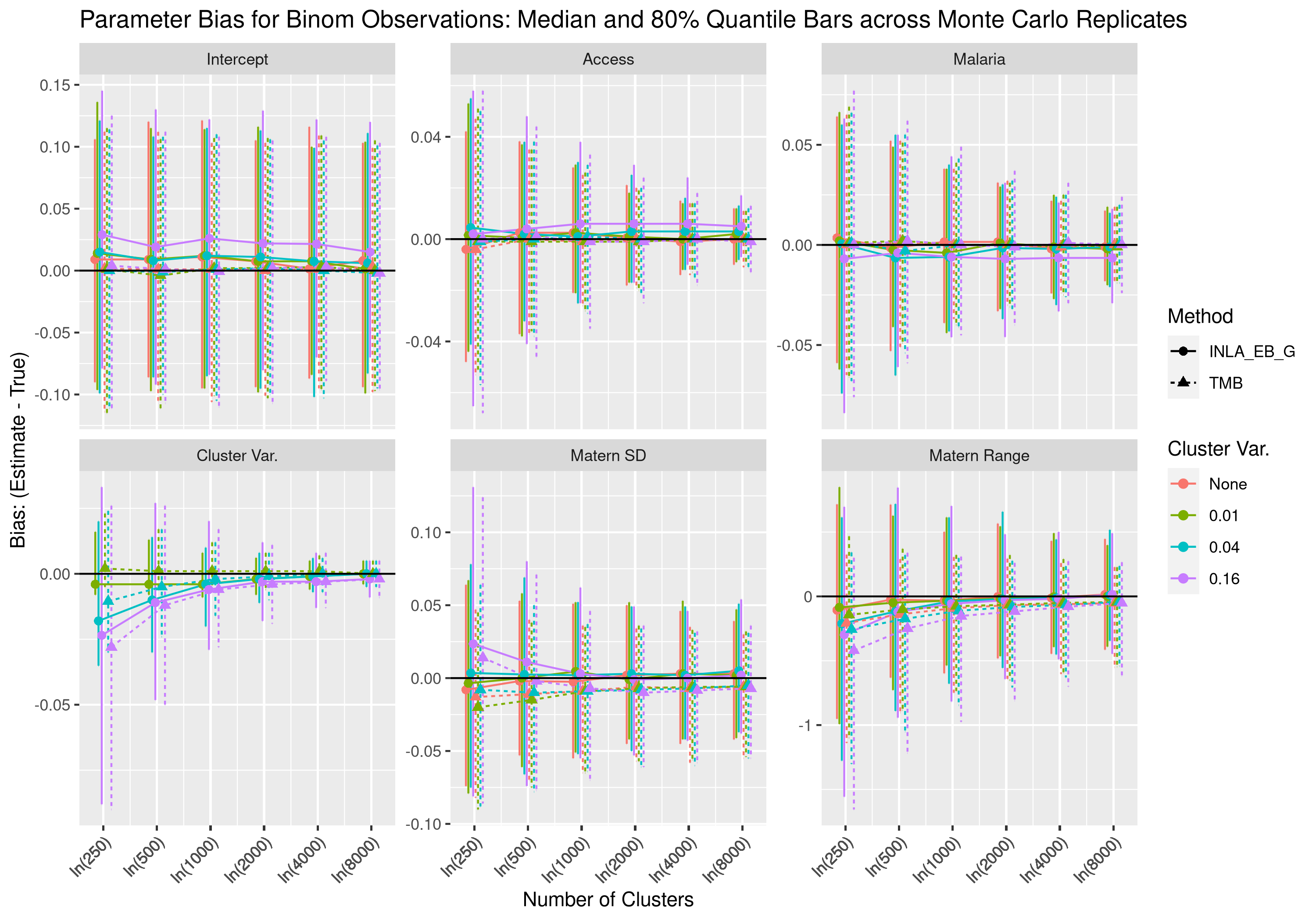} % \singlespace
\caption{ \footnotesize{ Comparison of the estimated parameter bias from TMB (dashed lines) and R-INLA using EB `integration' and Gaussian approximations (solid lines). \label{fig:binom.par.bias.eb.g}}}
\end{figure}

\begin{figure} \centering \includegraphics[width=.85\linewidth]{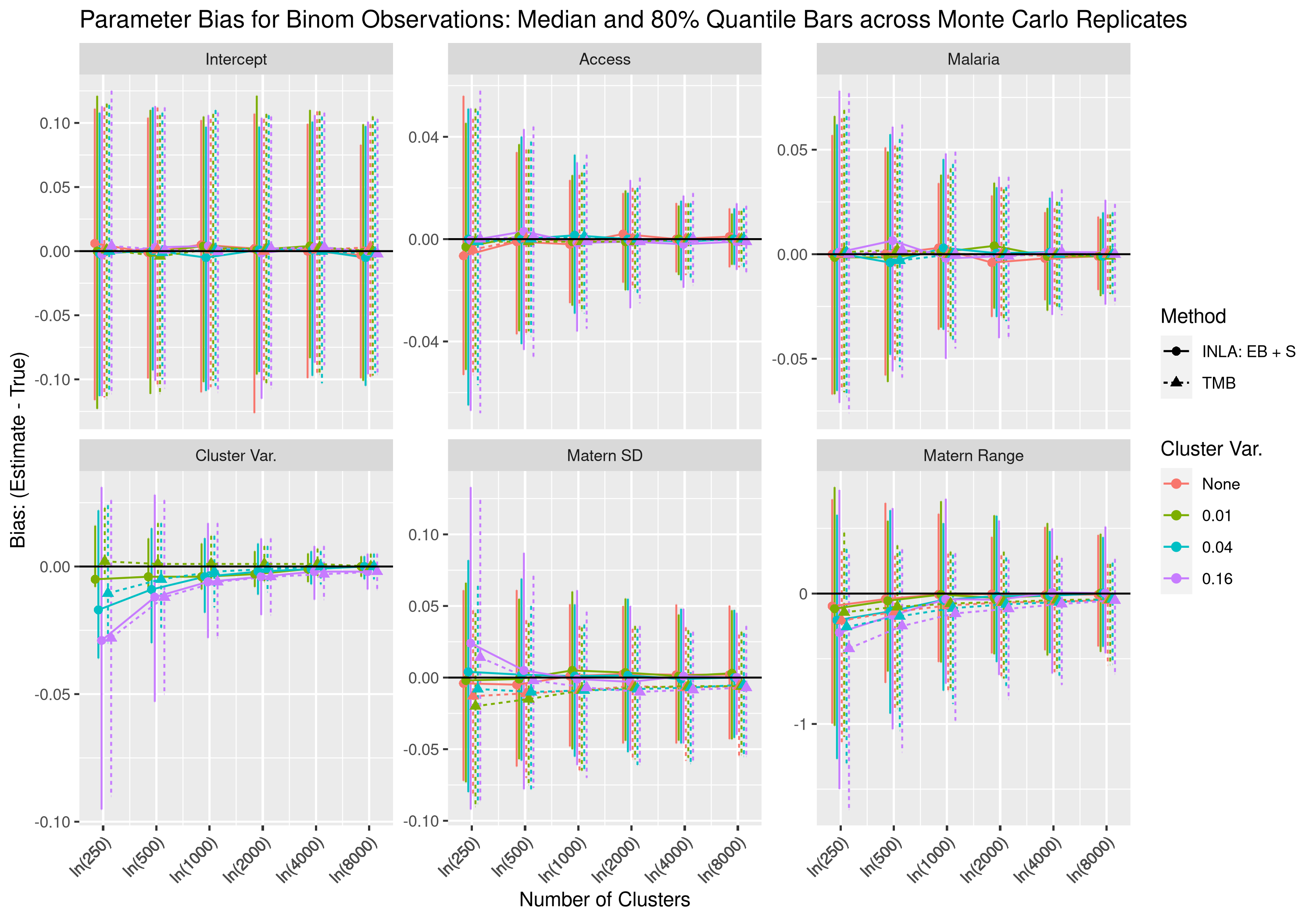} % \singlespace
\caption{ \footnotesize{ Comparison of the estimated parameter bias from TMB (dashed) and R-INLA using EB `integration' and simplified Laplace approximations (solid). \label{fig:binom.par.bias.eb.s}}}
\end{figure}

\begin{figure} \centering \includegraphics[width=.85\linewidth]{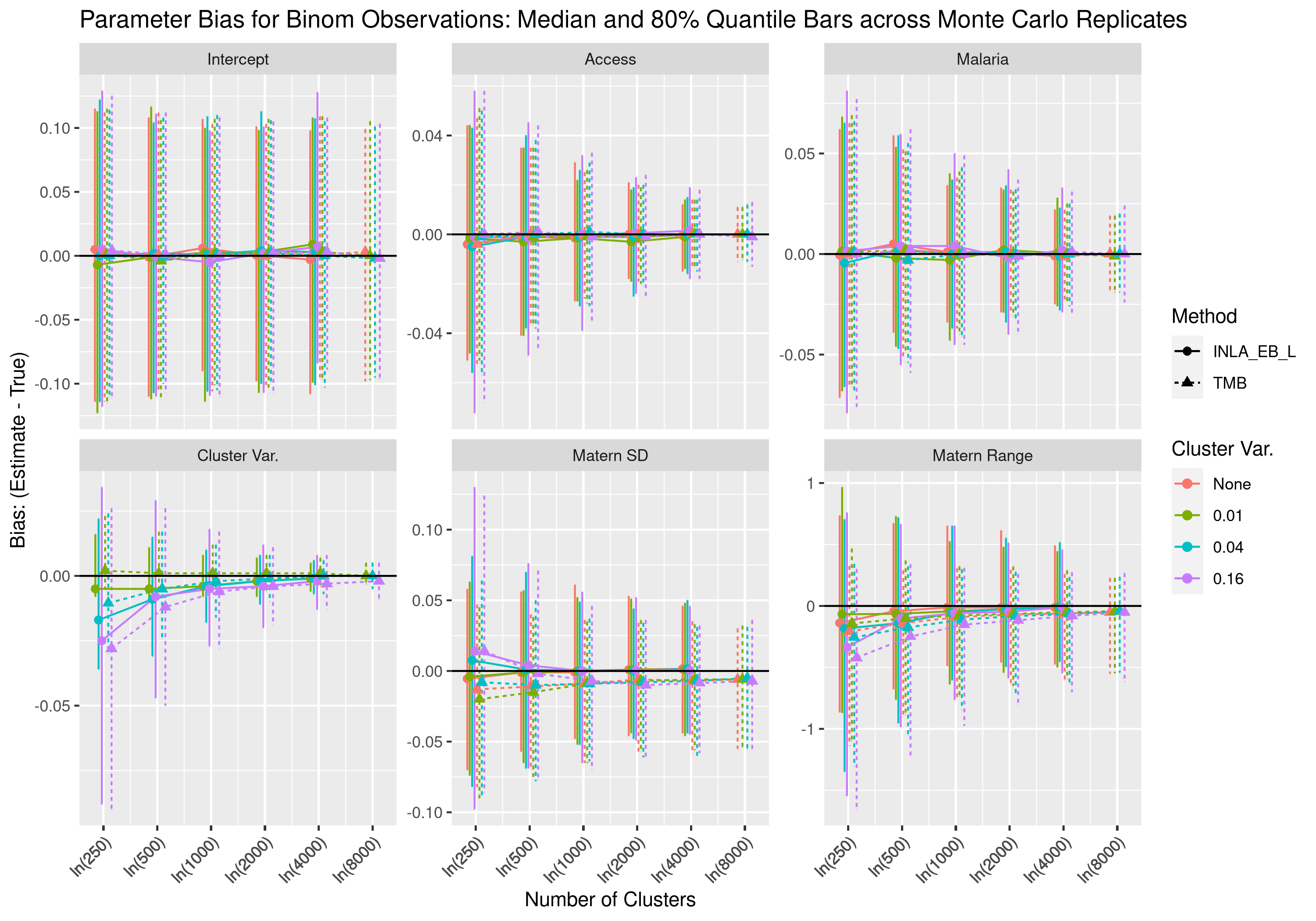} % \singlespace
\caption{ \footnotesize{ Comparison of the estimated parameter bias from TMB (dashed) and R-INLA using EB `integration' and full Laplace approximations (solid). \label{fig:binom.par.bias.eb.l}}}
\end{figure}

\begin{figure} \centering \includegraphics[width=.85\linewidth]{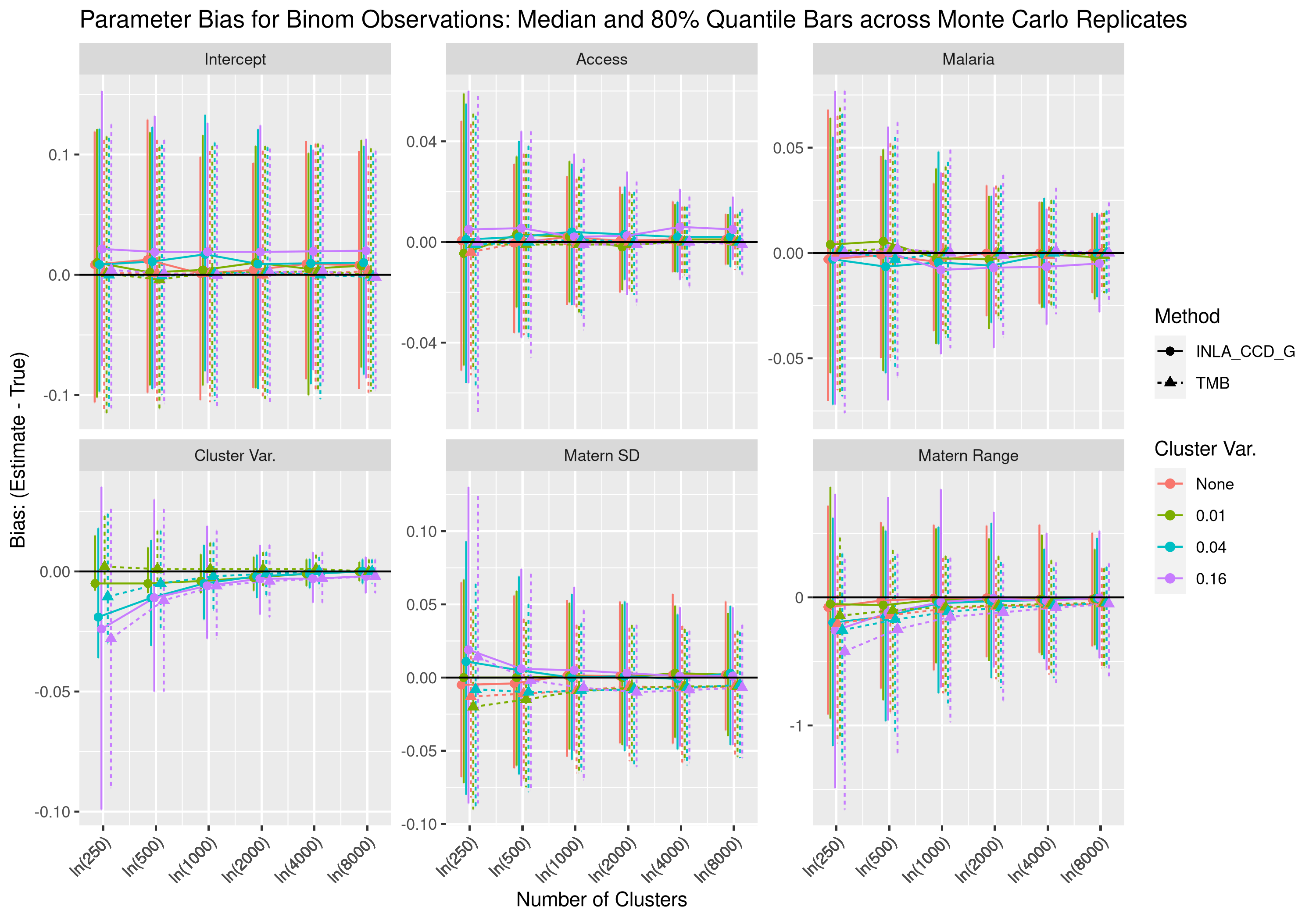} % \singlespace
\caption{ \footnotesize{ Comparison of the estimated parameter bias from TMB (dashed) and R-INLA using CCD integration and Gaussian approximations (solid). \label{fig:binom.par.bias.ccd.g}}}
\end{figure}

\begin{figure} \centering \includegraphics[width=.85\linewidth]{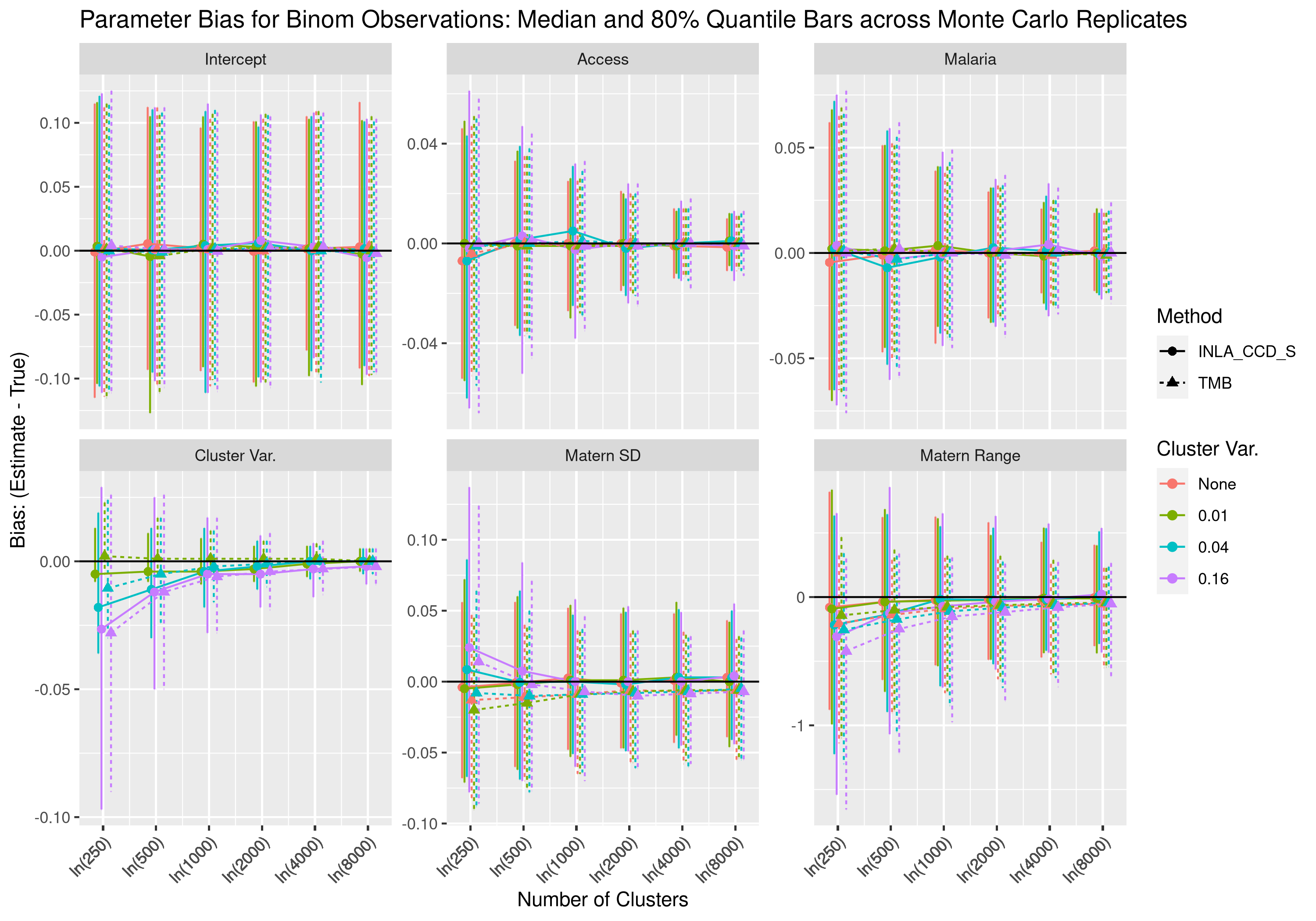} % \singlespace
\caption{ \footnotesize{ Comparison of the estimated parameter bias from TMB (dashed) and R-INLA using CCD integration and simplified Laplace approx. (solid). Same as Figure~\ref{fig:binom.par.bias}. \label{fig:binom.par.bias.ccd.s}}}
\end{figure}

\begin{figure} \centering \includegraphics[width=.85\linewidth]{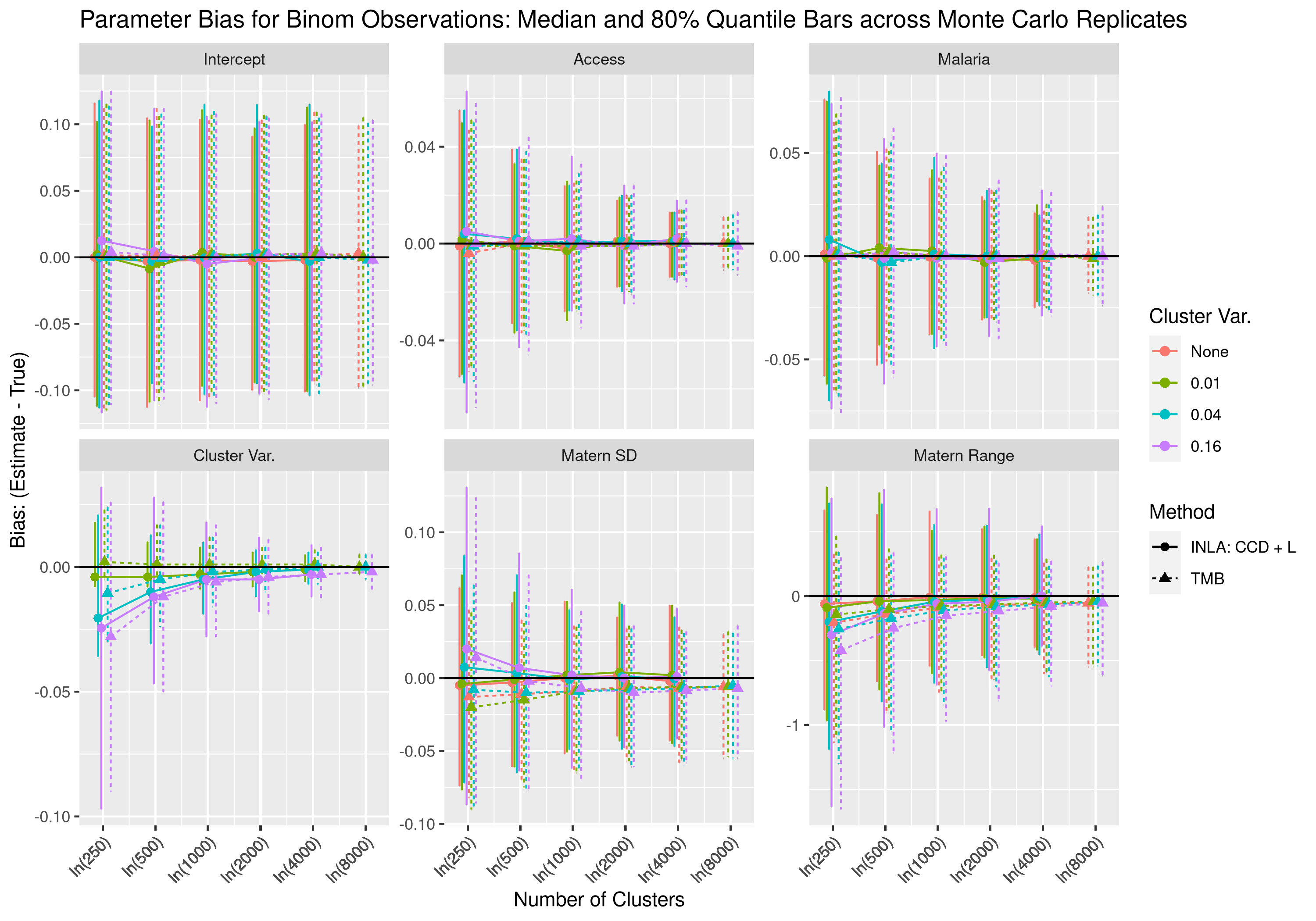} % \singlespace
\caption{ \footnotesize{  Comparison of the estimated parameter bias from TMB (dashed) and R-INLA using CCD integration and full Laplace approximations (solid).  \label{fig:binom.par.bias.ccd.l}}}
\end{figure}

\clearpage

\subsubsection{Differences due to SPDE triangulation resolution}
\label{app:sim.cont.mesh.diff}

Plots in this section show decreasing undercoverage of the spatial field of the true spatial field by the estimated field as the resolution of the triangulation mesh size increases (as the mesh becomes more dense with more vertices). The figures contrast results from TMB against those from INLA using the CCD integration and full Laplace approximations as the mesh density increases. These are the results from the best INLA approximations we evaluated, even though this pattern persists across all other INLA options evaluated, to demonstrate that this pattern is a functions of the SPDE approximation and not the INLA options or TMB algorithm.

\begin{figure}
  \vspace{-3em}
  \centering \includegraphics[width=\linewidth]{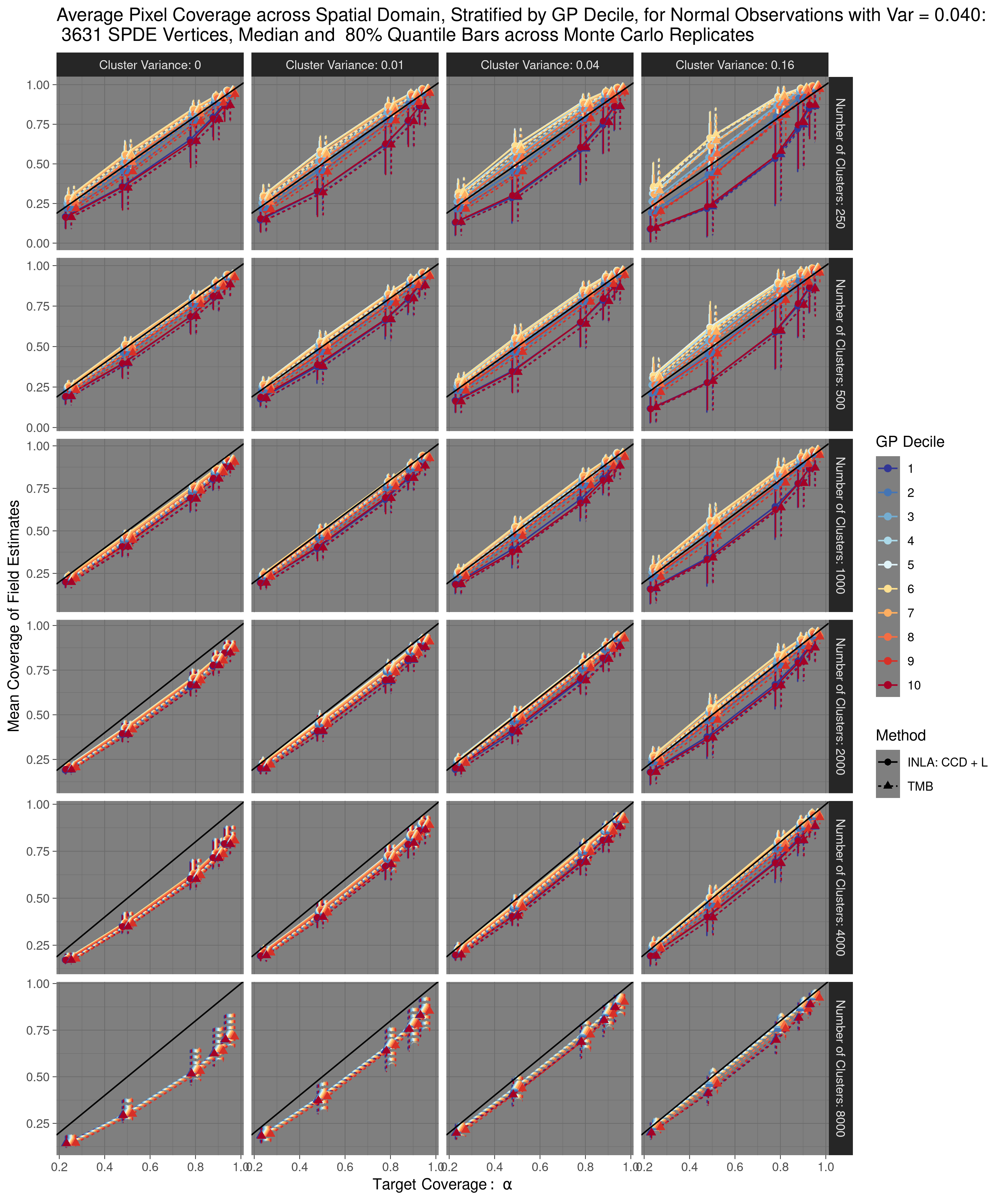} \singlespace \caption{ \footnotesize{ Comparison of the average estimated field coverage of the simulated truth, faceted by cluster (i.i.d. nugget) variance and the number of clusters, from TMB (dashed lines) and {\tt R-INLA} using CCD hyperparameter integration and full Laplace approximations (solid lines) plotted against the target nominal coverage, $\alpha$, for Gaussian observation experiments with $\sigma^2 =0.04$ and the coarse resolution SPDE triangulation. Colors stratify pixels included in the average coverage calculation by the decile of the true GP for the experiment replicate. Each point is the median average coverage of an experiment, calculated across 25 replicates, and the bars represent the middle 80\% quantile range of the average coverage across replicates.\label{fig:normal04.dist.cov.c}}}
\end{figure}

\begin{figure}
  \vspace{-3em}
  \centering \includegraphics[width=\linewidth]{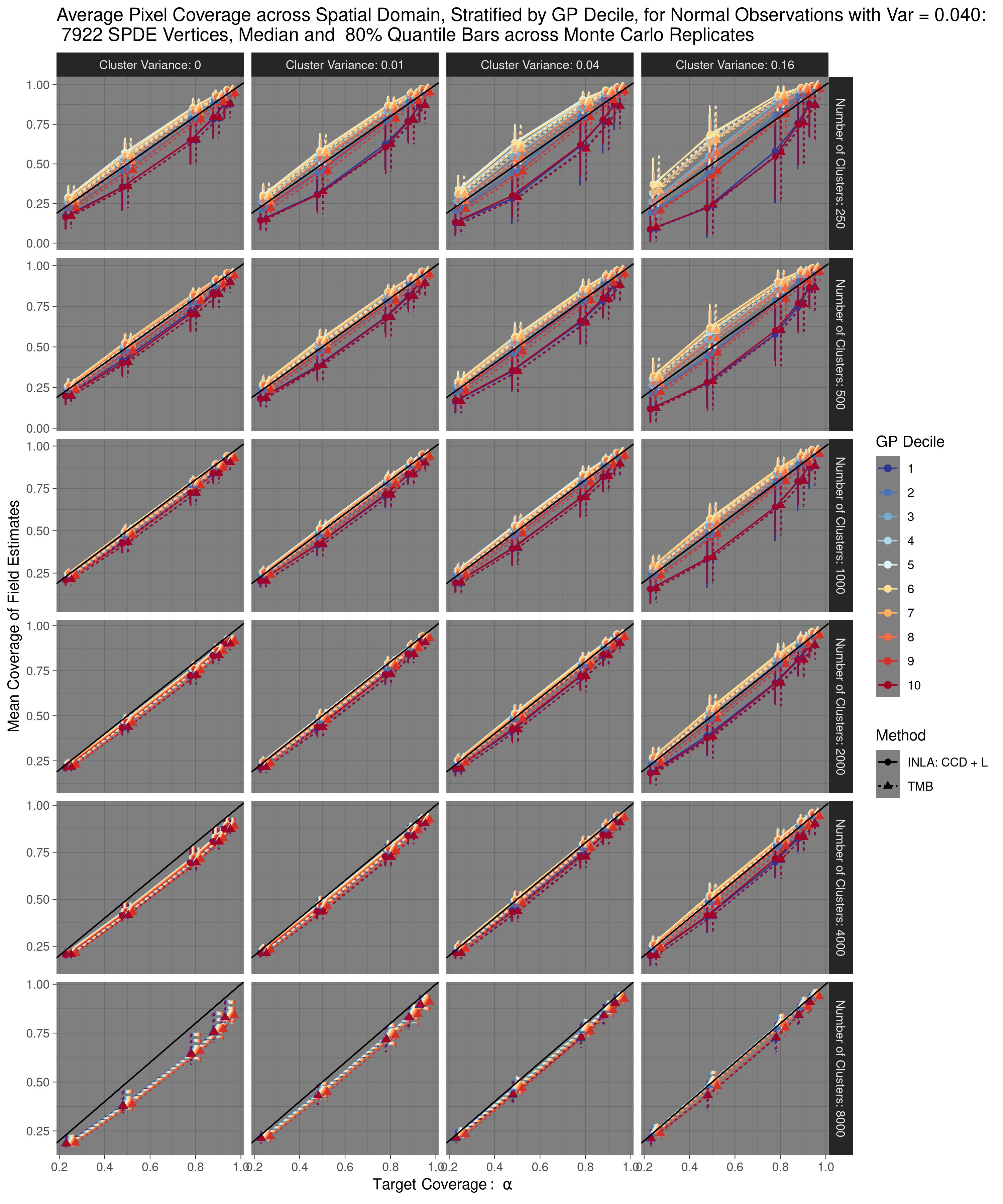} \singlespace \caption{ \footnotesize{ Comparison of the average estimated field coverage of the simulated truth, faceted by cluster (i.i.d. nugget) variance and the number of clusters, from TMB (dashed lines) and {\tt R-INLA} using CCD hyperparameter integration and full Laplace approximations (solid lines) plotted against the target nominal coverage, $\alpha$, for Gaussian observation experiments with $\sigma^2 =0.04$ and the medium resolution SPDE triangulation. Colors stratify pixels included in the average coverage calculation by the decile of the true GP for the experiment replicate. Each point is the median average coverage of an experiment, calculated across 25 replicates, and the bars represent the middle 80\% quantile range of the average coverage across replicates. This figure is shown in the main results section, but is replicated here for easy comparison against the other supplemental plots.\label{fig:normal04.dist.cov.m2}}}
\end{figure}

\begin{figure}
  \vspace{-3em}
  \centering \includegraphics[width=\linewidth]{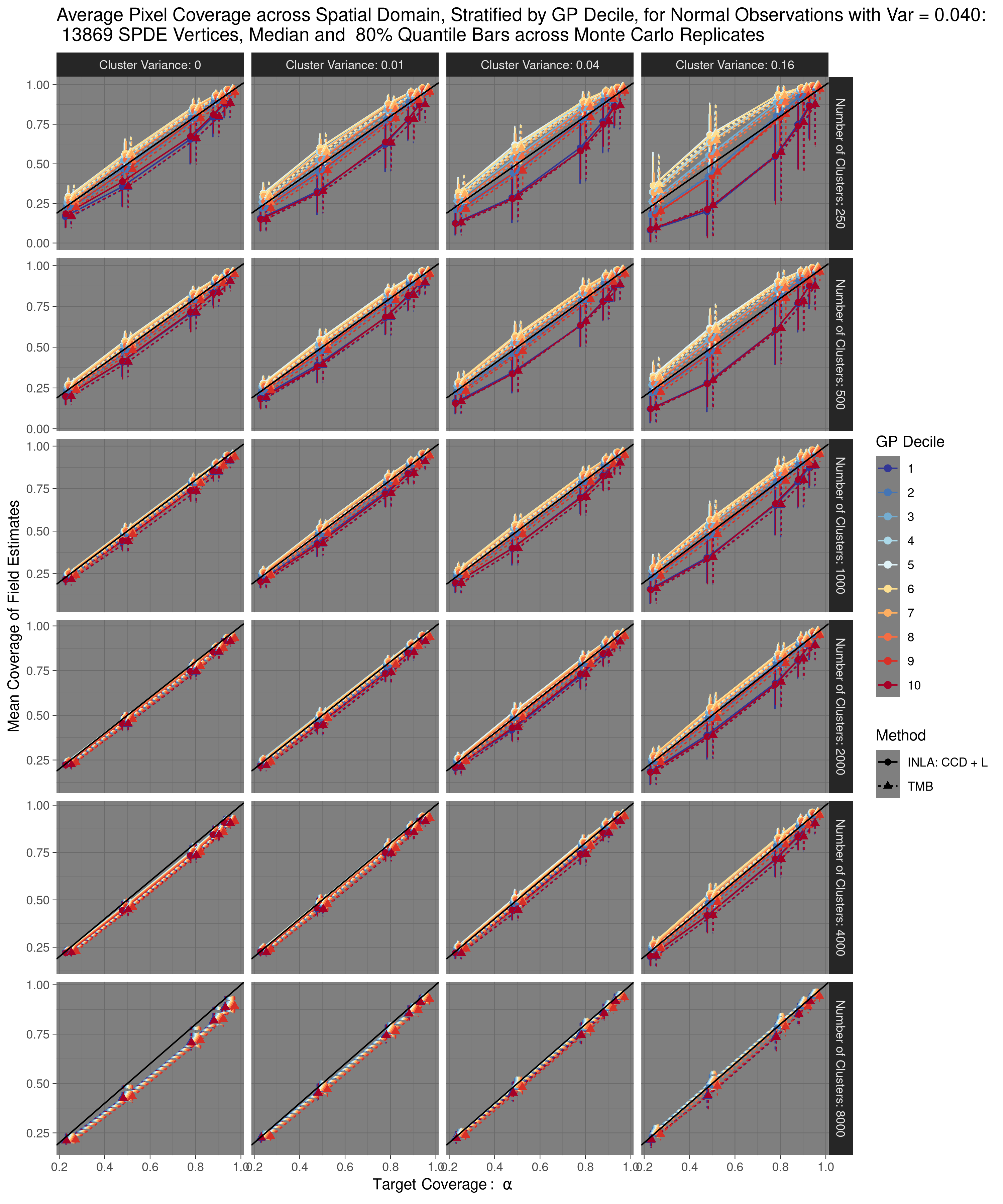} \singlespace \caption{ \footnotesize{ Comparison of the average estimated field coverage of the simulated truth, faceted by cluster (i.i.d. nugget) variance and the number of clusters, from TMB (dashed lines) and {\tt R-INLA} using CCD hyperparameter integration and full Laplace approximations (solid lines) plotted against the target nominal coverage, $\alpha$, for Gaussian observation experiments with $\sigma^2 =0.04$ and the fine resolution SPDE triangulation. Colors stratify pixels included in the average coverage calculation by the decile of the true GP for the experiment replicate. Each point is the median average coverage of an experiment, calculated across 25 replicates, and the bars represent the middle 80\% quantile range of the average coverage across replicates.\label{fig:normal04.dist.cov.f}}}
\end{figure}

\clearpage
\subsubsection{Timing comparisons}
\label{app:timing}

\begin{figure} \centering \includegraphics[width=\linewidth]{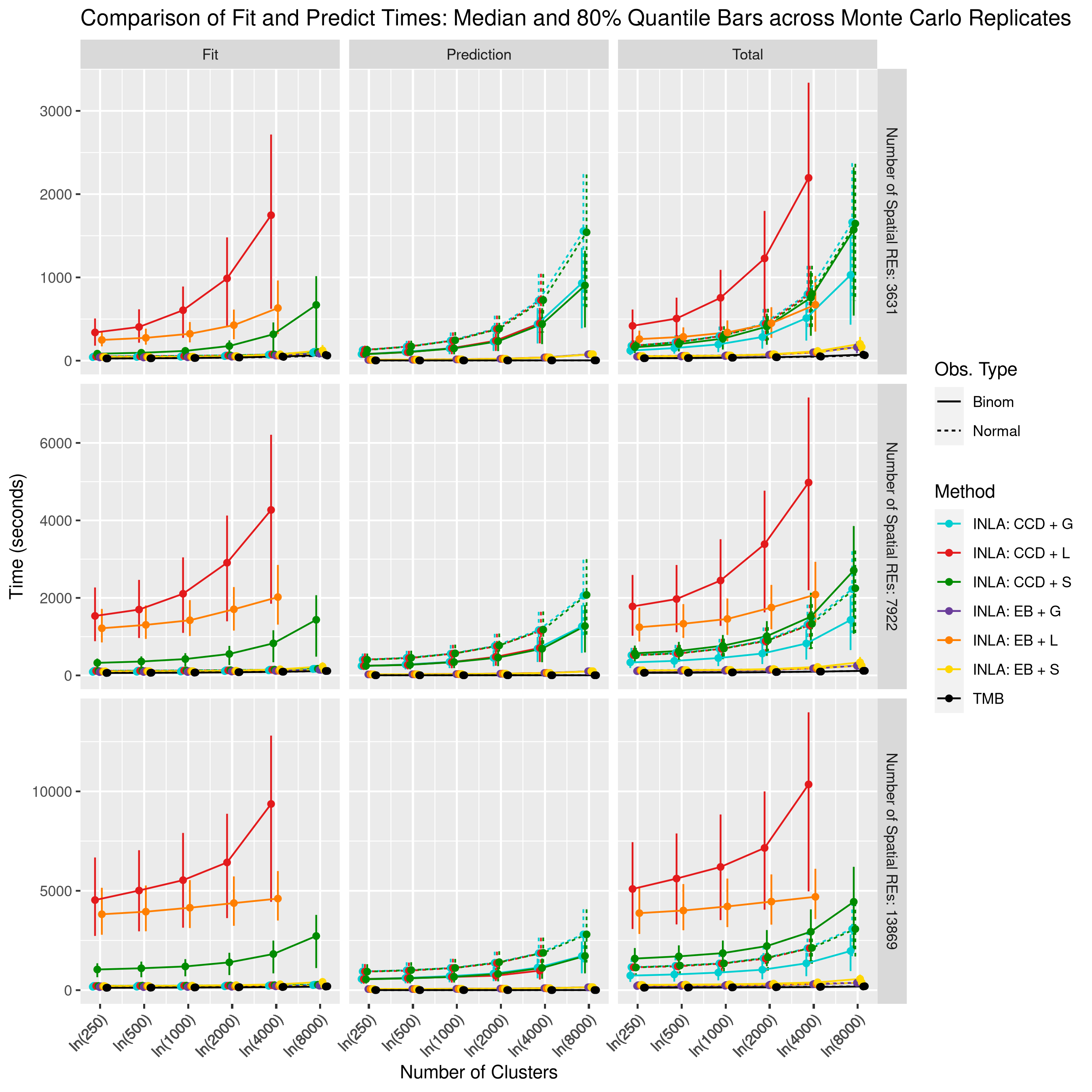} % \singlespace
 \caption{ \footnotesize{Comparison of the average fit, predict, and total times from TMB and R-INLA, faceted by the number of spatial random effects, plotted aginst the number of clusters.  Each point is the median time of an experiment, calculated across 25 replicates, and the bars represent the middle 80\% quantile range of the bias across replicates.\label{fig:time}}}
\end{figure}

The main set of continuous simulation experiments, shown in Table\ref{tab:sim_params}, were run restricting both TMB and R-INLA to use a single CPU thread. This was done in order to better leverage the particulars of the computing cluster which was used. The median and 80\% percent timing quantiles, taken across the 25 replicates of each experimental level, are shown in Figure \ref{fig:time}. The plot breaks down the timing by the number the method, the number of cluster observations and whether the observed data were Binomial or Gaussian, and the dimension of the spatial random effects, While it would be unusual for people under normal circumstances to restrict either method to use a single core, this plot nonetheless gives some indication of the total computational burden of each of the methods.

\begin{figure} \centering \includegraphics[width=\linewidth]{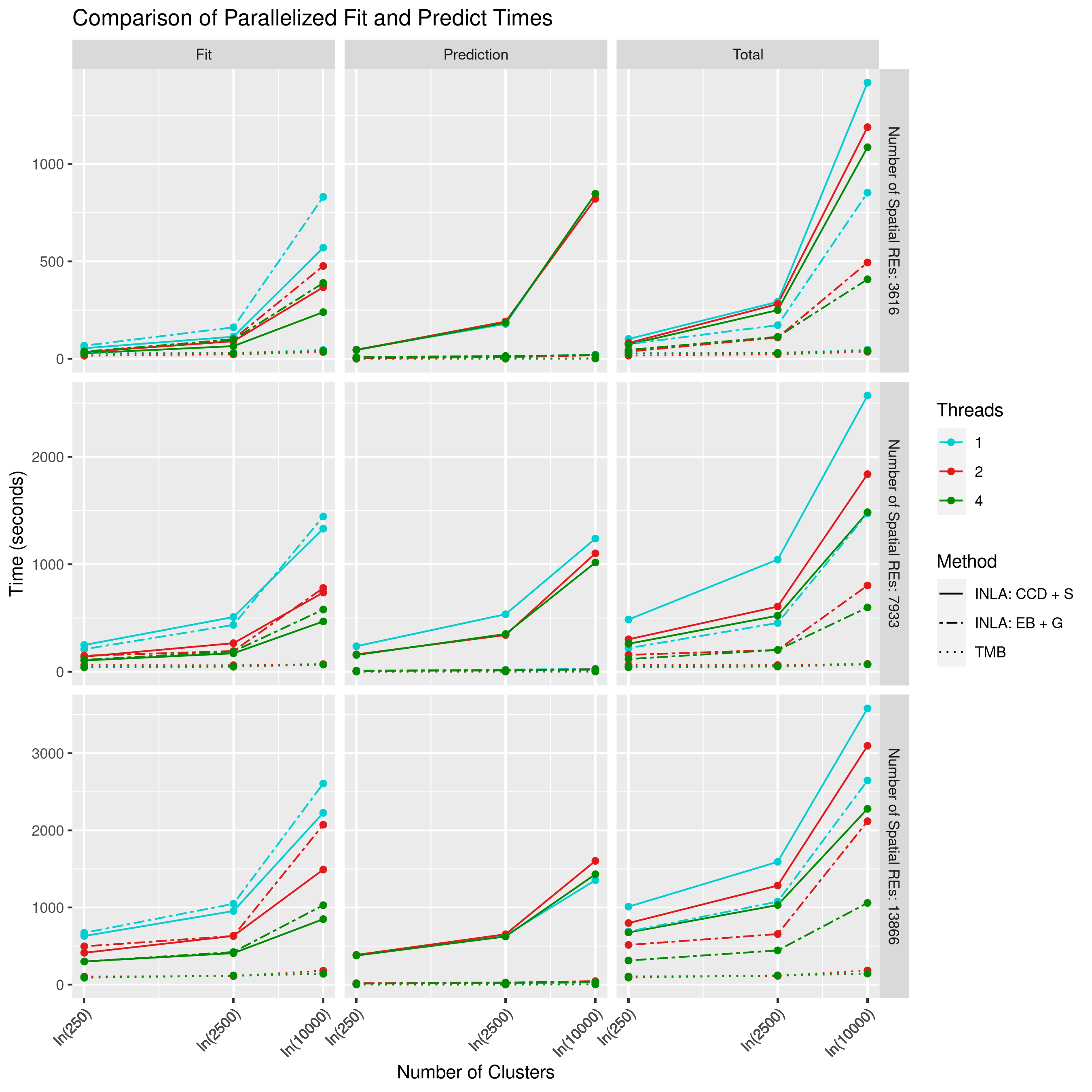} % \singlespace
 \caption{ \footnotesize{Comparison of the average fit, predict, and total times from TMB and R-INLA, faceted by the number of spatial random effects, plotted against the number of clusters.  Each point is the mean time across 5 replicates of a Binomial data experiment. The colors represent the number of threads available for parallelization within each method.\label{fig:time.parallel}}}
\end{figure}

In addition, a small experiment to give a sense of the timing under more usual, parallelized, scenarios was run. The design of this experiment was very similar to those of the Binomial data continuous GP experiments but only the number of clusters (250, 2500, 10000), the number of spatial random effects (low, medium, and high resolution SPDE triangulations with 3616, 7933, and 13866 vertices, respectively), and the number of CPU threads (1, 2, 4) were varied. This created 27 experimental levels, each which was replicated 5 times. The mean of the fit, prediction, and total times are summarized in Figure \ref{fig:time.parallel}. R-INLA was run using the PARDISO library (\cite{pardiso-7.2a}, \cite{pardiso-7.2b}, \cite{pardiso-7.2c}), and it was forced to use the requested number of threads (it has built-in logic that, by default, is capable of using fewer than the max threads if it believes that will be more efficient). TMB was run using METIS to create fill reducing orderings of sparse matrices and parallelization was enabled by setting the available OpenMP threads.

Details on the software versions and hardware used in this study can be found in \label{app:software}.

\clearpage

\section{Discrete Spatial Simulation Study}
\label{app:sim.disc}

An overview of the discrete simulation details is provided in Section \ref{sec:sim.disc}. Here we provide an in-depth version with all remaining details.

\subsection{Discrete Spatial Simulation Details}

In addition to the continuous simulations, this study also assessed discrete spatial models simulations using TMB and R-INLA. In order to correctly implement the BYM2 discrete spatial model (a modern formulation of the classic Besag-York-Mollie model developed by \cite{riebler2016}), a sum-to-zero constraint was implemented in TMB using appropriate conditional densities \cite[Section 12.1.7.4]{gelfand2010}.

The discrete model was implemented over the 37 regions (first-level administrative units) of Nigeria, and neighbors were defined by immediate adjacency. Within each region, Poisson observations were simulated from a population of size $n_s$, arising from the following hierarchical model:
\begin{eqnarray}
  y_i|n_s, \eta_i &\sim &\text{Poisson}(n_s\times \eta_i)   \label{eq:disc.sim.hier} \\
  \eta_i &= &\text{exp}\left(\alpha + \bs b_i \right)\nonumber\\
  \bs b &= &\frac{1}{\sqrt{\tau}}\left( \sqrt{1-\varphi}\boldsymbol v + \sqrt{\varphi}\boldsymbol u_{\star} \right) \nonumber \\
  \bs v &\sim &N\left(\bs 0, \bs I\right)\nonumber \\
  \bs u_\star &\sim &N\left(\bs 0, \bs Q_{\star}^{-1} \right), {\text s.t. }   \sum_{i=1}^{37} u_{\star_i} = \bs 0, \nonumber
\end{eqnarray}
\noindent with $\alpha$ the GMRF intercept, fixed at -3 across simulations, BYM2 field $\bs b$ with total variance $\tau^{-1}$, mixing parameter $\varphi$ controlling the contribution of $\bs v$, the unstructured i.i.d. portion of the BYM2 field, and $\bs u_{\star}$, the scaled spatially structured component of the BYM2. The structured portion of the BYM2 is specified with precision $\bs Q_\star$, a scaled version of the precision from the classic BYM ICAR model, and is constrained to sum to zero. The effect of the constraint was correctly included in TMB by conditioning on the constraint:
\begin{equation}
p(\bmu_\star | \bA \bmu_\star) = \frac{p(\bA \bmu_\star|\bmu_\star)p(\bmu_\star)}{p(\bA \bmu_\star)} \label{eq:lin.constr}
\end{equation}
where, generally, $\bA \bmu_\star = \bme$ encodes the linear constraints and specific to this example, $\bA \bmu_\star = \sum \bmu_\star = 0$.

To complete the model specification, the following priors are included:
\begin{eqnarray*}
  \alpha &\sim &\text{N}(0, 5^2) \\
  \varphi   &\sim &\text{Beta}(.5, .5) \\
  1/\sqrt{\tau} = \sigma &\sim &\text{N}(0, 5^2) \mathbbm{1}_{\sigma > 0}.
\end{eqnarray*}

Internally, TMB was coded to use the same internal representations of parameters for the optimization step: $\text{logit}(\varphi)$ and $\text{log}(\tau)$.

The full combinatorial grid of simulation parameters shown in Table~\ref{tab:disc_sim_params} comprised the set of 20 experiments and each experiment was replicated 25 times to obtain Monte Carlo errors on the validation metrics. For each replicate within each experiment, constrained BYM2 fields are generated, new observations are simulated, models are fit in both TMB and in R-INLA, and 500 joint estimator samples are drawn from each model to compare the estimates against the truth. The results shown in the following section use the empirical distribution taken across all replicates for each level of the experiment.

\section{European Breast Cancer Application Details}
\label{app:euro.bc}

 \subsection{Data}

\begin{figure}
    \includegraphics[width=\linewidth]{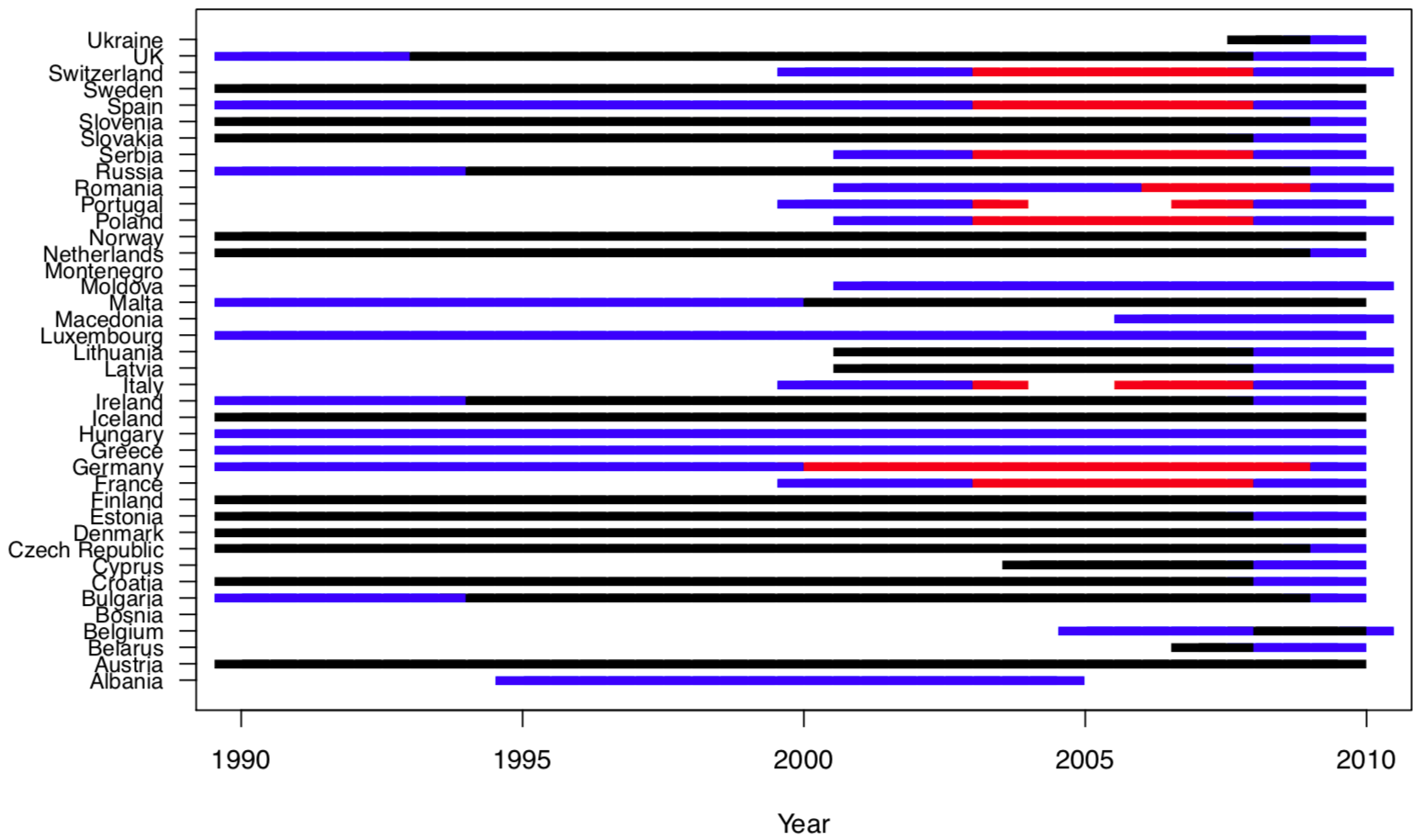}
    \caption{Data type by country from 1990–2010. Type IV (Montenegro) is blank.}
    \label{fig:datatype}
  \end{figure}

  We work with breast cancer incidence and mortality data from IARC for this report. The current implementation of the methods described in this paper rely on an aggregated version of the IARC scores and are limited to countries within Europe. We will refer to countries as having one of four types of data:
  \begin{itemize}
  \item (I) national incidence and mortality
  \item (II) sub-national incidence and mortality (from registries) and national mortality,
  \item (III) only national mortality, and
  \item (IV) no available data.
  \end{itemize}
  Figure~\ref{fig:datatype} shows the data type available for the 40 countries from 1990-2010.

  Although the data is available across age groups and time, this current project will focus on ages 50-54 (age group 11). While we will use data from all time periods, allowing the country type to vary in time, we assume that the underlying parameters are fixed in time and only estimate spatial variability.

   \subsection{Model}
  \label{sec:mod}

  We assume a probabilistic models for incidence, mortality, and mortality given incidence. For most countries an alternative would be to rely only on unconditional models for just incidence and mortality. The MI modeling approach facilitates estimating national incidence in countries without out national or without local incidence data by providing an explicit link between mortality and incidence.

  First, notation is defined. For a country, $c$, an age group, $a$, and a time period (year), $t$, we use $L$ to denote local registry data and $R$ to denote the remainder of the data. In countries with no local registry data ($L$), all the data will fall into the remainder category ($R$). With terms, we define:

  \begin{itemize}
  \item $N_{act}^L =$ Population for age group $a$ in country $c$ at time $t$ covered by the available registry,
  \item $Y_{act}^L =$ Total incident reported cases from all registries for age group $a$, country $c$, and time $t$,
  \item $Z_{act}^L =$ Total reported deaths (mortality) from all registries for age group $a$ in country $c$ at time $t$,
  \item $N_{act}^R =$ Population for age group $a$ in country $c$ at time $t$ not covered by the available registry,
  \item $Y_{act}^R =$ Total incident reported cases not covered by registries for age group $a$ in country $c$ at time $t$,
  \item $Z_{act}^R =$ Total reported deaths (mortality)  not covered by registries for age group $a$ in country $c$ at time $t$,
  \item $N_{act} = N_{act}^L + N_{act}^R$ is the total population for age group $a$ in country $c$ at time $t$,
  \item $Y_{act} = Y_{act}^L + Y_{act}^R$ is all reported cases for age group $a$ in country $c$ at time $t$,
  \item $Z_{act} = Z_{act}^L + Z_{act}^R$ is all reported deaths for age group $a$ in country $c$ at time $t$,
  \item $p_{act} = P(\text{ Reported incidence } | a,c,t)$,
  \item $r_{act} = P(\text{ Reported mortality } | a,c,t)$,
  \item $q_{act} = P(\text{ Reported mortality } | \text{ Reported incidence },a,c,t)$.
  \end{itemize}

  For countries that have both national mortality and incidence data (type I) we can assume a Poisson process for cancer incidence, and then conditional on having cancer, we model mortality as a binomial outcome. This also induces a Poisson process for mortality when incidence is unobserved. We suppress age and time notation. Our base model for type I countries is:
  \begin{align}
    Y_c|N_c,p_c &\sim \text{Poisson}(N_c p_c),\hspace{1em} p_c = \text{exp}(\alpha_c^I)\label{eq:1} \\
    Z_c|Y_c,r_c &\sim \text{Binomial}(Y_c r_c),\hspace{1em} r_c = \frac{\text{exp}(\alpha_c^{MI})}{1+\text{exp}(\alpha_c^{MI})}\label{eq:2}
  \end{align}
  which implies the unconditional mortality model:
  \begin{align}
    Z_c|N_c,p_c &\sim \text{Poisson}(N_c q_c),\hspace{1em} q_c = p_c\cdot r_c\label{eq:3}.
  \end{align}
  We assume a log- and logit-linear model for incidence and conditional mortality and we assume the following forms:
  \begin{align}
    \alpha_c^I &= \alpha^I + b_c^I \label{eq:4}\\
    \alpha_c^{MI} &= \alpha^{MI} + b_c^{MI} \label{eq:5}
  \end{align}
  where $\alpha^*$ are global intercepts, and $b_c^*$ are country random effects that are assumed to have BYM-2 structure comprising of a spatially correlated term as well as an unstructured (iid) country specific term. Specifically, each vector of the country random effects are assumed to independent from one another take the form of the BYM2 effects defined in (\ref{eq:disc.sim.hier}) and (\ref{eq:lin.constr}).
 %  \begin{align}
 %    \label{eq:bym2}
 %    \boldsymbol b^* &\sim MVN(\boldsymbol 0,  Q^*), \ \ \ \ \text{for}\ *\in\{I, MI\}\\
 %    Q^* &= \tau^* \big((1 - \lambda^*) \cdot I + \lambda^* \cdot K\big)
 % \end{align}
 % where $\tau^*$ is the overall precision of the country random effects, $\lambda^*$ determines the mixing between the unstructured and spatial random effects, $I$ is an identity matrix, and $K$ is a structure matrix which defines the adjacency relationship of the countries.

  The $\alpha^I$, $ \alpha^{MI}$, $\boldsymbol b^{I}$, and $\boldsymbol b^{MI}$ parameters are used for across all the country types, but the way that these parameters learn and leverage the information depend on the country data type and thus the way the data enter into the joint likelihood.

  For type II countries, those with local incidence and mortality data and national mortality data, we assume the same model used for type I countries for the local registry data and the implied mortality Poisson process for the remaining national mortality data.

  That is, for the local data we assume:
   \begin{align}
    Y_c^L|N_c^L,p_c &\sim \text{Poisson}(N_c^L p_c),\hspace{1em} p_c = \text{exp}(\alpha_c^I)\label{eq:6} \\
    Z_c^L|Y_c^L,r_c &\sim \text{Binomial}(Y_c r_c),\hspace{1em} r_c = \frac{\text{exp}(\alpha_c^{MI})}{1+\text{exp}(\alpha_c^{MI})}\label{eq:7}
   \end{align}
   where the intercept parameters are of the form shown in \ref{eq:4} and \ref{eq:5}.

   Furthermore, for the remaining non-registry mortality data, we assume that the MI ratio  is the same in the local registry and national remainder data and we model it as the implied (unconditional) Poisson process:
   \begin{align}
     \label{eq:8}
         Z_c^R|N_c^R,q_c &\sim \text{Poisson}(N_c^R q_c),\hspace{1em} q_c = \text{exp}(\alpha_c^I) \cdot \frac{\text{exp}(\alpha_c^{MI})}{1+\text{exp}(\alpha_c^{MI})}.
   \end{align}

   For type III countries, those with only national level mortality, we use the induced unconditional Poisson model as written in (\ref{eq:3}).

   Finally, for type IV countries, those with no data, we rely on the global intercept and the country random effect (both the iid and the spatially correlated and smoothed random components from the BYM-2) from the posterior distribution to estimate their incidence and mortality rates.

   To complete the Bayesian specification, we assign the following prior distributions:
   \begin{itemize}
   \item $aI,\ aMI \sim (iid)\ N(0,\sigma^2=100)$
   \item $\varphi_{*}  \sim  (iid)\ \text{Beta}(.5, .5)$
   \item  $1/\sqrt{\tau_*} = \sigma_* \sim (iid)\ \text{N}(0, 5^2) \mathbbm{1}_{\sigma > 0}$.
     % \item $\tau_{bI},\ \tau_{bMI} \sim (iid)\ Gamma(1, 1)$
   %\item $\lambda_{bI},\ \lambda_{bMI} \sim (iid)\ Beta(3, 1)$.
   \end{itemize}
   The model is fit in R using {\em Template Model Builder} and the nonlinear optimizer, {\em nlminb}.

   \subsection{Simulation}

   To assess the feasibility of this model, we first consider a small simulation restricting ourselves to a single year and age group. We set $\alpha^I =-6.5$,  $\alpha^{MI} =-1.0$ and the standard deviation of the spatial random effects to be $0.5$ (with no iid country effect - effectively setting the mixing term for the BYM-2 to be 1.0). We use the form of the data (country types and populations) from 2008 and age-group 50-54. This resulted in 14, 2, 21, and 3 type I, II, III, and IV countries respectively. Conditional on these true parameters, country data types, and the observed populations, data was simulated from the model outlined in Section~\ref{sec:mod}. Results for the fits are summarized in the appendix in Figure~\ref{fig:simres} and indicate that overall the model is performing well, even in the challenging situation with over half of the countries set to type III or IV. Notably, the precision for the MI country random effects has been estimated to be too high and this has resulted in some over-shrinkage of the MI estimates.

  \subsection{Results}
\begin{figure}
      \includegraphics[width=\linewidth]{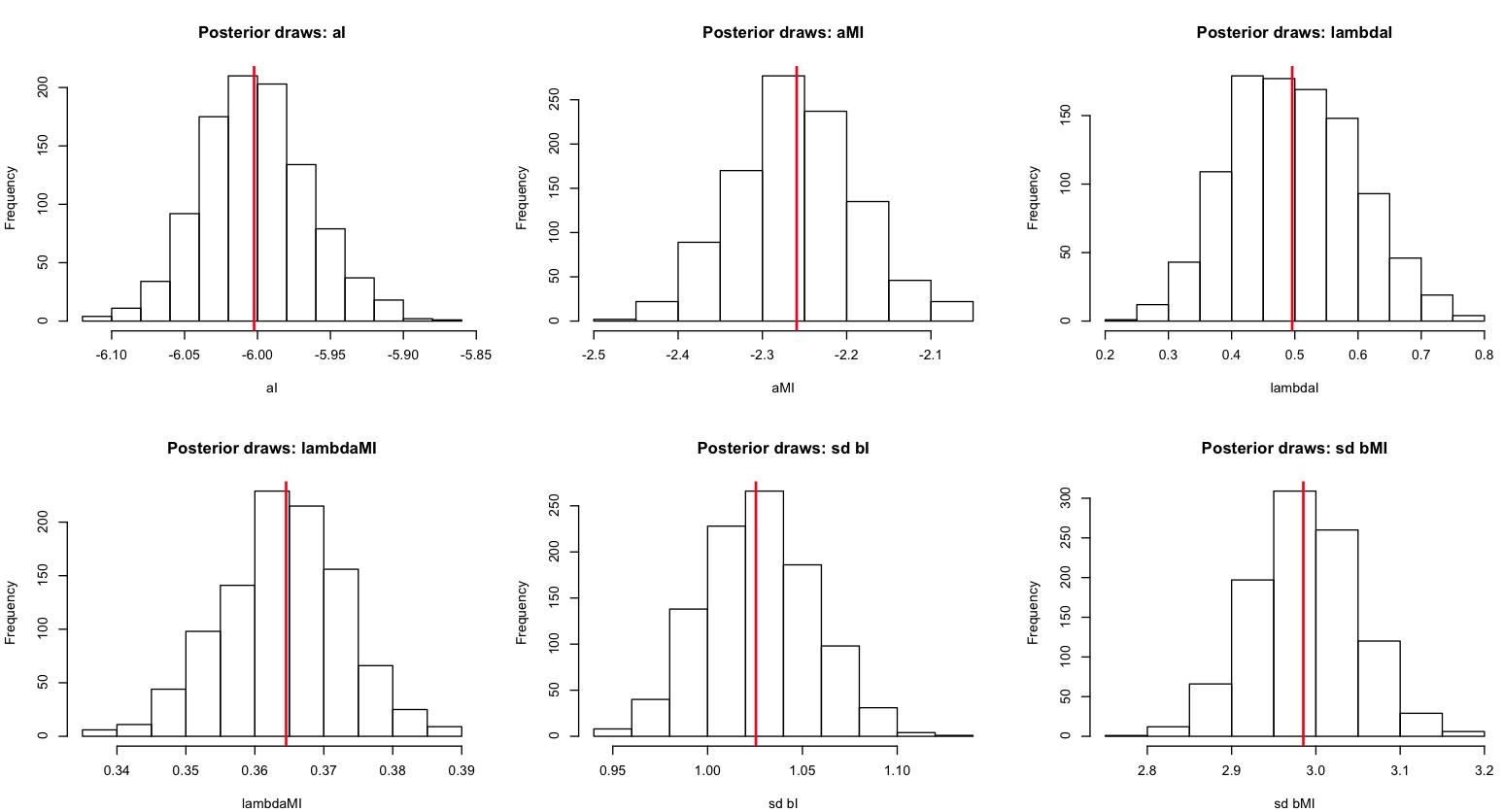} \singlespace \caption{\footnotesize{Top row: histograms of the incidence rate intercept $aI$, mortality-incidence ratio intercept $aMI$, and the BYM2 incidence mixture parameter $\lambda_I$. Top row: histograms of the BYM2 mortality-incidence mixture parameter $\lambda_{MI}$, and the standard deviations ($\tau_*^{-1/2}$) of the two BYM2 processes.}}
    \label{fig:fesum}
  \end{figure}

  The model can be run quite quickly in R and once it has finished fitting, 1000 multivariate normal draws are taken from the joint posterior of all parameters. These draws are then summarized and some relevant quantities are shown in Figures~\ref{fig:euro.bc.maps} and \ref{fig:fesum}.

  \begin{figure}
  \begin{centering}
    \includegraphics[width=.9\linewidth]{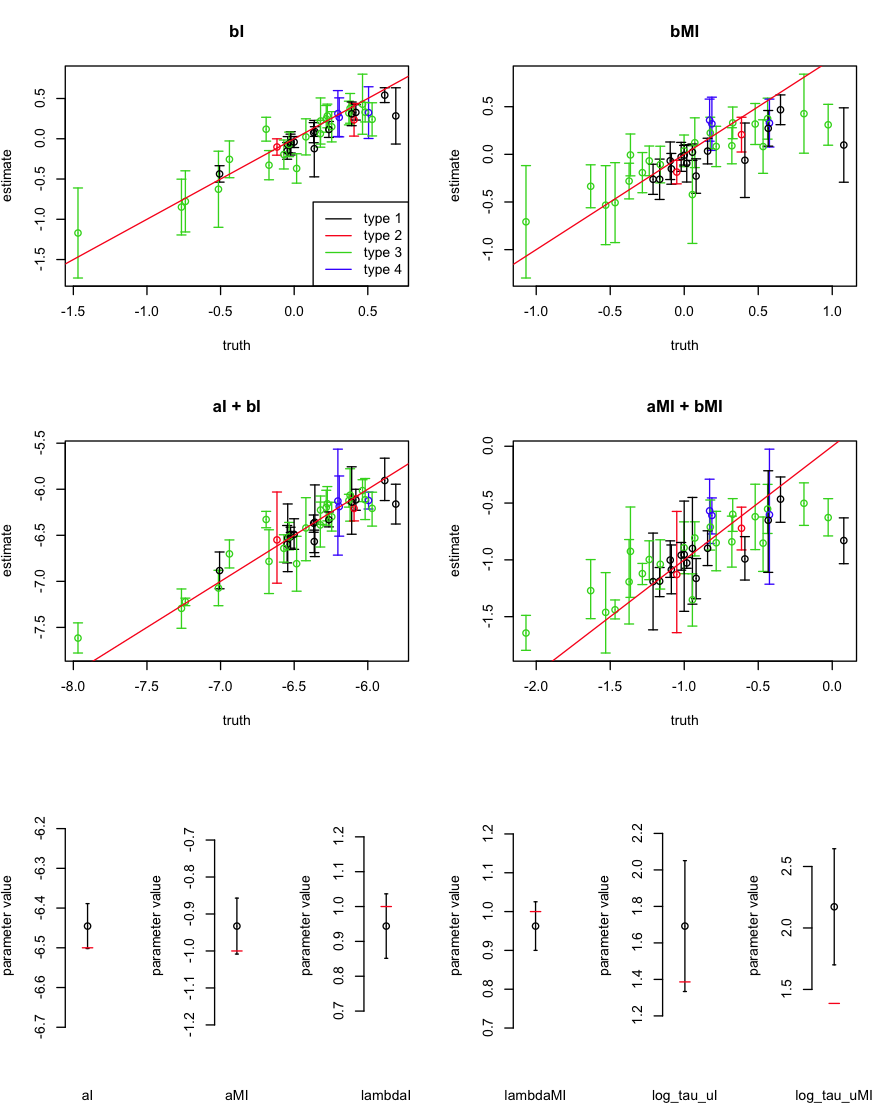} \singlespace \caption{\footnotesize{Fitted results from one run of the simulation study. The top row shows the simulated country random effects from the BYM2 specification plotted against the associated fitted median and 95\% credible intervals. The second row shows the simulated overall country effect (intercept plus country random effects) plotted against the associated fitted median and 95\% credible intervals. In the third row we have the true values for each of the fixed and hyperparameters (shown with the red line) plotted against the associated fitted median and 95\% credible intervals. }}
    \label{fig:simres}
    \end{centering}
  \end{figure}

\clearpage

\section{Example spatial model code}
\label{app:code}

The complete code used in this study is available online at:\\ \href{https://faculty.washington.edu/jonno/software.html}{https://faculty.washington.edu/jonno/software.html}. The following two sections provide succinct yet complete examples of continuous and discrete spatial model inference in both TMB and R-INLA. We start by simulating data on the unit square, and then use R-INLA functions to make the SPDE objects for fitting, some of which are re-used by TMB.

\subsection{Simulating data and generating SPDE objects}
\label{app:code.prep}

\begin{lstlisting}[language=R]

## install missing packages
pkgs <- c('data.table', 'ggplot2', 'RColorBrewer', 'RandomFields',
          'raster', 'TMB', 'viridis')
new.packages <- pkgs[!(pkgs %in% installed.packages()[,"Package"])]
if(length(new.packages)) install.packages(new.packages)
if(!('INLA' %in% installed.packages()[, 'Package'])){
  ## INLA is not on CRAN
  install.packages("INLA",
                   repos=c(getOption("repos"),
                     INLA="https://inla.r-inla-download.org/R/stable"),
                   dep=TRUE)
}
# load packages
invisible(lapply(c(pkgs, 'INLA'), library, character.only = TRUE))

## setup continuous domain
set.seed(413206)
x <- seq(0, 10, length = 200)
grid.pts <- expand.grid(x, x)

## set up matern params, also set param priors to be used in modeling
sp.alpha <- 2
sp.kappa <- 0.5
sp.var   <- 0.5
gp.int  <- -2
# prior on spde parameters: c(a, b, c, d), where
# P(sp.range < a) = b
# P(sp.sigma > c) = d
matern.pri <- c(10, .95, 1., .05) ## a, b, c, d
# mean and sd for normal prior on fixed effects (alpha and betas)
alpha.pri <- c(0, 3) ## N(mean, sd)

## sample from matern RF on our grid
model <- RMmatern(nu    = sp.alpha - 1, ## from INLA book
                  scale = sqrt(2 * (sp.alpha - 1)) / sp.kappa,
                  var   = 1)
true.gp <- RFsimulate(model, x = x, y = x, n =1, spConform = FALSE)

## insert into a raster
gp.rast <- raster(nrows=length(x), ncols=length(x),
                  xmn=0, xmx=10, ymn=0, ymx=10,
                  vals=(true.gp + gp.int))

## define cluster locations and sample size at each
n.clust <- 500
clust.mean.ss <- 35
dat <- data.table(x = runif(n.clust, min = min(x), max = max(x)),
                  y = runif(n.clust, min = min(x), max = max(x)),
                  n = rpois(n.clust, clust.mean.ss)
                  )

## extract value of raster at cluster locs and logit transform
##   to binom probs
dat[, latent.truth := raster::extract(x = gp.rast, y = cbind(x, y))]
dat[, p.truth := plogis(latent.truth)]

## sample binomial data
dat[, obs := rbinom(n = .N, size = n, p = p.truth)]

## make SPDE triangulation mesh over our domain
mesh.s <- inla.mesh.2d(loc.domain = grid.pts,
                       max.e = c(0.25, 5))
## check number of vertices
mesh.s[['n']]

## plot true latent field, the observed/empirical binom probs at
##   cluster locs, and the mesh
par(mfrow = c(3, 1))
plot(gp.rast, maxpixels = length(x) ^ 2,
     xlim = range(x), ylim = range(x), main = 'latent truth')
fields::quilt.plot(dat[, x], dat[, y], dat[, obs] / dat[, n],
                   main = 'empirical binom probs')
plot(mesh.s)
polygon(x = c(0, 0, 10, 10, 0), y = c(0, 10, 10, 0, 0),
col = NA, border = 2, lwd = 5)

## make the SPDE objects (including prec components)
spde <- inla.spde2.pcmatern(mesh = mesh.s, alpha = 2,
                            prior.range = matern.pri[1:2],
                            prior.sigma = matern.pri[3:4])

## make projector matrices to:
## 1) project data to mesh
## 2) project mesh to raster grid
A.proj <- inla.spde.make.A(mesh = mesh.s,
                           loc = dat[, as.matrix(x, y)])
A.pred <- inla.spde.make.A(mesh = mesh.s,
                           loc = as.matrix(grid.pts),
                           group = 1)
\end{lstlisting}

  \begin{figure}
    \begin{centering}
      \vspace{-10em}
    \includegraphics[width=.73\linewidth]{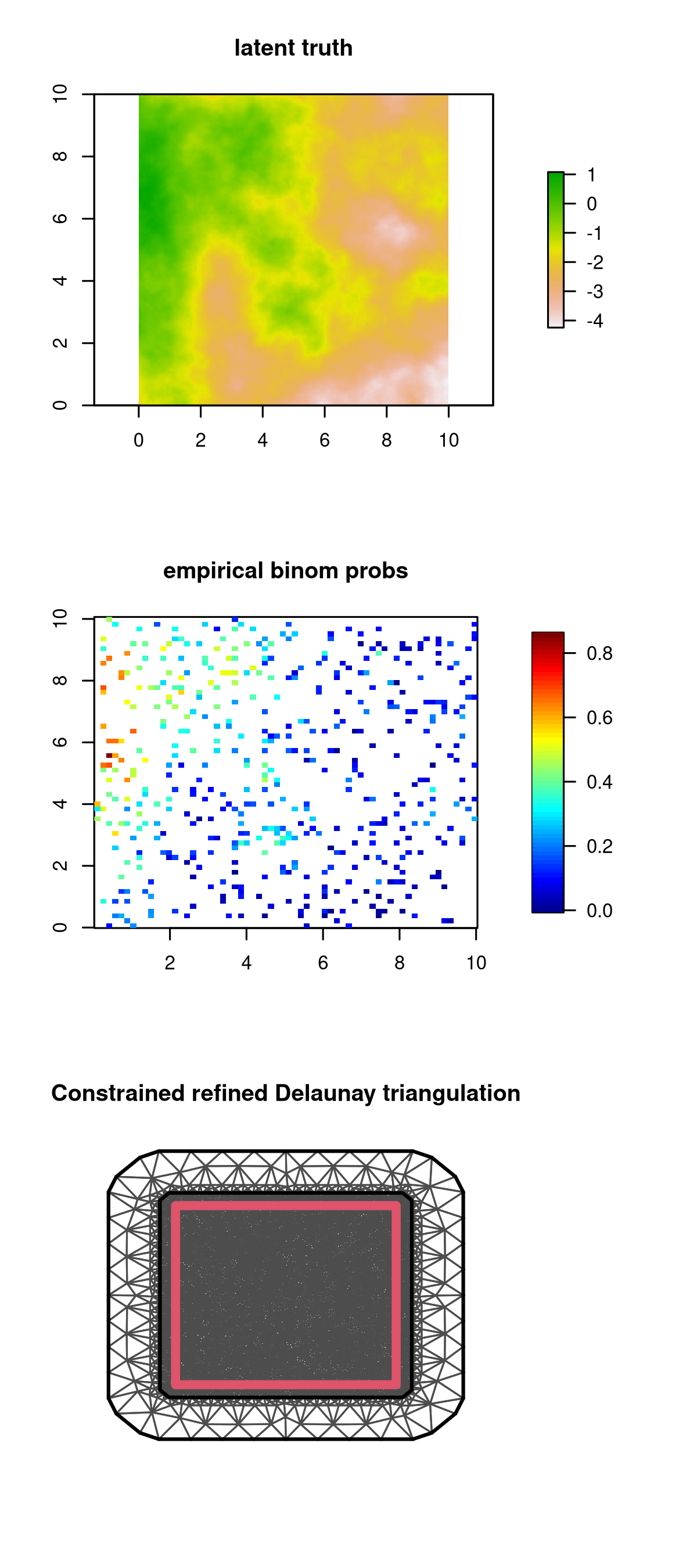} \singlespace \caption{\footnotesize{Simulated GP on 10$\times$10 grid, simulated data locations and empirical probabilities, and SPDE mesh from the preparation portion of the code example.}}
    \label{fig:example_prep}
    \end{centering}
  \end{figure}

\clearpage

\subsection{Continuous GP modeling with SPDE in R-INLA}
\label{app:code.inla}

\begin{lstlisting}[language=R]
## prep inputs for INLA
design_matrix <- data.frame(int = rep(1, nrow(dat)))
stack.obs <- inla.stack(tag='est',
                        data=list(Y = dat$obs, ## response
                                  N = dat$n), ## binom trials
                        A=list(A.proj, ## A.proj for space
                               1),     ## 1 for design.mat
                        effects=list(
                          space = 1:mesh.s[['n']],
                          design_matrix))

## define the INLA model
formula <- formula(Y ~ -1 + int + f(space, model = spde))

## run INLA
i.fit <- inla(formula,
              data = inla.stack.data(stack.obs),
              control.predictor = list(A = inla.stack.A(stack.obs),
                                       compute = FALSE),
              control.fixed = list(expand.factor.strategy = 'inla',
                               prec = list(default = 1 / alpha.pri[2] ^ 2)),
              control.inla = list(strategy = 'simplified.laplace',
                                  int.strategy = 'ccd'),
              control.compute=list(config = TRUE),
              family = 'binomial',
              Ntrials = N,
              verbose = FALSE,
              keep = FALSE)

## take draws from inla
i.draws <- inla.posterior.sample(n = 500, i.fit,
                                 use.improved.mean = TRUE,
                                 skew.corr = TRUE)

## summarize the draws
par_names <- rownames(i.draws[[1]][['latent']])
s_idx <- grep('^space.*', par_names)
a_idx <- which(!c(1:length(par_names)) %in%
                  grep('^space.*|Predictor|clust.id', par_names))

# project from mesh to raster, add intercept
pred_s <- sapply(i.draws, function (x) x[['latent']][s_idx])
pred_inla <- as.matrix(A.pred %*% pred_s)
alpha_inla_draws <- sapply(i.draws, function (x) x[['latent']][a_idx])
pred_inla <- sweep(pred_inla, 2, alpha_inla_draws, '+')


## find the median and sd across draws, as well as 90% intervals
summ_inla <- cbind(median = (apply(pred_inla, 1, median)),
                  sd     = (apply(pred_inla, 1, sd)),
                  lower = (apply(pred_inla, 1, quantile, .05)),
                  upper = (apply(pred_inla, 1, quantile, .95)))

## make summary rasters
ras_med_inla <- ras_sdv_inla <- ras_lower_inla <-
  ras_upper_inla <- ras_inInt_inla <- gp.rast
values(ras_med_inla)   <- summ_inla[, 1]
values(ras_sdv_inla)   <- summ_inla[, 2]
values(ras_lower_inla) <- summ_inla[, 3]
values(ras_upper_inla) <- summ_inla[, 4]
values(ras_inInt_inla) <- 0
ras_inInt_inla[gp.rast < ras_lower_inla | ras_upper_inla < gp.rast] <- 1

## plot truth, pixels falling within/without the 90% interval,
##  post. median, and post sd

# set the range for the truth and median
rast.zrange <- range(c(values(gp.rast), values(ras_med_inla)), na.rm = T)

# plot
par(mfrow = c(2, 2))
plot(gp.rast, main = 'Truth', zlim = rast.zrange, col = (viridis(100)))
points(dat[, .(x, y)])
plot(ras_inInt_inla, main = 'Pixels where 90% CIs did not cover Truth')
points(dat[, .(x, y)])
plot(ras_med_inla, main = 'INLA Posterior Median',
     zlim = rast.zrange, col = (viridis(100)))
points(dat[, .(x, y)])
plot(ras_sdv_inla, main = 'INLA Posterior Standard Deviation')
points(dat[, .(x, y)])
\end{lstlisting}

  \begin{figure}
  \begin{centering}
    \includegraphics[width=\linewidth]{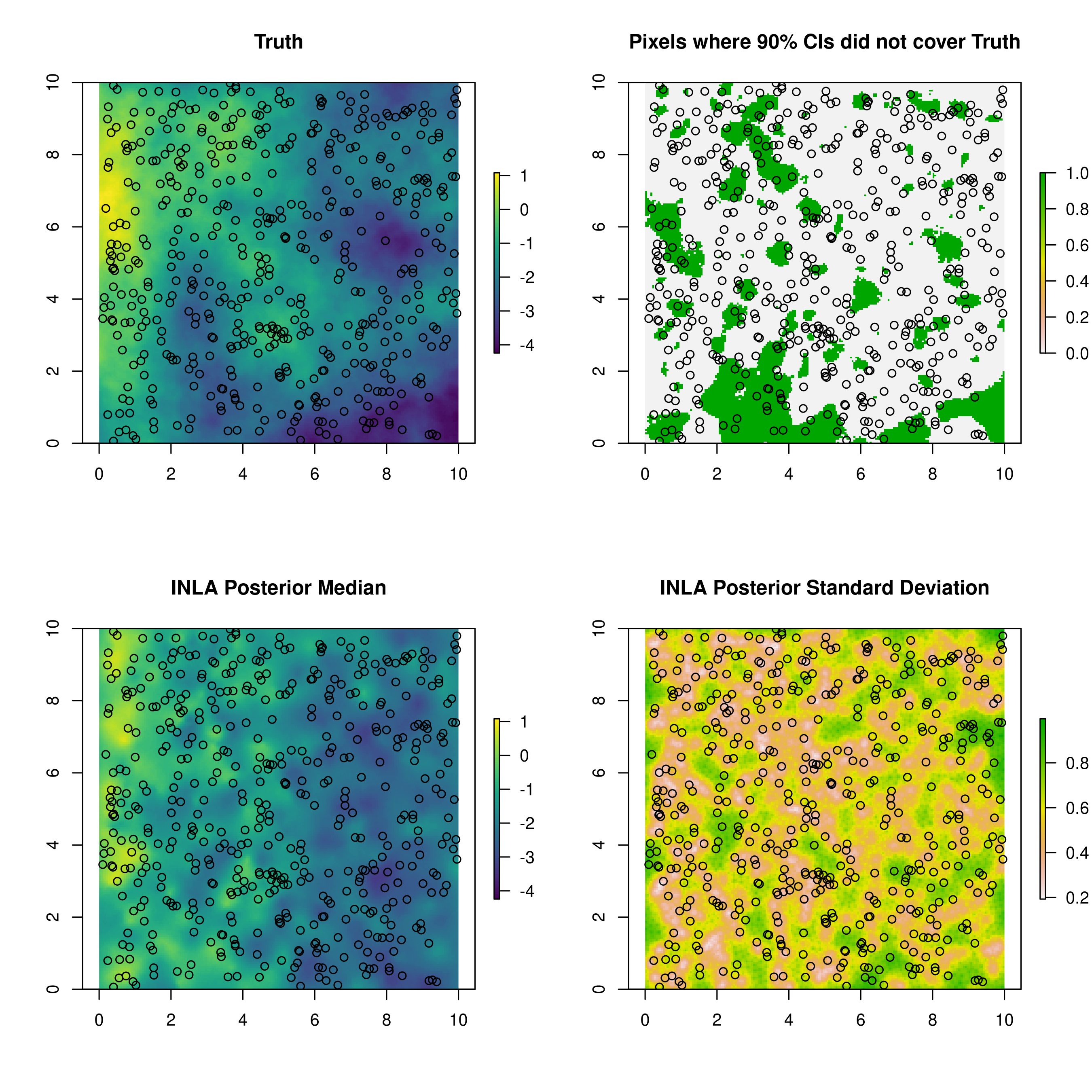} \singlespace \caption{\footnotesize{Simulated GP on 10$\times$10 grid, posterior median and standard deviations summarized from 500 draws from the approximate INLA posterior distribution, and a binary plot indicating which pixels fell outside the 90\% credible interval. Points indicate cluster locations.}}
    \label{fig:example_inla}
    \end{centering}
  \end{figure}

\clearpage

\subsection{Continuous GP modeling with SPDE in TMB}
\label{app:code.tmb}

\begin{lstlisting}[language=R]
## define the TMB model using c++ template code
##  this is usually done in a separate file,
##  but it can be done all from within 1 R script

tmb_spde <-
  "// include libraries
#include <TMB.hpp>
#include <Eigen/Sparse>
#include <vector>
using namespace density;
using Eigen::SparseMatrix;

// helper function for detecting NAs in the data supplied from R
template<class Type>
bool isNA(Type x){
  return R_IsNA(asDouble(x));
}

// helper function to make sparse SPDE precision matrix
// Inputs:
//    logkappa: log(kappa) parameter value
//    logtau: log(tau) parameter value
//    M0, M1, M2: these sparse matrices are output from:
//     R::INLA::inla.spde2.matern()$param.inla$M*
template<class Type>
SparseMatrix<Type> spde_Q(Type logkappa, Type logtau, SparseMatrix<Type> M0,
                          SparseMatrix<Type> M1, SparseMatrix<Type> M2) {
    SparseMatrix<Type> Q;
    Type kappa2 = exp(2. * logkappa);
    Type kappa4 = kappa2*kappa2;
    Q = pow(exp(logtau), 2.)  * (kappa4*M0 + Type(2.0)*kappa2*M1 + M2);
    return Q;
}

// helper function to use the same penalized complexity prior on
//  matern params that is used in INLA

template<class Type>
Type dPCPriSPDE(Type logtau, Type logkappa,
                Type matern_par_a, Type matern_par_b,
                Type matern_par_c, Type matern_par_d,
                //vector<Type> matern_pri(4),
                int give_log=0)
{

  // matern_pri = c(a, b, c, d): P(range < a) = b; P(sigma > c) = d

  Type penalty; // prior contribution to jnll

  Type d = 2.;  // dimension
  Type lambda1 = -log(matern_par_b) * pow(matern_par_a, d/2.);
  Type lambda2 = -log(matern_par_d) / matern_par_c;
  Type range   = sqrt(8.0) / exp(logkappa);
  Type sigma   = 1.0 / sqrt(4.0 * 3.14159265359 * exp(2.0 * logtau) *
                            exp(2.0 * logkappa));

  penalty = (-d/2. - 1.) * log(range) - lambda1 * pow(range, -d/2.) -
               lambda2 * sigma;
  // Note: (rho, sigma) --> (x=log kappa, y=log tau) -->
  //  transforms: rho = sqrt(8)/e^x & sigma = 1/(sqrt(4pi)*e^x*e^y)
  //  --> Jacobian: |J| propto e^(-y -2x)
  Type jacobian = - logtau - 2.0*logkappa;
  penalty += jacobian;

  if(give_log)return penalty; else return exp(penalty);
}

///////////////////////////
// the main function     //
// to calculate the jnll //
///////////////////////////
template<class Type>
Type objective_function<Type>::operator() ()
{

  // ~~~~~~~~~------------------------------------------------------~~
  // FIRST, we define params/values/data that will be passed in from R
  // ~~~~~~~~~~~------------------------------------------------------

  // normalization flag - used for speed-up
  DATA_INTEGER( flag ); // flag == 0 => no data contribution added to jnll

  // Indices
  DATA_INTEGER( num_i );   // Number of data points in space
  DATA_INTEGER( num_s );   // Number of mesh points in space mesh

  // Data (all except for X_ij is a vector of length num_i)
  DATA_VECTOR( y_i );   // obs per binomial experiment at point i (clust)
  DATA_VECTOR( n_i );   // Trials per cluster
  DATA_MATRIX( X_alpha );  // 'design matrix' for just int

  // SPDE objects
  DATA_SPARSE_MATRIX( M0 );
  DATA_SPARSE_MATRIX( M1 );
  DATA_SPARSE_MATRIX( M2 );
  DATA_SPARSE_MATRIX( Aproj );

  // Options
  DATA_VECTOR( options );
  // options[0] == 1 : use normalization trick
  // options[1] == 1 : adreport transformed params

  // Prior specifications
  DATA_VECTOR( alpha_pri );
  DATA_VECTOR( matern_pri);
  // matern_pri = c(a, b, c, d): P(range < a) = b; P(sigma > c) = d
  Type matern_par_a = matern_pri[0]; // range limit:    rho0
  Type matern_par_b = matern_pri[1]; // range prob:     alpha_rho
  Type matern_par_c = matern_pri[2]; // field sd limit: sigma0
  Type matern_par_d = matern_pri[3]; // field sd prob:  alpha_sigma

  // Fixed effects
  PARAMETER( alpha ); // Intercept
  // Log of INLA tau param (precision of space covariance matrix)
  PARAMETER( log_tau );
  // Log of INLA kappa (related to spatial correlation and range)
  PARAMETER( log_kappa );

  // Random effects for each spatial mesh vertex
  PARAMETER_VECTOR( Epsilon_s );

  // ~~~~~~~~~------------------------------------------------~~
  // SECOND, we define all other objects that we need internally
  // ~~~~~~~~~------------------------------------------------~~

  // objective function -- joint negative log-likelihood
  Type jnll = 0;

  // Make spatial precision matrix
  SparseMatrix<Type> Q_ss = spde_Q(log_kappa, log_tau, M0, M1, M2);

  // Transform some of our parameters
  Type sp_range = sqrt(8.0) / exp(log_kappa);
  Type sp_sigma = 1.0 / sqrt(4.0 * 3.14159265359 *
                  exp(2.0 * log_tau) * exp(2.0 * log_kappa));

  // Define objects for derived values
  vector<Type> fe_i(num_i); // main effect: alpha
  // Logit estimated prob for each cluster i
  vector<Type> latent_field_i(num_i);
  // value of gmrf at data points
  vector<Type> projepsilon_i(num_i);

  // fixed effects is just alpha in this example
  fe_i = X_alpha * Type(alpha); // initialize

  // Project GP approx from mesh points to data points
  projepsilon_i = Aproj * Epsilon_s.matrix();

  // ~~~~~~~~~------------------------------------------------~~-
  // THIRD, we calculate the contribution to the likelihood from:
  // 1) priors
  // 2) GP field
  // 3) data
  // ~~~~~~~~~------------------------------------------------~~-

  /////////
  // (1) //
  /////////
  // the random effects. we do this first so to do the
  //   normalization outside of every optimization step
  // NOTE: likelihoods from namespace 'density' already return NEGATIVE
  //       log-liks so we add other likelihoods return positive log-liks
  if(options[0] == 1){
    // then we are not calculating the normalizing constant in the inner opt
    // that norm constant means taking an expensive determinant of Q_ss
    jnll += GMRF(Q_ss, false)(Epsilon_s);
    // return without data ll contrib to avoid unneccesary log(det(Q)) calcs
    if (flag == 0) return jnll;
  }else{
    jnll += GMRF(Q_ss)(Epsilon_s);
  }

  /////////
  // (2) //
  /////////
  // Prior contributions to joint likelihood (if options[1]==1)

  // add in priors for spde gp
  jnll -= dPCPriSPDE(log_tau, log_kappa,
                     matern_par_a, matern_par_b, matern_par_c, matern_par_d,
                     true);

  // prior for intercept
  jnll -= dnorm(alpha, alpha_pri[0], alpha_pri[1], true); // N(mean, sd)

  /////////
  // (3) //
  /////////
  // jnll contribution from each datapoint i

  for (int i = 0; i < num_i; i++){

    // latent field estimate at each obs
    latent_field_i(i) = fe_i(i) + projepsilon_i(i);

    // and add data contribution to jnll
    if(!isNA(y_i(i))){

     // Uses the dbinom_robust function, which takes the logit probability
      	jnll -= dbinom_robust( y_i(i), n_i(i), latent_field_i(i), true );

    } // !isNA

  } // for( i )


  // ~~~~~~~~~~~
  // ADREPORT: used to return estimates and cov for transforms?
  // ~~~~~~~~~~~
  if(options[1]==1){
    ADREPORT(sp_range);
    ADREPORT(sp_sigma);
  }

  return jnll;

}"

## write model to file, compile, and load it into R
dir.create('TMB_spde_example')
write(tmb_spde,file="TMB_spde_example/tmb_spde.cpp")
compile( "TMB_spde_example/tmb_spde.cpp")
dyn.load( dynlib("TMB_spde_example/tmb_spde") )

## prep inputs for TMB
data_full <- list(num_i = nrow(dat),  # Total number of observations
                  num_s = mesh.s[['n']], # num. of vertices in SPDE mesh
                  y_i   = dat[, obs],# num. of pos. obs in the cluster
                  n_i   = dat[, n],  # num. of exposures in the cluster
                  X_alpha  = matrix(1, nrow = nrow(dat), ncol = 1),# des.mat
                  M0    = spde[['param.inla']][['M0']], # SPDE sparse matrix
                  M1    = spde[['param.inla']][['M1']], # SPDE sparse matrix
                  M2    = spde[['param.inla']][['M2']], # SPDE sparse matrix
                  Aproj = A.proj,             # Projection matrix
                  options = c(1, ## if 1, use normalization trick
                              1), ## if 1, run adreport
                  # normalization flag.
                  flag = 1,
                  alpha_pri = alpha.pri, ## normal
                  matern_pri = matern.pri
                  )

## Specify starting values for TMB params
tmb_params <- list(alpha = 0.0, # intercept
                   log_tau = 0, # Log inverse of tau (Epsilon)
                   log_kappa = 0, # Matern range parameter
                   Epsilon_s = rep(0, mesh.s[['n']]) # RE on mesh vertices
                   )

## make a list of things that are random effects
rand_effs <- c('Epsilon_s')

## make the autodiff generated liklihood func & gradient
obj <- MakeADFun(data=data_full,
                 parameters=tmb_params,
                 random=rand_effs,
                 hessian=TRUE,
                 DLL='tmb_spde')

## we can normalize the GMRF outside of the nested optimization,
## avoiding unnecessary and expensive cholesky operations.
obj <- normalize(obj, flag="flag", value = 0)

## run TMB
opt0 <- nlminb(start       =    obj[['par']],
               objective   =    obj[['fn']],
               gradient    =    obj[['gr']],
               lower = rep(-10, length(obj[['par']])),
               upper = rep( 10, length(obj[['par']])),
               control     =    list(trace=1))

## Get standard errors
SD0 <- TMB::sdreport(obj, getJointPrecision=TRUE,
                     bias.correct = TRUE,
                     bias.correct.control = list(sd = TRUE))
## summary(SD0, 'report')

## take samples from fitted model
mu <- c(SD0$par.fixed,SD0$par.random)

## simulate draws
rmvnorm_prec <- function(mu, chol_prec, n.sims) {
  z <- matrix(rnorm(length(mu) * n.sims), ncol=n.sims)
  L <- chol_prec #Cholesky(prec, super=TRUE)
  z <- Matrix::solve(L, z, system = "Lt") ## z = Lt^-1 %*% z
  z <- Matrix::solve(L, z, system = "Pt") ## z = Pt    %*% z
  z <- as.matrix(z)
  mu + z
}

L <- Cholesky(SD0[['jointPrecision']], super = T)
t.draws <- rmvnorm_prec(mu = mu , chol_prec = L, n.sims = 500)

## summarize the draws
parnames <- c(names(SD0[['par.fixed']]), names(SD0[['par.random']]))
epsilon_tmb_draws  <- t.draws[parnames == 'Epsilon_s',]
alpha_tmb_draws    <- matrix(t.draws[parnames == 'alpha',], nrow = 1)

# project from mesh to raster, add intercept
pred_tmb <- as.matrix(A.pred %*% epsilon_tmb_draws)
pred_tmb <- sweep(pred_tmb, 2, alpha_tmb_draws, '+')

## find the median and sd across draws, as well as 90% intervals
summ_tmb <- cbind(median = (apply(pred_tmb, 1, median)),
                  sd     = (apply(pred_tmb, 1, sd)),
                  lower = (apply(pred_tmb, 1, quantile, .05)),
                  upper = (apply(pred_tmb, 1, quantile, .95)))

## make summary rasters
ras_med_tmb <- ras_sdv_tmb <- ras_lower_tmb <-
  ras_upper_tmb <- ras_inInt_tmb <- gp.rast
values(ras_med_tmb)   <- summ_tmb[, 1]
values(ras_sdv_tmb)   <- summ_tmb[, 2]
values(ras_lower_tmb) <- summ_tmb[, 3]
values(ras_upper_tmb) <- summ_tmb[, 4]
values(ras_inInt_tmb) <- 0
ras_inInt_tmb[gp.rast < ras_lower_tmb | ras_upper_tmb < gp.rast] <- 1

## plot truth, pixels falling within/without the 90% interval,
##  post. median, and post sd

# set the range for the truth and median
rast.zrange <- range(c(values(gp.rast), values(ras_med_tmb)), na.rm = T)

# plot tmb
png(file='figures/example_tmb.png', width=9, height=9, units='in', res=300)
par(mfrow = c(2, 2))
plot(gp.rast, main = 'Truth', zlim = rast.zrange, col = (viridis(100)))
points(dat[, .(x, y)])
plot(ras_inInt_tmb, main = 'Pixels where 90% CIs did not cover Truth')
points(dat[, .(x, y)])
plot(ras_med_tmb, main = 'TMB Posterior Median',
     zlim = rast.zrange, col = (viridis(100)))
points(dat[, .(x, y)])
plot(ras_sdv_tmb, main='TMB Posterior Standard Deviation')
points(dat[, .(x, y)])
dev.off()


## compare INLA and TMB meds and stdevs
med.zrange <- range(c(values(ras_med_tmb), values(ras_med_inla)), na.rm = T)
sdv.zrange <- range(c(values(ras_sdv_tmb), values(ras_sdv_inla)), na.rm = T)

png(file='figures/example_tmb_v_inla.png', width=9, height=9, units='in', res=300)
par(mfrow = c(2, 2))
plot(ras_med_inla, main = 'INLA Posterior Median',
     zlim = med.zrange, col = (viridis(100)))
points(dat[, .(x, y)])
plot(ras_sdv_inla, main = 'INLA Posterior Standard Deviation',
     zlim = sdv.zrange)
points(dat[, .(x, y)])
plot(ras_med_tmb, main = 'TMB Posterior Median',
     zlim = med.zrange, col = (viridis(100)))
points(dat[, .(x, y)])
plot(ras_sdv_tmb, main = 'TMB Posterior Standard Deviation',
     zlim = sdv.zrange)
points(dat[, .(x, y)])
dev.off()
\end{lstlisting}

  \begin{figure}
  \begin{centering}
    \includegraphics[width=\linewidth]{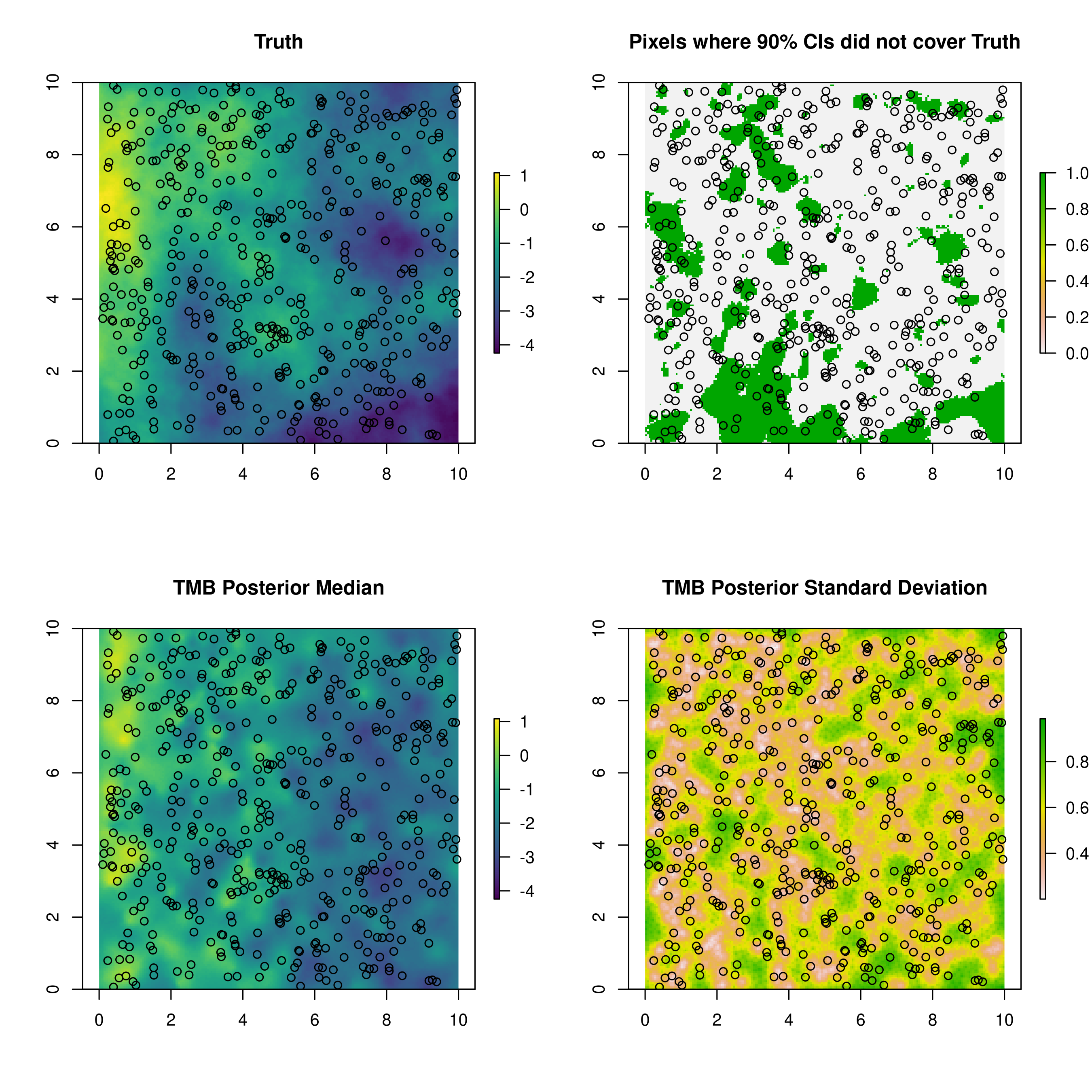} \singlespace \caption{\footnotesize{Simulated GP on 10$\times$10 grid, posterior median and standard deviations summarized from 500 draws from the approximate TMB posterior distribution, and a binary plot indicating which pixels fell outside the 90\% credible interval. Points indicate cluster locations.}}
    \label{fig:example_tmb}
    \end{centering}
  \end{figure}

  \begin{figure}
  \begin{centering}
    \includegraphics[width=\linewidth]{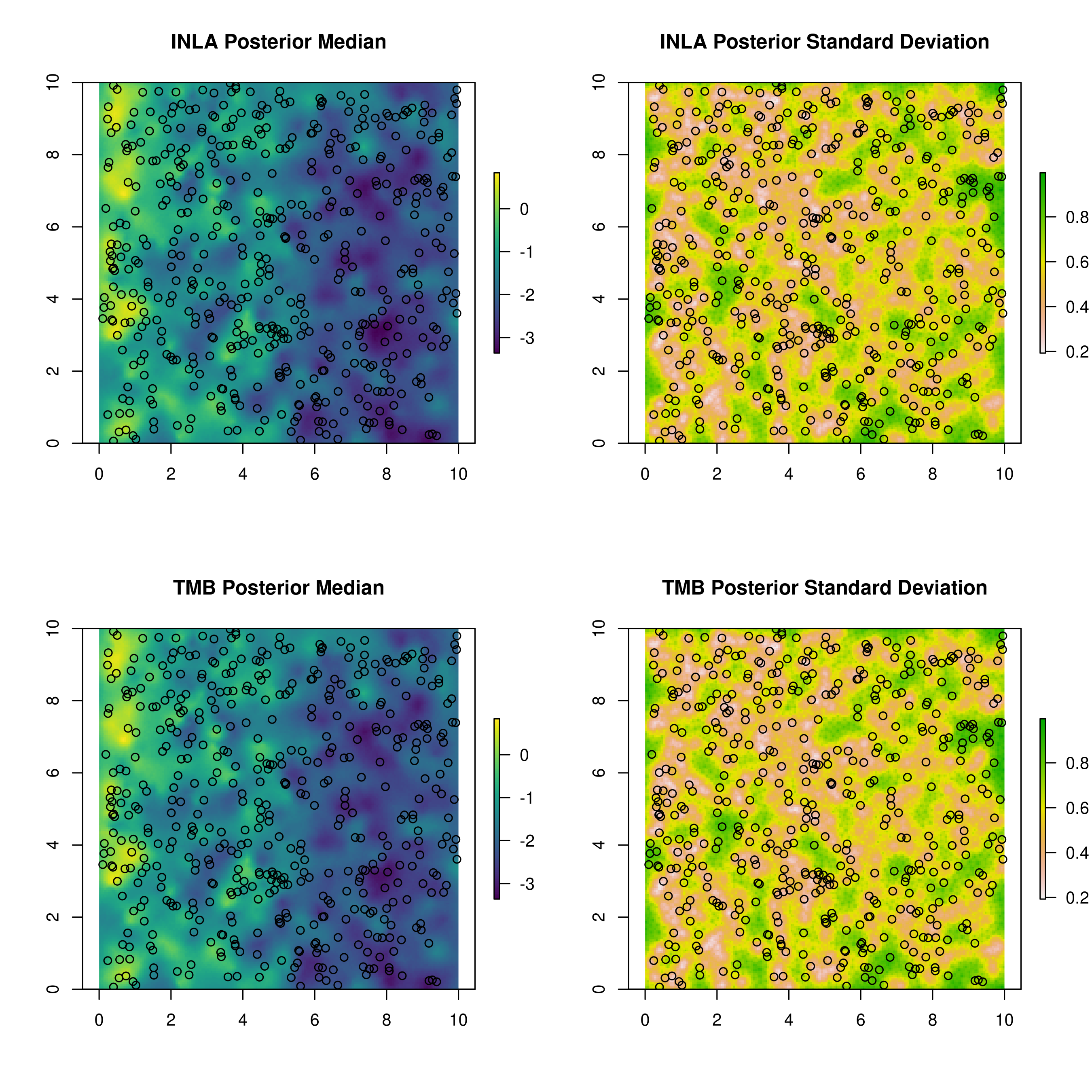} \singlespace \caption{\footnotesize{Comparison of the posterior median and standard deviations from the example R-INLA and TMB code. Points indicate cluster locations.}}
    \label{fig:example_tmb_inla}
    \end{centering}
  \end{figure}

\clearpage

\section{Software and hardware details}
\label{app:software}
The large scale continuous simulations of Section \ref{sec:sim.cont} were run on a large computing cluster containing a variety of hardware types. The jobs were randomly placed onto different machines as space became available and the effects of the various hardware were averaged over in producing all results, including the serial timing results in Figure \ref{fig:time}. A singularity image was used to ensure consistent software versions across the nodes. Specifically, these simulations were run on:

\begin{itemize}
\item R 3.6.1,
\item R-INLA 20.01.29.9000,
\item TMB 1.7.16.
\end{itemize}

All other work was performed on a laptop with an Intel Core i7-8550U CPU (4 cores, 8 threads @ 1.8GHz) and 16Gb of RAM. This machine used the following software versions:

\begin{itemize}
\item R 4.0.4,
\item R-INLA 21.02.23, with PARDISO solver enabled,
\item TMB 1.7.18, with METIS reordering enabled.
\end{itemize}

\end{document}